\documentclass[english,aps, prb, superscriptaddress, twocolumn]{revtex4}
\usepackage[T1]{fontenc}
\usepackage[latin9]{inputenc}
\usepackage{color}
\usepackage{babel}
\usepackage{units}
\usepackage{mathrsfs}
\usepackage{bm}
\usepackage{amsmath}
\usepackage{amssymb}
\usepackage{graphicx}
\usepackage{esint}
\usepackage[unicode=true,pdfusetitle,
 bookmarks=true,bookmarksnumbered=false,bookmarksopen=false,
 breaklinks=false,pdfborder={0 0 0},backref=false,colorlinks=true]
 {hyperref}

\makeatletter
\@ifundefined{textcolor}{}
{%
 \definecolor{BLACK}{gray}{0}
 \definecolor{WHITE}{gray}{1}
 \definecolor{RED}{rgb}{1,0,0}
 \definecolor{GREEN}{rgb}{0,1,0}
 \definecolor{BLUE}{rgb}{0,0,1}
 \definecolor{CYAN}{cmyk}{1,0,0,0}
 \definecolor{MAGENTA}{cmyk}{0,1,0,0}
 \definecolor{YELLOW}{cmyk}{0,0,1,0}
}

\makeatother

\begin{document}

\title{Mott physics and spin fluctuations: a functional viewpoint}

\author{Thomas Ayral}

\email{thomas.ayral@polytechnique.edu}

\affiliation{Centre de Physique Théorique, Ecole Polytechnique, CNRS-UMR7644,
91128 Palaiseau, France}

\affiliation{Institut de Physique Théorique (IPhT), CEA, CNRS, URA 2306, 91191
Gif-sur-Yvette, France}

\author{Olivier Parcollet}

\affiliation{Institut de Physique Théorique (IPhT), CEA, CNRS, URA 2306, 91191
Gif-sur-Yvette, France}
\begin{abstract}
We present a formalism for strongly correlated systems with fermions
coupled to bosonic modes. We construct the three-particle irreducible
functional $\mathcal{K}$ by successive Legendre transformations of
the free energy of the system. We derive a closed set of equations
for the fermionic and bosonic self-energies for a given $\mathcal{K}$.
We then introduce a local approximation for $\mathcal{K}$, which
extends the idea of dynamical mean field theory (DMFT) approaches
from two- to three-particle irreducibility. This approximation entails
the locality of the three-leg electron-boson vertex $\Lambda(i\omega,i\Omega)$,
which is self-consistently computed using a quantum impurity model
with dynamical charge and spin interactions. This local vertex is
used to construct frequency- and momentum-dependent electronic self-energies
and polarizations. By construction, the method interpolates between
the spin-fluctuation or GW approximations at weak coupling and the
atomic limit at strong coupling. We apply it to the Hubbard model
on two-dimensional square and triangular lattices. We complement the
results of Ref. \onlinecite{Ayral2015} by (i) showing that, at half-filling,
as DMFT, the method describes the Fermi-liquid metallic state and
the Mott insulator, separated by a first-order interacting-driven
Mott transition at low temperatures, (ii) investigating the influence
of frustration and (iii) discussing the influence of the bosonic decoupling
channel.
\end{abstract}
\maketitle

\section{Introduction}

Systems with strong Coulomb correlations such as high-temperature
superconductors pose a difficult challenge to condensed-matter theory.

One class of theoretical approaches to this problem emphasizes long-ranged
bosonic fluctuations e.g. close to a quantum critical point as the
main ingredient to account for the experimental facts. This is the
starting point of methods such as spin fluctuation theory \cite{Chubukov2002,Efetov2013,Wang2014,Metlitski2010,Onufrieva2009,Onufrieva2012},
two-particle self-consistent theory \cite{Vilk1994,Dare1996,Vilk1996,Vilk1997,Tremblay2012}
or the fluctuation-exchange approximation \cite{Bickers1989}. These
methods typically rely on an approximation of the electronic self-energy
as a one-loop diagram with a suitably constructed bosonic propagator,
neglecting vertex corrections.

Another class of approaches focuses instead, following Anderson \cite{Anderson1987},
on the fact that the parent compounds of high-temperature superconductors
are Mott insulators and assumes that Mott physics is essential to
describe the doped compounds. In recent years, dynamical mean-field
theory (DMFT) \cite{Georges1996} and its cluster extensions like
cellular DMFT \cite{Lichtenstein2000,Kotliar2001} or the dynamical
cluster approximation \cite{Hettler1998,Hettler1999,Maier2005a} have
emerged as powerful tools to capture the physics of doped Mott insulators.
Formally based on a local approximation of the two particle-irreducible
(2PI, or Luttinger-Ward) functional $\Phi$, they consist in self-consistently
mapping the extended lattice problem onto an impurity problem describing
the coupling of a small number ($N_{c}$) of correlated sites with
a noninteracting bath. The coarse-grained (short-ranged) self-energy
obtained by solving the impurity model is used as an approximation
of the lattice self-energy. 

Cluster DMFT methods have given valuable insights into the physics
of cuprate superconductors, in particular via the study of the Hubbard
model: they have allowed to map out the main features of its phase
diagram, to characterize $d$-wave superconductivity or investigate
its pseudogap phase with realistic values of the interaction strength
\cite{Kyung2009,Sordi2012,Civelli2008,Ferrero2010,Gull2013,Macridin2004,Maier2004,Maier2005,Maier2006,Gull2010,Yang2011,Macridin2008,Macridin2006,Jarrell2001,Bergeron2011,Kyung2004,Kyung2006a,Okamoto2010,Sordi2010,Sordi2012a,Civelli2005,Ferrero2008,Ferrero2009,Gull2009}.
Moreover, they come with a natural control parameter, the size $N_{c}$
of the impurity cluster, which can a priori be used to assess quantitatively
the accuracy of a given prediction as it interpolates between the
single-site DMFT solution ($N_{c}=1$) and the exact solution of the
lattice problem ($N_{c}=\infty$). Systematic comparisons with other
approaches, in certain parameter regimes, have started to appear.\cite{Leblanc2015}
Yet, cluster methods suffer from three major flaws, namely (i) they
cannot describe the effect of long-range bosonic fluctuations beyond
the size of the cluster, which can be experimentally relevant (e.g.
in neutron scattering\cite{Rossat-Mignod1991,Keimer1992,Bourges1996})
; (ii) the negative Monte-Carlo sign problem precludes the solution
of large impurity clusters, (iii) the cluster self-energy is still
quite coarse-grained (typically up to 8 or 16 patches in regimes of
interest \cite{Gull2009,Gull2010,Vidhyadhiraja2009,Macridin2008})
or relies on uncontrolled periodization or interpolation schemes (see
e.g. Ref. \onlinecite{Kotliar2001}). 

Recent attempts at incorporating some long-range correlations in the
DMFT framework include the GW+EDMFT method \cite{Sun2002,Biermann2003,Sun2008,Ayral2013,Biermann2014}
(which has been so far restricted to the charge channel only), the
dynamical vertex approximation (D$\Gamma$A\cite{Toschi2007,Katanin2009,Schafer2014,Valli2014})
and the dual fermion\cite{Rubtsov2008} and dual boson\cite{Rubtsov2011,VanLoon2014}
methods. D$\Gamma$A consists in approximating the fully irreducible
two-particle vertex by a local, four-leg vertex $\Gamma_{\mathrm{fir}}(i\omega,i\nu,i\Omega)$
computed with a DMFT impurity model. This idea has so far been restricted
to very simple systems\cite{Valli2014} (``parquet D$\Gamma$A'')
or further simplified so as to avoid the costly solution of the parquet
equations (``ladder D$\Gamma$A''\cite{Katanin2009}). This makes
$D\Gamma A$ either (for parquet D$\Gamma$A) difficult to implement
for realistic calculations, at least in the near future (the existing
``parquet solvers'' have so far been restricted to very small systems
only \cite{Yang2009,Tam2013}), or (for the ladder variant) dependent
on the choice of a given channel to solve the Bethe-Salphether equation.
In either case, (i) rigorous and efficient parametrizations of the
vertex functions only start to appear\cite{Li2015a}, (ii) two-particle
observables do not feed back on the impurity model in the current
implementations\cite{Held2014}, and (iii) most importantly, achieving
control like in cluster DMFT is very arduous: since both D$\Gamma$A
and the dual fermion method require the manipulation of functions
of three frequencies, their extension to cluster versions\cite{Yang2011c}
raises serious practical questions in terms of storage and speed. 

The TRILEX (TRiply-Irreducible Local EXpansion) method, introduced
in Ref. \onlinecite{Ayral2015}, is a simpler approach. It approximates
the three-leg electron-boson vertex by a local impurity vertex and
hence interpolates between the spin-fluctuation and the atomic limit.
This vertex evolves from a constant in the spin-fluctuation regime
to a strongly frequency-dependent function in the Mott regime. The
method yields frequency and momentum-dependent self-energies and polarizations
which, upon doping, lead to a momentum-differentiated Fermi surface
similar to the Fermi arcs seen in cuprates.

In this paper, we provide a complete derivation of the TRILEX method
as a local approximation of the three-particle irreducible functional
$\mathcal{K}$, as well as additional results of its application to
the Hubbard model (i) in the frustrated square lattice case and (ii)
on the triangular lattice.

In section \ref{sec:Formalism}, we derive the TRILEX formalism and
describe the corresponding algorithm. In section \ref{sec:impurity},
we elaborate on the solution of the impurity model. In section \ref{sec:Discussion},
we apply the method to the two-dimensional Hubbard model and discuss
the results. We give a few conclusions and perspectives in section
\ref{sec:Conclusions-and-perspectives}.

\section{Formalism\label{sec:Formalism}}

In this section, we derive the TRILEX formalism. Starting from a generic
electron-boson problem, we derive a functional scheme based on a Legendre
transformation with respect to not only the fermionic and bosonic
propagators, but also the fermion-boson coupling vertex (subsection
\ref{sub:Functional-derivation}). In subsection \ref{sub:From-an-electron-electron},
we show that electron-electron interaction problems can be studied
in the three-particle irreducible formalism by introducing an auxiliary
boson. Finally, in subsection \ref{sub:Local-approximation}, we introduce
the main approximation of the TRILEX scheme, which allows us to write
down the complete set of equations (subsection \ref{sub:Equations-scheme}). 

Our starting point is a generic mixed electron-boson action with a
Yukawa-type coupling between the bosonic and the fermionic field:
\begin{equation}
S_{\mathrm{eb}}=\bar{c}_{\bar{u}}\left[-G_{0}^{-1}\right]_{\bar{u}v}c_{v}+\frac{1}{2}\phi_{\alpha}\left[-W_{0}^{-1}\right]_{\alpha\beta}\phi_{\beta}+\lambda_{\bar{u}v\alpha}\bar{c}_{\bar{u}}c_{v}\phi_{\alpha}\label{eq:general_action_eb}
\end{equation}
$\bar{c}_{\bar{u}}$ and $c_{u}$ are Grassmann fields describing
fermionic degrees of freedom, while $\phi_{\alpha}$ is a real bosonic
field describing bosonic degrees of freedom. Latin indices gather
space, time, spin and possibly orbital or spinor indices: $u\equiv(\mathbf{R}_{u},\tau_{u},\sigma_{u},\dots)$,
where $\mathbf{R}_{u}$ denotes a site of the Bravais lattice,$\tau_{u}$
denotes imaginary time and $\sigma_{u}$ is a spin (or orbital) index
($\sigma_{u}\in\{\uparrow,\downarrow\}$ in a single-orbital context).
Barred indices denote outgoing points, while indices without a bar
denote ingoing points. Greek indices denote $\alpha\equiv(\mathbf{R}_{\alpha},\tau_{\alpha},I_{\alpha})$,
where $I_{\alpha}$ indexes the bosonic channels. These are for instance
the charge ($I_{\alpha}=0$) and the spin ($I_{\alpha}=x,y,z$) channels.
Repeated indices are summed over. Summation $\sum_{u}$ is shorthand
for $\sum_{\mathbf{R}\in\mathrm{BL}}\sum_{\sigma}\int_{0}^{\beta}\mathrm{d}\tau$.
$G_{0,u\bar{v}}$ (resp. $W_{0,\alpha\beta}$) is the non-interacting
fermionic (resp. bosonic) propagator. 

The action (\ref{eq:general_action_eb}) describes a broad spectrum
of physical problems ranging from electron-phonon coupling problems
to spin-fermion models. As will be elaborated on in subsection \ref{sub:From-an-electron-electron},
it may also stem from an exact rewriting of a problem with only electron-electron
interactions such as the Hubbard model or an extension thereof via
a Hubbard-Stratonovich transformation.

\subsection{Three-particle irreducible formalism\label{sub:Functional-derivation}}

In this subsection, we construct the three-particle irreducible (3PI)
functional $\mathcal{K}[G,W,\Lambda]$. This construction has first
been described in the pioneering works of de Dominicis and Martin.\cite{Dominicis1964,Dominicis1964a}
It consists in successive Legendre transformations of the free energy
of the interacting system.

Let us first define the free energy of the system in the presence
of linear ($h_{\alpha}$), bilinear ($B_{\alpha\beta}$, $F_{\bar{u}v}$)
and trilinear sources ($\lambda_{\bar{u}v\alpha}$) coupled to the
bosonic and fermionic operators,
\begin{align}
 & \Omega[h,B,F,\lambda]\label{eq:free_energy}\\
 & \equiv-\log\int\mathcal{D}[\bar{c},c,\phi]e^{-S_{\mathrm{eb}}+h_{\alpha}\phi_{\alpha}-\frac{1}{2}\phi_{\alpha}B_{\alpha\beta}\phi_{\beta}-\bar{c}_{\bar{u}}F_{\bar{u}v}c_{v}}\nonumber 
\end{align}
We do not need any additional trilinear source term (similar to $h$,
$B$ and $F$) since the electron-boson coupling term already plays
this role.

$\Omega[h,B,F,\lambda]$ is the generating functional of correlation
functions, \emph{viz}.:\begin{subequations}

\begin{eqnarray}
\varphi_{\alpha} & \equiv & \langle\phi_{\alpha}\rangle=-\frac{\partial\Omega}{\partial h_{\alpha}}\Bigg|_{B,F,\lambda}\label{eq:def_varphi}\\
W_{\alpha\beta}^{\mathrm{nc}} & \equiv & -\langle\phi_{\alpha}\phi_{\beta}\rangle=-2\frac{\partial\Omega}{\partial B_{\beta\alpha}}\Bigg|_{h,F,\lambda}\label{eq:def_W}\\
G_{u\bar{v}} & \equiv & -\langle c_{u}\bar{c}_{\bar{v}}\rangle=\frac{\partial\Omega}{\partial F_{\bar{v}u}}\Bigg|_{h,B,\lambda}\label{eq:def_G}
\end{eqnarray}
\end{subequations}The above correlators contain disconnected terms
as denoted by the superscript ``nc'' (non-connected).

\subsubsection{First Legendre transform: with respect to propagators}

Let us now perform a first Legendre transform with respect to $h$,
$B$ and $F$ :
\begin{eqnarray}
\Gamma_{2}[\varphi,G,W^{\mathrm{nc}},\lambda] & \equiv & \Omega[h,F,B,\lambda]-\mathrm{Tr}\left(FG\right)\nonumber \\
 &  & +\frac{1}{2}\mathrm{Tr}\left(BW^{\mathrm{nc}}\right)+h_{\alpha}\varphi_{\alpha}\label{eq:Legendre_def}
\end{eqnarray}
with $\mathrm{Tr}AB\equiv A_{\bar{u}v}B_{v\bar{u}}$. By construction
of the Legendre transformation, the sources are related to the derivatives
of $\Gamma$ through:\begin{subequations} 
\begin{align}
\frac{\partial\Gamma}{\partial G_{v\bar{u}}}\Bigg|_{\varphi,W^{nc},\lambda} & =-F_{\bar{u}v}\label{eq:dGamma_dG}\\
\frac{\partial\Gamma}{\partial W_{\beta\alpha}^{\mathrm{nc}}}\Bigg|_{\varphi,G,\lambda} & =\frac{1}{2}B_{\alpha\beta}\label{eq:dGamma_dW}\\
\frac{\partial\Gamma}{\partial\varphi_{\alpha}}\Bigg|_{G,W^{nc},\lambda} & =h_{\alpha}\label{eq:dGamma_dh}
\end{align}
 \end{subequations}In a fermionic context, $\Gamma_{2}$ is often
called the Baym-Kadanoff functional.\cite{Baym1961,Baym1962} We can
decompose it in the following way:
\begin{equation}
\Gamma_{2}[\varphi,G,W^{\mathrm{nc}},\lambda]=\Gamma_{2}[\varphi,G,W^{\mathrm{nc}},\lambda=0]+\Psi[\varphi,G,W^{\mathrm{nc}},\lambda]\label{eq:Gamma_decomposition}
\end{equation}
The computation of the noninteracting contribution $\Gamma_{2}[\varphi,G,W,\lambda=0]$
is straightforward since in this case relations (\ref{eq:def_varphi}-\ref{eq:def_W}-\ref{eq:def_G})
are easily invertible (as shown in Appendix \ref{sec:Details-of-Legendre}),
so that
\begin{eqnarray}
\Gamma_{2}[\varphi,G,W,\lambda] & = & -\mathrm{Tr}\log\left[G^{-1}\right]+\mathrm{Tr}\left[\left(G^{-1}-G_{0}^{-1}\right)G\right]\nonumber \\
 &  & +\frac{1}{2}\mathrm{Tr}\log\left[W^{-1}\right]+\frac{1}{2}\mathrm{Tr}\left[\left(W-\varphi^{2}\right)W_{0}^{-1}\right]\nonumber \\
 &  & +\Psi[\varphi,G,W,\lambda]\label{eq:Gamma_result}
\end{eqnarray}
where we have defined the connected correlation function:
\begin{equation}
W_{\alpha\beta}\equiv-\Big\langle\left(\phi_{\alpha}-\varphi\right)\left(\phi_{\beta}-\varphi\right)\Big\rangle=W_{\alpha\beta}^{\mathrm{nc}}+\varphi^{2}\label{eq:Wc_def}
\end{equation}
and $\varphi^{2}$ denotes the matrix of elements $[\varphi^{2}]_{\alpha,\beta}=\varphi_{\alpha}\varphi_{\beta}$.
The physical Green's functions (obtained by setting $F=B=0$ in Eqs(\ref{eq:dGamma_dG}-\ref{eq:dGamma_dW}))
obey Dyson equations:\begin{subequations}

\begin{eqnarray}
\Sigma_{\bar{u}v} & = & \left[G_{0}^{-1}\right]_{\bar{u}v}-\left[G^{-1}\right]_{\bar{u}v}\label{eq:Dyson_Sigma}\\
P_{\alpha\beta} & = & \left[W_{0}^{-1}\right]_{\alpha\beta}-\left[W^{-1}\right]_{\alpha\beta}\label{eq:Dyson_P}
\end{eqnarray}
\end{subequations}where we have defined the fermionic and bosonic
self-energies $\Sigma$ and $P$ as functional derivatives with respect
to $\Psi$:\begin{subequations} 
\begin{eqnarray}
\Sigma_{\bar{u}v} & \equiv & \frac{\partial\Psi}{\partial G_{v\bar{u}}}\label{eq:Sigma_dPsi}\\
P_{\alpha\beta} & \equiv & -2\frac{\partial\Psi}{\partial W_{\beta\alpha}}\label{eq:P_dPsi}
\end{eqnarray}
\end{subequations}

The two Dyson equations (\ref{eq:Dyson_Sigma}-\ref{eq:Dyson_P})
and the functional derivative equations (\ref{eq:Sigma_dPsi}-\ref{eq:P_dPsi})
form a closed set of equations that can be solved self-consistently
once the dependence of $\Psi$ on $G$ and $W$ is specified.

\begin{center}
\begin{figure}
\begin{centering}
\includegraphics[width=0.55\columnwidth]{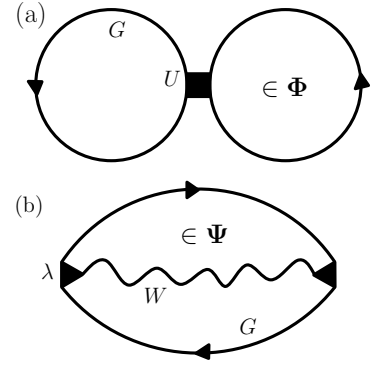}
\par\end{centering}

\caption{\label{fig:diagrams_functionals-2PI}Simplest contribution to 2PI
functionals: (a) Luttinger-Ward functional $\Phi$ (b) Almbladh functional
$\Psi$}
\end{figure}

\par\end{center}

The functional $\Psi[\varphi,G,W,\lambda]$ is the Almbladh functional.\cite{Barth1999}
It is the extension of the Luttinger-Ward functional $\Phi[G]$,\cite{Luttinger1960,Baym1962}
which is defined for fermionic actions, to mixed electron-boson actions.
While $\Phi[G]$ contains two-particle irreducible graphs with fermionic
lines $G$ and bare interactions $U$ (see \emph{e.g.} diagram (a)
of Fig. \ref{fig:diagrams_functionals-2PI}), $\Psi[\varphi,G,W,\lambda]$
contains two-particle irreducible graphs with fermionic ($G$) and
bosonic ($W$) lines, and bare electron-boson interactions vertices
$\lambda$ (see \emph{e.g.} diagram (b) of Fig. \ref{fig:diagrams_functionals-2PI}). 

Both $\Phi$ and $\Psi$ can be approximated in various ways, which
in turn leads to an approximate form for the self-energies, through
Eqs (\ref{eq:Sigma_dPsi}-\ref{eq:P_dPsi}). Any such approximation,
if performed self-consistently, will obey global conservation rules.\cite{Baym1961}
A simple example is the $GW$ approximation,\cite{Hedin1965} which
consists in approximating $\Psi$ by its most simple diagram (diagram
(b) of Fig. \ref{fig:diagrams_functionals-2PI}). The DMFT (resp.
extended DMFT, EDMFT\cite{Sengupta1995,Kajueter1996,Si1996}) approximation,
on the other hand, consists in approximating $\Phi[G]$ (resp. $\Psi[\varphi,G,W,\lambda]$)
by the local diagrams of the exact functional:\begin{subequations}

\begin{align}
\Phi^{\mathrm{DMFT}}[G] & =\sum_{\mathbf{R}}\Phi[G_{\mathbf{RR}}]\label{eq:Phi_DMFT}\\
\Psi^{\mathrm{EDMFT}}[\varphi,G,W] & =\sum_{\mathbf{R}}\Psi[\varphi,G_{\mathbf{RR}},W_{\mathbf{RR}}]\label{eq:Psi_EDMFT}
\end{align}
\end{subequations}

The DMFT approximation becomes exact in the limit of infinite dimensions.\cite{Georges1996}
Motivated by this link between irreducibility and reduction to locality
in high dimensions, we perform an additional Legendre transform to
go one step further in terms of irreducibilty.

\begin{figure}

\begin{centering}
\includegraphics[width=0.8\columnwidth]{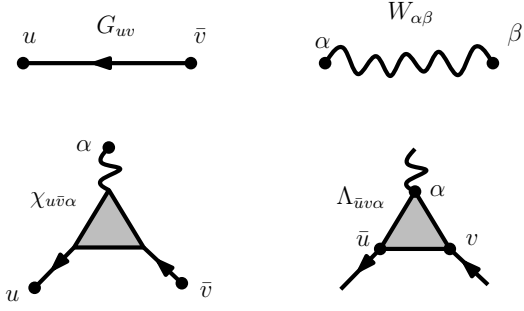}
\par\end{centering}

\caption{Graphical representation of the diagrammatic objects of the electron-boson
model (Eq. \ref{eq:general_action_eb}): the fermionic propagator
$G_{u\bar{v}}$ (Eq. (\ref{eq:def_G})), the bosonic propagator $W_{\alpha\beta}$
(Eq. (\ref{eq:Wc_def})), the three-point correlation function $\chi_{u\bar{v}\alpha}$
(Eq. (\ref{eq:chi3_nc_def})) and the three-leg vertex $\Lambda_{\bar{u}v\alpha}$
(Eq. (\ref{eq:Lambda_def})). \label{fig:Graphical-representation-of-eb-action}}

\end{figure}

\subsubsection{Second Legendre transform: with respect to the three-leg vertex}

We introduce the Legendre transform of $\Gamma_{2}$ with respect
to $\lambda$:
\begin{equation}
\Gamma_{3}[\varphi,G,W,\chi^{\mathrm{nc}}]\equiv\Gamma_{2}[\varphi,G,W,\lambda]+\lambda_{\bar{v}u\alpha}\chi_{u\bar{v}\alpha}^{\mathrm{nc}}\label{eq:Gamma_3_def}
\end{equation}
where $\chi_{u\bar{v}\alpha}^{\mathrm{nc}}$ is the three-point correlator:
\begin{equation}
\chi_{u\bar{v}\alpha}^{\mathrm{nc}}\equiv\langle c_{u}\bar{c}_{\bar{v}}\phi_{\alpha}\rangle=-\frac{\partial\Omega}{\partial\lambda_{\bar{v}u\alpha}}\Bigg|_{h,F,B}\label{eq:chi3_nc_def}
\end{equation}
We also define the connected three-point function $\chi$ and the
three-leg vertex $\Lambda$ as:

\begin{align}
\chi_{u\bar{v}\alpha} & \equiv\langle c_{u}\bar{c}_{\bar{v}}\left(\phi_{\alpha}-\varphi_{\alpha}\right)\rangle=\chi_{u\bar{v}\alpha}^{\mathrm{nc}}+G_{u\bar{v}}\varphi_{\alpha}\label{eq:chi3_def}\\
\Lambda_{\bar{v}u\alpha} & \equiv G_{\bar{x}u}^{-1}G_{\bar{v}w}^{-1}W_{\alpha\beta}^{-1}\chi_{w\bar{x}\beta}\label{eq:Lambda_def}
\end{align}
$\Lambda$ is the amputated, connected correlation function. It is
the renormalized electron-boson vertex. These objects are shown graphically
in Fig. \ref{fig:Graphical-representation-of-eb-action}. $G_{\bar{u}v}^{-1}$
is a shorthand notation for $\left[G^{-1}\right]_{\bar{u}v}$.

We now define the three-particle irreducible functional $\mathcal{K}$
as:
\begin{eqnarray}
\mathcal{K}[\varphi,G,W,\Lambda] & \equiv & \Psi[\varphi,G,W,\lambda]+\lambda_{\bar{v}u\alpha}\chi_{u\bar{v}\alpha}^{\mathrm{nc}}\nonumber \\
 &  & -\frac{1}{2}\Lambda_{\bar{x}u\alpha}G_{w\bar{x}}G_{u\bar{v}}W_{\alpha\beta}\Lambda_{\bar{v}w\beta}\label{eq:Kappa_def}
\end{eqnarray}
Note that in the right-hand site, $\lambda$ is determined by $\Lambda$,
$G$, $W$ and $\varphi$ (by the Legendre construction). $\mathcal{K}$
is the generalization of the functional $\mathscr{K}^{3/2}$ introduced
in Ref. \onlinecite{Dominicis1964a} to mixed fermionic and bosonic
fields. We will come back to its diagrammatic interpretation in the
next subsection.

Differentiating $\mathcal{K}$ with respect to the three-point vertex
$\Lambda$ yields $K$, the generalization of the self-energy at the
three-particle irreducible level, defined as:

\begin{equation}
K_{\bar{v}u\alpha}\equiv-G_{\bar{x}u}^{-1}G_{\bar{v}w}^{-1}W_{\alpha\beta}^{-1}\frac{\partial\mathcal{K}}{\partial\Lambda_{\bar{x}w\beta}}\label{eq:K_def}
\end{equation}

\begin{figure}

\begin{centering}
\includegraphics[width=1\columnwidth]{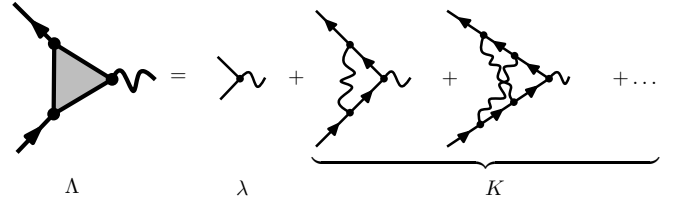}
\par\end{centering}

\caption{Graphical representation of the diagrammatic content of $\Lambda$.\label{fig:Lambda_expansion}}
\end{figure}

Before proceeding with the derivation, let us first state the main
results: $K$ and $\Lambda$ are related by the following relation: 

\begin{equation}
\Lambda_{\bar{v}u\alpha}=\lambda_{\bar{v}u\alpha}+K_{\bar{v}u\alpha}\label{eq:Dyson_Lambda}
\end{equation}
This is the equivalent of Dyson's equations at the 3PI level. This
relation is remarkably simple: it does not involve any inversion,
contrary to the Dyson equations (\ref{eq:Dyson_Sigma}-\ref{eq:Dyson_P}).
This relation is illustrated in Figure \ref{fig:Lambda_expansion}.

\begin{figure}

\begin{centering}
\includegraphics[width=0.7\columnwidth]{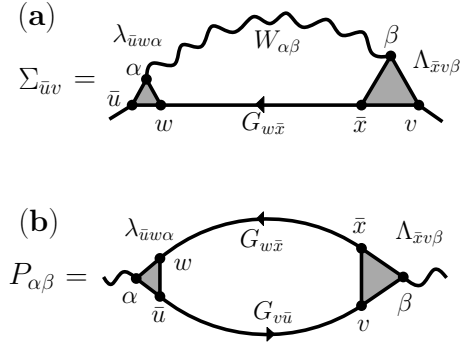}
\par\end{centering}

\caption{Graphical representation of the self-energy (beyond the Hartree term)
(panel (a)) and polarization (panel (b)).\label{fig:Self-energies}}

\end{figure}

The fermionic and bosonic self-energies $\Sigma$ and $P$ are related
to $\Lambda$ by the following exact relations:\begin{subequations}

\begin{eqnarray}
\Sigma_{\bar{u}v} & = & -\lambda_{\bar{u}w\alpha}G_{w\bar{x}}W_{\alpha\beta}\Lambda_{\bar{x}v\beta}+\lambda_{\bar{u}v\alpha}\varphi_{\alpha}\label{eq:Sigma_Hedin}\\
P_{\alpha\beta} & = & \lambda_{\bar{u}w\alpha}G_{v\bar{u}}G_{w\bar{x}}\Lambda_{\bar{x}v\beta}\label{eq:P_Hedin}
\end{eqnarray}
\end{subequations}The second term in $\Sigma$ is nothing but the
Hartree contribution. These expressions will be derived later. The
graphical representation of these equations is shown in Figure \ref{fig:Self-energies}.

\subsubsection{Discussion}

\begin{center}
\begin{figure}
\begin{centering}
\includegraphics[width=1\columnwidth]{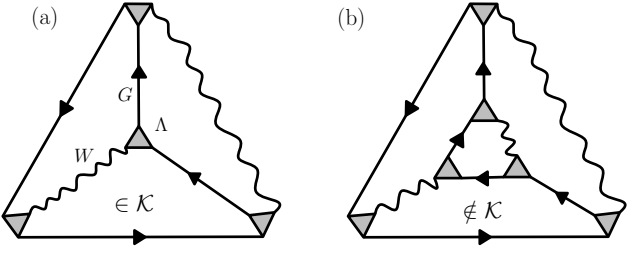}
\par\end{centering}

\caption{\label{fig:diagrams_functionals-3PI}Three-particle irreducibility
(a) Simplest contribution to the 3PI functional $\mathcal{K}$ (b)
an example of diagram not contributing to $\mathcal{K}$.}
\end{figure}

\par\end{center}

The above equations, Eqs (\ref{eq:K_def}-\ref{eq:Dyson_Lambda}-\ref{eq:Sigma_Hedin}-\ref{eq:P_Hedin}-\ref{eq:Dyson_Sigma}-\ref{eq:Dyson_P}),
form a closed set of equations for $G$, $W$, $\Lambda$, $\Sigma$,
$P$ and $K$. The central quantity is the three-particle irreducible
functional $\mathcal{K}[G,W,\Lambda]$, obtained from the 2PI functional
algebraically by a Legendre transformation with respect to the bare
vertex $\lambda$, or diagrammatically by a 'boldification' of the
bare vertex.

$\mathcal{K}$ has been shown to be made up of all three-particle
irreducible (3PI) diagrams by de Dominicis and Martin\cite{Dominicis1964a}
in the bosonic case. A 3PI diagram is defined as follows: for any
set of three lines whose cutting leads to a separation of the diagram
in two parts, one and only one of those parts is a simple three-leg
vertex $\Lambda$. The simplest 3PI diagram is shown in Fig. \ref{fig:diagrams_functionals-3PI}(a).
Conversely, neither diagram (b) of Fig. \ref{fig:diagrams_functionals-2PI}
nor diagram (b) of Fig. \ref{fig:diagrams_functionals-3PI} are 3PI
diagrams. 

Most importantly, the hierarchy is closed once the functional form
of $\mathcal{K}$ is specified: there is no a priori need for a higher-order
vertex. This contrasts with e.g. the functional renormalization group
(fRG\cite{Metzner2012}) formalism (which requires the truncation
of the flow equations) or the Hedin formalism\cite{Hedin1965,Aryasetiawan1998,Aryasetiawan2008}
which involves the four-leg vertex $\delta\Sigma/\delta G$ via the
following Bethe-Salpether-like expression for $K$:

\begin{equation}
K=\frac{\delta\Sigma}{\delta G}GG\Lambda\label{eq:K_ladder}
\end{equation}

Of course, one must devise approximation strategies for $\mathcal{K}$
in order to solve this set of equations. In particular, any approximation
involving the neglect of vertex corrections, like the FLEX approximation\cite{Bickers1989},
spin fluctuation theory\cite{Monthoux1991,Schmalian1998,Chubukov2002},
the GW approximation\cite{Hedin1965} or the Migdal-Eliashberg theory
of superconductivity\cite{Migdal1958,Eliashberg1960} corresponds
to the approximation
\begin{equation}
\mathcal{K}[\varphi,G,W,\Lambda]\approx0\label{eq:fluctuation_exchange_approx}
\end{equation}
which yields, in particular, the simple one-loop form for the self-energy:\begin{subequations}

\begin{eqnarray}
\Sigma_{\bar{u}v} & = & -\lambda_{\bar{u}w\alpha}G_{w\bar{x}}W_{\alpha\beta}\lambda_{\bar{x}v\beta}+\lambda_{\bar{u}v\alpha}\varphi_{\alpha}\label{eq:Sigma_GW}\\
P_{\alpha\beta} & = & \lambda_{\bar{u}w\alpha}G_{v\bar{u}}G_{w\bar{x}}\lambda_{\bar{x}v\beta}\label{eq:P_GG}
\end{eqnarray}
\end{subequations}These approximations only differ in the type of
fermionic and bosonic fields in the initial action, Eq. (\ref{eq:general_action_eb}):
normal/superconducting fermions, bosons in the particle-hole/particle-particle
sector, in the spin/charge channel... 

The core idea of the DMFT and descendent methods is to make an approximation
of $\Phi$ (or $\Psi$) around the atomic limit. TRILEX is a similar
approximation for $\mathcal{K}$, as will be discussed in section
\ref{sub:Local-approximation}.

\subsubsection{Derivation of the main equations}

In this subsection, we derive Eqs (\ref{eq:Dyson_Lambda}-\ref{eq:Sigma_Hedin}-\ref{eq:P_Hedin}).
Combining (\ref{eq:Gamma_result}), (\ref{eq:Gamma_3_def}) and (\ref{eq:Kappa_def})
leads to:
\begin{eqnarray}
\Gamma_{3}[\varphi,G,W,\Lambda] & = & \Gamma_{2}\left[\varphi,G,W,\lambda=0\right]+\mathcal{K}[\varphi,G,W,\Lambda]\nonumber \\
 &  & +\frac{1}{2}\Lambda_{\bar{x}u\alpha}G_{w\bar{x}}G_{u\bar{v}}W_{\alpha\beta}\Lambda_{\bar{v}w\beta}\label{eq:Gamma3_result-1}
\end{eqnarray}
By construction of the Legendre transform $\Gamma_{3}$ (Eq. (\ref{eq:Gamma_3_def})),

\[
\lambda_{\bar{v}u\alpha}=\frac{\partial\Gamma_{3}}{\partial\chi_{u\bar{v}\alpha}}\Bigg|_{\varphi,G,W}
\]
We note that at fixed $G$ and $\varphi$, this is equivalent to differentiating
with respect to $\chi_{u\bar{v}\alpha}^{\mathrm{nc}}$. Using the
the chain rule and then (\ref{eq:Gamma3_result-1}) and (\ref{eq:Lambda_def})
to decompose both factors yields: 
\begin{eqnarray*}
\lambda_{\bar{v}u\alpha} & = & \frac{\partial\Gamma_{3}}{\partial\Lambda_{\bar{x}w\beta}}\Bigg|_{\varphi,G,W}\frac{\partial\Lambda_{\bar{x}w\beta}}{\partial\chi_{u\bar{v}\alpha}}\Bigg|_{\varphi,G,W}\\
 & = & \left(\frac{\partial\mathcal{K}}{\partial\Lambda_{\bar{x}w\beta}}+G_{w\bar{s}}G_{r\bar{x}}W_{\beta\gamma}\Lambda_{\bar{s}r\gamma}\right)\left(G_{\bar{v}w}^{-1}G_{\bar{x}u}^{-1}W_{\beta\alpha}^{-1}\right)\\
 & = & G_{\bar{v}w}^{-1}G_{\bar{x}u}^{-1}W_{\beta\alpha}^{-1}\frac{\partial\mathcal{K}}{\partial\Lambda_{\bar{x}w\beta}}+\Lambda_{\bar{v}u\alpha}
\end{eqnarray*}
Using the definition of $K_{uv\alpha}$ (Eq. (\ref{eq:K_def})), this
proves Eq (\ref{eq:Dyson_Lambda}). 

Let us now derive Eqs (\ref{eq:Sigma_Hedin}-\ref{eq:P_Hedin}). They
are well-known from a diagrammatic point of view, but the point of
this section is to derive them analytically from the properties of
$\mathcal{K}$. In order to obtain the self-energy $\Sigma$, we use
Eq. (\ref{eq:Sigma_dPsi}). We first need to reexpress $\Psi$ in
terms of $\mathcal{K}$ using (\ref{eq:Kappa_def}): thus
\begin{align*}
\Psi[\varphi,G,W,\lambda] & =\tilde{\Psi}[\varphi,G,W,\lambda,\Lambda]\\
 & \equiv\mathcal{K}[\varphi,G,W,\Lambda]\\
 & +\lambda_{\bar{v}u\alpha}\chi_{u\bar{v}\alpha}^{\mathrm{nc}}-\frac{1}{2}\Lambda_{\bar{x}u\alpha}G_{w\bar{x}}G_{u\bar{v}}W_{\alpha\beta}\Lambda_{\bar{v}w\beta}
\end{align*}
where $\Lambda$ is a function of $\varphi$, $G$, $W$, $\lambda$.
Thus, Eq. (\ref{eq:Sigma_dPsi}) becomes:

\begin{eqnarray*}
\Sigma_{\bar{u}v} & = & \frac{\partial\tilde{\Psi}[\varphi,G,W,\lambda,\Lambda]}{\partial G_{v\bar{u}}}\Bigg|_{\varphi,W,\lambda}
\end{eqnarray*}
This derivative must be performed with care since the electron-boson
vertex now appears in its interacting form $\Lambda$. This yields:

\begin{eqnarray}
\Sigma_{\bar{u}v} & = & \frac{\partial\tilde{\Psi}}{\partial G_{v\bar{u}}}\Bigg|_{\varphi,W,\lambda,\Lambda}+\frac{\partial\tilde{\Psi}}{\partial\Lambda_{\bar{q}p\gamma}}\Bigg|_{\varphi,W,G,\lambda}\frac{\partial\Lambda_{\bar{q}p\gamma}}{\partial G_{v\bar{u}}}\Bigg|_{\varphi,W,\lambda}
\end{eqnarray}
The second term vanishes by construction of the Legendre transform.
Indeed, using (\ref{eq:Kappa_def}), (\ref{eq:Dyson_Lambda}) and
(\ref{eq:chi3_def}):

\begin{eqnarray*}
\frac{\partial\tilde{\Psi}}{\partial\Lambda_{\bar{q}p\delta}}\Bigg|_{\varphi,W,G,\lambda} & = & \frac{\partial}{\partial\Lambda_{\bar{q}p\delta}}\Bigg[\mathcal{K}-\lambda_{\bar{s}r\gamma}\left(\chi_{r\bar{s}\gamma}-G_{r\bar{s}}\varphi_{\gamma}\right)\\
 &  & \;+\frac{1}{2}\Lambda_{\bar{x}u\alpha}G_{w\bar{x}}G_{u\bar{v}}W_{\alpha\beta}\Lambda_{\bar{v}w\beta}\Bigg]\\
 & = & -G_{r\bar{q}}G_{p\bar{s}}W_{\gamma\delta}K_{\bar{s}r\gamma}-G_{r\bar{q}}G_{p\bar{s}}W_{\gamma\delta}\lambda_{\bar{s}r\gamma}\\
 &  & \;+G_{r\bar{q}}G_{p\bar{s}}W_{\gamma\delta}\Lambda_{\bar{s}r\gamma}\\
 & = & 0
\end{eqnarray*}
As a result,
\begin{eqnarray*}
\Sigma_{\bar{u}v} & = & \frac{\partial}{\partial G_{v\bar{u}}}\Bigg[\mathcal{K}-\lambda_{\bar{s}r\gamma}\left(\chi_{r\bar{s}\gamma}-G_{r\bar{s}}\varphi_{\gamma}\right)\\
 &  & +\frac{1}{2}\Lambda_{\bar{x}y\alpha}G_{w\bar{x}}G_{y\bar{z}}W_{\alpha\beta}\Lambda_{\bar{z}w\beta}\Bigg]\Bigg|_{\varphi,W,\lambda,\Lambda}\\
 & = & \frac{\partial\mathcal{K}}{\partial G_{v\bar{u}}}-\lambda_{\bar{u}y\alpha}G_{y\bar{z}}W_{\alpha\beta}\Lambda_{\bar{z}v\beta}-\Lambda_{\bar{u}y\alpha}G_{y\bar{z}}W_{\beta\alpha}\lambda_{\bar{z}v\beta}\\
 &  & +\lambda_{\bar{u}v\alpha}\varphi_{\alpha}+\Lambda_{\bar{u}y\alpha}G_{y\bar{z}}W_{\alpha\beta}\Lambda_{\bar{z}v\beta}
\end{eqnarray*}
Finally, using (\ref{eq:Dyson_Lambda}), we obtain:

\begin{eqnarray}
\Sigma_{\bar{u}v} & = & -\lambda_{\bar{u}w\alpha}G_{w\bar{x}}W_{\alpha\beta}\Lambda_{\bar{x}v\beta}+\lambda_{\bar{v}u\alpha}\varphi_{\alpha}\label{eq:Sigma_interm}\\
 &  & +\left[\frac{\partial\mathcal{K}}{\partial G_{v\bar{u}}}+\Lambda_{\bar{u}w\alpha}G_{w\bar{x}}W_{\alpha\beta}K_{\bar{x}v\beta}\right]\nonumber 
\end{eqnarray}
Similarly, using (\ref{eq:P_dPsi}), one gets for $P$:

\begin{eqnarray*}
P_{\alpha\beta} & = & -2\frac{\partial\tilde{\Psi}[\varphi,G,W,\lambda,\Lambda]}{\partial W_{\beta\alpha}}\Bigg|_{\varphi,G,\lambda,\Lambda}\\
 & = & -2\frac{\partial}{\partial W_{\beta\alpha}}\Bigg[\mathcal{K}-\lambda_{\bar{s}r\gamma}\left(\chi_{r\bar{s}\gamma}-G_{r\bar{s}}\varphi_{\gamma}\right)\\
 &  & +\frac{1}{2}\Lambda_{\bar{x}y\delta}G_{w\bar{x}}G_{y\bar{z}}W_{\delta\gamma}\Lambda_{\bar{z}w\gamma}\Bigg]\Bigg|_{\varphi,G,\lambda,\Lambda}\\
 & = & -2\frac{\partial\mathcal{K}}{\partial W_{\beta\alpha}}+2\lambda_{\bar{u}w\alpha}G_{v\bar{u}}G_{w\bar{x}}\Lambda_{\bar{x}v\beta}\\
 &  & -\Lambda_{\bar{u}w\alpha}G_{v\bar{u}}G_{w\bar{x}}\Lambda_{\bar{x}v\beta}
\end{eqnarray*}
Thus, using (\ref{eq:Dyson_Lambda}), we get 
\begin{eqnarray}
P_{\alpha\beta} & = & \lambda_{\bar{u}w\alpha}G_{v\bar{u}}G_{w\bar{x}}\Lambda_{\bar{x}v\beta}\label{eq:P_interm}\\
 &  & -\left[2\frac{\partial\mathcal{K}}{\partial W_{\beta\alpha}}+K_{\bar{u}w\alpha}G_{v\bar{u}}G_{w\bar{x}}\Lambda_{\bar{x}v\beta}\right]\nonumber 
\end{eqnarray}

\begin{figure}
\begin{centering}
\includegraphics[width=0.95\columnwidth]{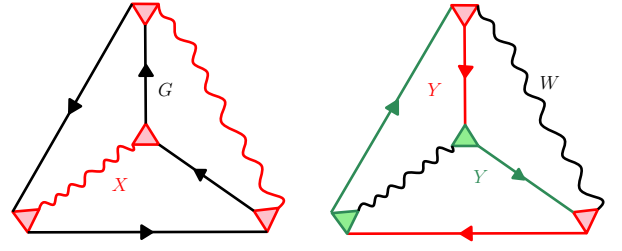}
\par\end{centering}

\caption{(color online) Homogeneity properties of the simplest diagram of $\mathcal{K}$.
\emph{Left}: dependence on $X$ (red). \emph{Right}: dependence on
$Y$ (red and green). The lines are defined in Fig. \ref{fig:Graphical-representation-of-eb-action}\label{fig:Homogeneity}}
\end{figure}

Let now prove that the bracketed terms in Eqs (\ref{eq:Sigma_interm}-\ref{eq:P_interm})
vanish. We first note from the diagrammatic interpretation of $\mathcal{K}$
that $\mathcal{K}$ is a homogeneous function of\begin{subequations}

\begin{eqnarray}
Y_{uw\alpha} & \equiv & G_{w\bar{v}}\Lambda_{\bar{v}u\alpha}\label{eq:homogeneity_G-1-3}\\
X_{u\bar{p}w\bar{z}} & \equiv & \Lambda_{\bar{p}u\alpha}W_{\alpha\beta}\Lambda_{\bar{z}w\beta}\label{eq:homogeneity_W-1-3}
\end{eqnarray}
\end{subequations}i.e. $\mathcal{K}$ can be written as:\begin{subequations}

\begin{eqnarray}
\mathcal{K} & = & f_{1}(Y_{uw\alpha},W_{\alpha\beta})\label{eq:homogeneity_G-1}\\
\mathcal{K} & = & f_{2}(X_{u\bar{p}w\bar{z}},G_{u\bar{v}})\label{eq:homogeneity_W-1}
\end{eqnarray}
\end{subequations}where $f_{1}$ and $f_{2}$ are two functions.
This is illustrated in Fig. \ref{fig:Homogeneity} for the simplest
diagram of $\mathcal{K}$.

From (\ref{eq:homogeneity_G-1}), one gets:\begin{subequations}

\begin{eqnarray}
\frac{\partial\mathcal{K}}{\partial G_{x\bar{y}}}G_{x\bar{z}} & = & \Lambda_{\bar{y}u\alpha}G_{w\bar{z}}\frac{\partial f_{1}}{\partial Y_{uw\alpha}}\label{eq:homogeneity_G-1-1}\\
\Lambda_{\bar{y}x\alpha}\frac{\partial\mathcal{K}}{\partial\Lambda_{\bar{z}x\alpha}} & = & \Lambda_{\bar{y}u\alpha}G_{w\bar{z}}\frac{\partial f_{1}}{\partial Y_{uw\alpha}}\label{eq:homogeneity_W-1-1}
\end{eqnarray}
\end{subequations}From (\ref{eq:homogeneity_W-1}), in turn, one
gets:\begin{subequations}

\begin{eqnarray}
\Lambda_{\bar{x}v\delta}\frac{\partial\mathcal{K}}{\partial\Lambda_{\bar{x}v\gamma}} & = & \Lambda_{\bar{x}v\delta}\Bigg(W_{\gamma\beta}\Lambda_{\bar{z}w\beta}\frac{\partial f_{2}}{\partial X_{v\bar{x},w\bar{z}}}\nonumber \\
 &  & +\Lambda_{\bar{p}u\alpha}W_{\alpha\gamma}\frac{\partial f_{2}}{\partial X_{u\bar{p},v\bar{x}}}\Bigg)\nonumber \\
 & = & 2\Lambda_{\bar{x}v\delta}W_{\alpha\gamma}\Lambda_{\bar{p}u\alpha}\left(\frac{\partial f_{2}}{\partial X_{v\bar{x},u\bar{p}}}\right)\label{eq:homogeneity_G-1-2}\\
\frac{\partial\mathcal{K}}{\partial W_{\delta\mu}}W_{\mu\gamma} & = & \Lambda_{\bar{x}v\delta}W_{\alpha\gamma}\Lambda_{\bar{p}u\alpha}\frac{\partial f_{2}}{\partial X_{v\bar{x},u\bar{p}}}\label{eq:homogeneity_W-1-2}
\end{eqnarray}
\end{subequations}where we have used the property that $W$ is symmetric
twice: first by trivially using $W_{\alpha\beta}=W_{\beta\alpha}$,
and second to prove that 
\[
\frac{\partial f_{2}}{\partial X_{v\bar{x},u\bar{p}}}=\frac{\partial f_{2}}{\partial X_{u\bar{p},v\bar{x}}}
\]
This latter property can be proven by noticing that when $W$ is symmetric,
$f_{2}$ is a homogeneous function of the symmetrized $X$: $f_{2}(X_{AB})=\tilde{f}_{2}\left(X_{AB}^{s}\right)$,
with $A=(v,\bar{x})$, $B=(u,\bar{p})$, and $X_{AB}^{s}\equiv\frac{1}{2}\left[X_{AB}+X_{BA}\right]$.
Then, one has $\frac{\partial f_{2}}{\partial X_{CD}}=\frac{1}{2}\left(\delta_{CA}\delta_{DB}+\delta_{CB}\delta_{DA}\right)\frac{\partial\tilde{f}_{2}}{\partial X_{AB}^{s}}=\frac{\partial f_{2}}{\partial X_{DC}}$. 

We thus obtain the following relations:\begin{subequations}

\begin{eqnarray}
\frac{\partial\mathcal{K}}{\partial G_{v\bar{u}}}G_{v\bar{r}} & = & \Lambda_{\bar{u}w\alpha}\frac{\partial\mathcal{K}}{\partial\Lambda_{\bar{r}w\alpha}}\label{eq:homogeneity_G}\\
2\frac{\partial\mathcal{K}}{\partial W_{\beta\alpha}}W_{\alpha\delta} & = & \frac{\partial\mathcal{K}}{\partial\Lambda_{\bar{x}v\delta}}\Lambda_{\bar{x}v\beta}\label{eq:homogeneity_W}
\end{eqnarray}
\end{subequations}Right-multiplying (\ref{eq:homogeneity_G}) by
$G^{-1}$ and (\ref{eq:homogeneity_W}) by $W^{-1}$ and replacing
$\partial\mathcal{K}/\partial\Lambda$ using the definition of $K$
(Eq. (\ref{eq:K_def})) shows that the bracketed terms in Eqs (\ref{eq:Sigma_interm}-\ref{eq:P_interm})
vanish. Thus, these expressions simplify to the final expressions
for the self-energy and polarization, Eqs (\ref{eq:Sigma_Hedin}-\ref{eq:P_Hedin}). 

Finally, these exact expressions can be derived in alternative fashion
using equations of motion, as shown in Appendix \ref{sec:Alternative-derivation-EOMS}.

\subsection{Transposition to electron-electron problems\label{sub:From-an-electron-electron}}

In this section, we show how the formalism described above can be
used to study electron-electron interaction problems. We shall focus
on the two-dimensional Hubbard model, which reads: 
\begin{equation}
H=\sum_{\mathbf{RR'}\sigma}t_{\mathbf{RR'}}c_{\mathbf{R}\sigma}^{\dagger}c_{\mathbf{R'}\sigma}+U\sum_{\mathbf{R}}n_{\mathbf{R}\uparrow}n_{\mathbf{R}\downarrow}\label{eq:Hubbard_hamiltonian}
\end{equation}
$\mathbf{R}$ denotes a point of the Bravais lattice, $\sigma=\uparrow,\downarrow$,
$t_{\mathbf{R}\mathbf{R'}}$ is the tight-binding hopping matrix (its
Fourier transform is $\varepsilon(\mathbf{k})$), $U$ is the local
Hubbard repulsion, $c_{\mathbf{R}\sigma}^{\dagger}$ and $c_{\mathbf{R}\sigma}$
are creation and annihilation operators, $n\equiv n_{\uparrow}+n_{\downarrow}$,
with $n_{\sigma}=c_{\sigma}^{\dagger}c_{\sigma}$. In the path-integral
formalism, the corresponding action reads:

\begin{equation}
S_{\mathrm{ee}}=\bar{c}_{\bar{u}}\left[-G_{0}^{-1}\right]_{\bar{u}v}c_{v}+Un_{\mathbf{R}\uparrow\tau}n_{\mathbf{R}\downarrow\tau}\label{eq:general_action_ee}
\end{equation}
Here, $G_{0,\sigma}^{-1}(\mathbf{k},i\omega)=i\omega+\mu-\varepsilon(\mathbf{k})$,
where $i\omega$ denotes fermionic Matsubara frequencies, $\mu$ the
chemical potential and the bare dispersion reads $\varepsilon(\mathbf{k})=2t\left(\cos(k_{x})+\cos(k_{y})\right)$
in the case of nearest-neighbor hoppings. $\overline{c}_{u}$ and
$c_{u}$ are Grassmann fields. We remind that $u=(\mathbf{R},\tau,\sigma)$.

The Hubbard interaction in Eq. (\ref{eq:general_action_ee}) can be
decomposed in various ways. Defining $s^{I}\equiv\bar{c}_{u}\sigma_{uv}^{I}c_{v}$
(where $\sigma^{0}=\mathbf{1}$ and $\sigma^{x/y/z}$ denotes the
Pauli matrices), the following expressions hold, up to a density term:\begin{subequations}

\begin{align}
Un_{\uparrow}n_{\downarrow} & =\frac{1}{2}U^{\mathrm{ch}}nn+\frac{1}{2}U^{\mathrm{sp}}s^{z}s^{z}\label{eq:Ising_rewriting}\\
Un_{\uparrow}n_{\downarrow} & =\frac{1}{2}\tilde{U}^{\mathrm{ch}}nn+\frac{1}{2}\tilde{U}^{\mathrm{sp}}\vec{s}\cdot\vec{s}\label{eq:Heisenberg_rewriting}
\end{align}
\end{subequations}with the respective conditions:\begin{subequations}

\begin{align}
U & =U^{\mathrm{ch}}-U^{\mathrm{sp}}\label{eq:Fierz_relation_z_only}\\
U & =\tilde{U}^{\mathrm{ch}}-3\tilde{U}^{\mathrm{sp}}\label{eq:Fierz_relation}
\end{align}
\end{subequations}In Eq. (\ref{eq:Ising_rewriting}), the Hubbard
interaction is decomposed on the charge and longitudinal spin channel
(``Ising'', or ``$z$''-decoupling), while in Eq. (\ref{eq:Heisenberg_rewriting})
it is decomposed on the charge and full spin channel (``Heisenberg'',
or ``$xyz$''-decoupling). The Heisenberg decoupling preserves rotational
invariance, contrary to the Ising one. In addition to this freedom
of decomposition comes the choice of the ratio of the charge to the
spin channel, which is encoded in Eqs. (\ref{eq:Fierz_relation_z_only}-\ref{eq:Fierz_relation}).

The two equalities (\ref{eq:Ising_rewriting}-\ref{eq:Heisenberg_rewriting})
can be derived by writing that for any value of the unspecified parameters
$U^{\mathrm{ch}}$ and $U^{\mathrm{sp}}$:

\begin{align*}
\frac{1}{2}\left(U^{\mathrm{ch}}nn+U^{\mathrm{sp}}s^{z}s^{z}\right) & =\frac{1}{2}U^{\mathrm{ch}}\left(n_{\uparrow}+n_{\downarrow}\right)^{2}\\
 & \;\;+\frac{1}{2}U^{\mathrm{sp}}\left(n_{\uparrow}-n_{\downarrow}\right)^{2}\\
 & =\left(n_{\uparrow}n_{\downarrow}+\frac{n}{2}\right)(U^{\mathrm{ch}}-U^{\mathrm{sp}})
\end{align*}
where we have used: $n_{\sigma}^{2}=n_{\sigma}$. Similarly, we can
write:

\begin{eqnarray*}
\frac{1}{2}\tilde{U}^{\mathrm{ch}}nn+\frac{1}{2}\tilde{U}^{\mathrm{sp}}\vec{s}\cdot\vec{s} & = & \frac{1}{2}\bar{c}_{u}c_{v}\bar{c}_{w}c_{l}\delta_{uv}\delta_{wl}\tilde{U}^{\mathrm{ch}}\\
 &  & +\frac{1}{2}\bar{c}_{u}c_{v}\bar{c}_{w}c_{l}\left(2\delta_{ul}\delta_{vw}-\delta_{uv}\delta_{wl}\right)\tilde{U}^{\mathrm{sp}}\\
 & = & \frac{1}{2}\bar{c}_{u}c_{u}\bar{c}_{v}c_{v}\tilde{U}^{\mathrm{ch}}\\
 &  & +\frac{1}{2}\left(2\bar{c}_{u}c_{v}\bar{c}_{v}c_{u}-\bar{c}_{u}c_{u}\bar{c}_{v}c_{v}\right)\tilde{U}^{\mathrm{sp}}\\
 & = & \left(n_{\uparrow}n_{\downarrow}+\frac{n}{2}\right)(\tilde{U}^{\mathrm{ch}}-3\tilde{U}^{\mathrm{sp}})
\end{eqnarray*}

Based on Eq.(\ref{eq:Fierz_relation}-\ref{eq:Fierz_relation_z_only}),
the ratio of the bare interaction in the charge and spin channels
may be parametrized by a number $\alpha$. In the Heisenberg decoupling,\begin{subequations}

\begin{eqnarray}
\tilde{U}^{\mathrm{ch}} & = & (3\alpha-1)U\label{eq:U_ch_fierz}\\
\tilde{U}^{\mathrm{sp}} & = & \left(\alpha-\nicefrac{2}{3}\right)U\label{eq:U_sp_fierz}
\end{eqnarray}
\end{subequations}In the Ising decoupling,\begin{subequations}

\begin{eqnarray}
U^{\mathrm{ch}} & = & \alpha U\label{eq:U_ch_fierz-zonly}\\
U^{\mathrm{sp}} & = & \left(\alpha-1\right)U\label{eq:U_sp_fierz-zonly}
\end{eqnarray}
\end{subequations}

In the following, we adopt a more compact and general notation for
Eqs (\ref{eq:Ising_rewriting}-\ref{eq:Heisenberg_rewriting}), namely
we write the interacting part of the action as:

\begin{eqnarray}
S_{\mathrm{int}} & = & \frac{1}{2}U_{\alpha\beta}n_{\alpha}n_{\beta}\label{eq:Hubbard_Pauli_decomposition}
\end{eqnarray}
with 

\begin{equation}
n_{\alpha}\equiv\bar{c}_{\bar{u}}\lambda_{\bar{u}v\alpha}c_{v}\label{eq:def_n}
\end{equation}
We remind that $u=(\mathbf{R},\tau,\sigma)$ and $\alpha=(\mathbf{R},\tau,I)$.
The parameter $I$ may take the values $I=0,x,y,z$ (Heisenberg decoupling)
or a subset thereof (\emph{e.g} $I=0,z$ for the Ising decoupling). 

In the Hubbard model (Eq.(\ref{eq:Hubbard_hamiltonian})),
\begin{equation}
\lambda_{uv\alpha}\equiv\sigma_{\sigma_{u}\sigma_{v}}^{I}\delta_{\mathbf{R}_{u}-\mathbf{R}_{v}}\delta_{\mathbf{R}_{v}-\mathbf{R}_{\alpha}}\delta_{\tau_{u}-\tau_{v}}\delta_{\tau_{v}-\tau_{\alpha}}\label{eq:eb_coupling_Hubbard}
\end{equation}
and
\begin{equation}
U_{\alpha\beta}=U^{I_{\alpha}}\delta_{I_{\alpha}I_{\beta}}\delta_{\mathbf{R_{\alpha}-R_{\beta}}}\delta_{\tau_{\alpha}-\tau_{\beta}}\label{eq:interaction_Hubbard}
\end{equation}
In the paramagnetic phase, one can define $U^{0}\equiv U^{\mathrm{ch}}$
and $U^{x}=U^{y}=U^{z}\equiv U^{\mathrm{sp}}$, which gives back Eqs.
(\ref{eq:Ising_rewriting}-\ref{eq:Heisenberg_rewriting}). 

We now decouple the interaction (\ref{eq:Hubbard_Pauli_decomposition})
with a real\footnote{In principle, the interaction kernel $\left[-W_{0}^{-1}\right]_{\alpha\beta}\equiv\left[-U^{-1}\right]_{\alpha\beta}$
should be positive definite for this integral to be convergent. Should
it be negative definite, positive definiteness can be restored by
redefining $\phi\rightarrow i\phi$ and $\lambda\rightarrow i\lambda$,
which leaves the final equations unchanged. After this transformation,
the electron-electron action (\ref{eq:general_action_ee}) becomes
Eq. \ref{eq:general_action_eb}, where we have chosen the minus sign
for the Yukawa coupling in Eq. (\ref{eq:generic_HS_decoupling}).} bosonic Hubbard-Stratonovich field $\phi_{\alpha}$:
\begin{align}
 & e^{-\frac{1}{2}U_{\alpha\beta}\left(\bar{c}_{\bar{u}}\lambda_{\bar{u}v\alpha}c_{v}\right)\left(\bar{c}_{\bar{w}}\lambda_{\bar{w}x\beta}c_{x}\right)}\nonumber \\
 & =\int\mathcal{D}\left[\phi\right]e^{-\frac{1}{2}\phi_{\alpha}\left[-U^{-1}\right]_{\alpha\beta}\phi_{\alpha}\pm\lambda_{\bar{u}v\alpha}\phi_{\alpha}\bar{c}_{\bar{u}}c_{v}}\label{eq:generic_HS_decoupling}
\end{align}
We have thus cast the electron-electron interaction problem in the
form of Eq. (\ref{eq:general_action_eb}), namely an electron-boson
coupling problem. We can therefore apply the formalism developed in
the previous section to the Hubbard model and similar electronic problems.
The only caveat resides with the freedom in choosing the electron-boson
problem for a given electronic problem: we discuss this at greater
length in subsection \ref{sub:Choice-of-the-decoupling-Fierz}.

For later purposes, let us now specify the equations presented in
the previous section for the Hubbard model in the normal, paramagnetic
case.

In the absence of symmetry breaking, 
\begin{align}
G_{uv} & =G_{i_{u}i_{v}}\delta_{\sigma_{u}\sigma_{v}}\label{eq:G_PM}\\
W_{\alpha\beta} & =W_{i_{\alpha}i_{\beta}}^{\eta(I_{\alpha})}\delta_{I_{\alpha}I_{\beta}}\label{eq:W_PM}
\end{align}
with $i_{u}=\left(\mathbf{R}_{u},\tau_{u}\right)$, and $\eta(0)\equiv\mathrm{ch}$,
$\eta(x)=\eta(y)=\eta(z)\equiv\mathrm{sp}$. In particular, $W^{0}\equiv W^{\mathrm{ch}}$
and $W^{x}=W^{y}=W^{z}\equiv W^{\mathrm{sp}}$. The vertex can be
parametrized as: 
\begin{equation}
\Lambda_{\bar{u}v\alpha}=\Lambda_{i_{u}i_{v}i_{\alpha}}^{\eta(I_{\alpha})}\sigma_{\sigma_{u}\sigma_{v}}^{I_{\alpha}}\label{eq:vertex_factorization}
\end{equation}
 $\Lambda_{ijk}^{\mathrm{ch}}$ and $\Lambda_{ijk}^{\mathrm{sp}}$
can thus be computed e.g. from\begin{subequations}
\begin{align}
\Lambda_{ijk}^{\mathrm{ch}} & =\Lambda_{i\uparrow,j\uparrow,k0}\label{eq:Lambda_charge}\\
\Lambda_{ijk}^{\mathrm{sp}} & =\Lambda_{i\uparrow,j\uparrow,kz}\label{eq:Lambda_spin}
\end{align}
\end{subequations}Hence, in the Heisenberg decoupling, Eqs (\ref{eq:Sigma_Hedin}-\ref{eq:P_Hedin})
simplify to (as shown in Appendix \ref{sub:Simplification-of-Sigma_and_P_homog}):\begin{subequations}

\begin{align}
\Sigma_{ij} & =-G_{il}W_{in}^{\mathrm{ch}}\Lambda_{ljn}^{\mathrm{ch}}-3G_{il}W_{in}^{\mathrm{sp}}\Lambda_{ljn}^{\mathrm{sp}}+\varphi_{j,\mathrm{ch}}\delta_{ij}\label{eq:Sigma_Hedin_homogeneous}\\
P_{mn}^{\eta} & =2G_{ml}G_{jm}\Lambda_{ljn}^{\eta}\label{eq:P_Hedin_homogeneous}
\end{align}
\end{subequations}We recall that the latin indices $i,j\dots$ stand
for space-time indices: $i=\left(\mathbf{R},\tau\right)$. The factor
of 3 in the self-energy comes from the rotation invariance, while
the factor of 2 in the polarization comes from the spin degree of
freedom. Note that $\varphi_{\mathrm{ch}}$ can be related to $\langle n\rangle$
via (see Appendix \ref{sub:Link-between-bosonic_and_fermionic}, Eq.
(\ref{eq:varphi_n})):
\begin{equation}
\varphi_{\mathrm{ch}}=U^{\mathrm{ch}}\langle n\rangle\label{eq:varphi_n_rel}
\end{equation}
This is the Hartree term. In the following, we shall omit this term
in the expressions for $\Sigma$ as it can be absorbed in the chemical
potential term.

\subsection{A local approximation to $\mathcal{K}$\label{sub:Local-approximation}}

In this subsection, we introduce an approximation to $\mathcal{K}$
for the specific case discussed in the previous subsection (subsection
\ref{sub:From-an-electron-electron}).

\subsubsection{The TRILEX approximation}

The functional derivation discussed in subsection \ref{sub:Functional-derivation}
suggests a natural extension of the local approximations on the 2PI
functionals $\Phi$ (DMFT) or $\Psi$ (EDMFT) to the 3PI functional
$\mathcal{K}$. Such an approximation reads, in the case when $\mathcal{K}$
is considered as functional of $\chi_{uv\alpha}$ (instead of $\Lambda_{uv\alpha}$):
\begin{equation}
\mathcal{K}^{\mathrm{TRILEX}}[G,W,\chi]\approx\sum_{\mathbf{R}}\mathcal{K}[G_{\mathbf{RR}},W_{\mathbf{RR}},\chi_{\mathbf{RRR}}]\label{eq:TRILEX_approximation_functional}
\end{equation}
The TRILEX functional thus contains only local diagrams. This approximation
is exact in two limits: 
\begin{itemize}
\item in the atomic limit, all correlators become local and thus $\mathcal{K}[G_{\mathbf{RR'}},W_{\mathbf{RR'}},\chi_{\mathbf{RR'R''}}]=\mathcal{K}[G_{\mathbf{RR}},W_{\mathbf{RR}},\chi_{\mathbf{RRR}}]=\mathcal{K}^{\mathrm{TRILEX}}[G,W,\chi]$;
\item in the weak-interaction limit, $W$ becomes small and thus $\mathcal{K}\approx0$,
corresponding to the absence of vertex corrections and thus to the
spin-fluctuation approximation.
\end{itemize}
The local approximation of the 3PI functional leads to a local approximation
of the 3PI analog of the self-energy, $K$ (defined in Eq. (\ref{eq:K_def})).
Indeed, noticing that $\partial\mathcal{K}/\partial\chi_{vu\alpha}=K_{vu\alpha}$,
Eq. (\ref{eq:TRILEX_approximation_functional}) leads to: 
\begin{equation}
K(\mathbf{k},\mathbf{q},i\omega,i\Omega)\approx K(i\omega,i\Omega)\label{eq:TRILEX_local_approximation}
\end{equation}

As in DMFT, we will use an effective impurity model as an auxiliary
problem to sum these local diagrams. Its fermionic Green's function,
bosonic Green's function and three-point function are denoted as $G_{\mathrm{imp}}(i\omega)$,
$W_{\mathrm{imp}}(i\Omega)$ and $\chi_{\mathrm{imp}}(i\omega,i\Omega)$
respectively. The action of the auxiliary problem is chosen such that
$\mathcal{K}^{\mathrm{imp}}[G_{\mathrm{imp}},W_{\mathrm{imp}},\chi_{\mathrm{imp}}]$
is equal (up to a factor equal to the number of sites) to $\mathcal{K}^{\mathrm{TRILEX}}$
evaluated for :\begin{subequations}
\begin{eqnarray}
\chi_{\mathbf{R}\mathbf{R}\mathbf{R}}^{\eta}(i\omega,i\Omega) & = & \chi_{\mathrm{imp}}^{\eta}(i\omega,i\Omega)\label{eq:chi_3pt_sc}\\
G_{\mathrm{\mathbf{R}\mathbf{R}}}(i\omega) & = & G_{\mathrm{imp}}(i\omega)\label{eq:G_sc}\\
W_{\mathrm{\mathbf{R}\mathbf{R}}}^{\eta}(i\Omega) & = & W_{\mathrm{imp}}^{\eta}(i\Omega)\label{eq:W_sc}
\end{eqnarray}
\end{subequations}This prescription, by imposing that the diagrams
of the impurity model have the same topology as the diagrams corresponding
to the lattice action, sets the form of impurity action as follows:
\begin{eqnarray}
S_{\mathrm{imp}} & = & \iint_{\tau\tau'}\sum_{\sigma}\bar{c}_{\sigma\tau}\left[-\mathcal{G}^{-1}(\tau-\tau')\right]c_{\sigma\tau'}\nonumber \\
 &  & +\frac{1}{2}\iint_{\tau\tau'}\phi_{I\tau}\left[-\left[\mathcal{U}^{I}\right]^{-1}(\tau-\tau')\right]\phi_{I\tau'}\label{eq:impurity_model_TRILEX_el_bo}\\
 &  & +\iiint_{\tau\tau'\tau''}\lambda_{\mathrm{imp},\sigma\sigma'}^{I}(\tau-\tau',\tau'-\tau'')\bar{c}_{\sigma\tau}c_{\sigma\tau'}\phi_{I\tau''}\nonumber 
\end{eqnarray}
The three self-consistency equations (\ref{eq:chi_3pt_sc}-\ref{eq:G_sc}-\ref{eq:W_sc})
completely determine the dynamical mean fields $\mathcal{G}(\tau)$,
$\mathcal{U}(\tau)$ and $\lambda_{\mathrm{imp}}(\tau,\tau')$. Note
that the bare vertex $\lambda_{\mathrm{imp}}(\tau,\tau')$ of the
impurity problem is \emph{a priori} different from $\lambda$, the
lattice's bare vertex, and is \emph{a priori} time-dependent. Indeed,
in addition to the two baths $\mathcal{G}$ and $\mathcal{U}$ present
in (extended) DMFT, one needs a third adjustable quantity (akin to
a Lagrange multiplier\cite{Georges2008}) in the impurity model to
enforce the third constraint, (\ref{eq:chi_3pt_sc}). This third Weiss
field is a time-dependent electron-boson interaction. As for DMFT,
the existence of Weiss fields fulfilling (\ref{eq:chi_3pt_sc}-\ref{eq:G_sc}-\ref{eq:W_sc})
is not obvious from a mathematical point of view. In practice, we
will try to construct such a model by solving iteratively the TRILEX
equations.

A direct consequence of Eq. (\ref{eq:TRILEX_local_approximation})
and (\ref{eq:chi_3pt_sc}-\ref{eq:G_sc}-\ref{eq:W_sc}) is the locality
of the lattice vertex:
\begin{equation}
\Lambda^{\eta}(\mathbf{k},\mathbf{q},i\omega,i\Omega)=\lambda^{\eta}+K_{\mathrm{imp}}^{\eta}(i\omega,i\Omega)\label{eq:Lambda_K_imp_rel}
\end{equation}
where the three-particle irreducible vertex $K_{\mathrm{imp}}^{\eta}(i\omega,i\Omega)$
is related to $\Lambda_{\mathrm{imp}}^{\eta}(i\omega,i\Omega)$ through
(see Eq. (\ref{eq:Dyson_Lambda})),
\begin{equation}
\Lambda_{\mathrm{imp}}^{\eta}(i\omega,i\Omega)=\lambda_{\mathrm{imp}}^{\eta}(i\omega,i\Omega)+K_{\mathrm{imp}}^{\eta}(i\omega,i\Omega)\label{eq:impurity_3PI_Dyson}
\end{equation}

\subsubsection{Equation for the impurity bare vertex\label{sub:Construction-of-the-impurity-vertex}}

In TRILEX, the impurity's bare vertex $\lambda_{\mathrm{imp}}^{\eta}(\tau,\tau')$
is a priori different from $\lambda^{\eta}$, the bare vertex of the
lattice problem. Like $\mathcal{G}(i\omega)$ and $\mathcal{U}(i\Omega)$
in EDMFT, it must be determined self-consistently. This can be contrasted
with D$\Gamma$A where the bare vertex of the impurity is not renormalized
and kept equal to the lattice's bare vertex, $U$.

Let us now determine the equation for $\lambda_{\mathrm{imp}}$. Using
Eq. (\ref{eq:chi_3pt_sc}) and (\ref{eq:Lambda_def}), $\Lambda_{\mathrm{imp}}$
is given by 
\[
\Lambda_{\mathrm{imp}}^{\eta}(i\omega,i\Omega)=\frac{\sum_{\mathbf{kq}}\chi^{\eta}(\mathbf{k},\mathbf{q},i\omega,i\Omega)}{G_{\mathrm{imp}}(i\omega+i\Omega)G_{\mathrm{imp}}(i\omega)W_{\mathrm{imp}}^{\eta}(i\Omega)}
\]
$\chi^{\eta}(\mathbf{k},\mathbf{q},i\omega,i\Omega)$ is given as
function of $K_{\mathrm{imp}}^{\eta}(i\omega,i\Omega)$ (after using
Eqs. (\ref{eq:Dyson_Lambda}) and (\ref{eq:TRILEX_local_approximation})),
by 
\[
\chi_{\mathbf{k},\mathbf{q},i\omega,i\Omega}^{\eta}=G_{\mathbf{k}+\mathbf{q},i\omega+i\Omega}G_{\mathbf{k},i\omega}W_{\mathbf{q},i\Omega}^{\eta}\left(\lambda^{\eta}+K_{\mathrm{imp}}^{\eta}(i\omega,i\Omega)\right)
\]
where we recall that $\lambda$ is the bare vertex \emph{on the lattice}.
Thus, $\lambda_{\mathrm{imp}}^{\eta}(i\omega,i\Omega)$ is found to
be given, as a function of $K_{\mathrm{imp}}^{\eta}$, as:

\begin{equation}
\lambda_{\mathrm{imp}}^{\eta}(i\omega,i\Omega)=\lambda^{\eta}+\zeta^{\eta}(i\omega,i\Omega)\left\{ \lambda^{\eta}+K_{\mathrm{imp}}^{\eta}(i\omega,i\Omega)\right\} \label{eq:impurity_bare_vertex_general_case}
\end{equation}
with
\begin{equation}
\zeta^{\eta}(i\omega,i\Omega)\equiv\frac{\sum_{\mathbf{k}\mathbf{q}}\tilde{G}_{\mathbf{k}+\mathbf{q},i\omega+i\Omega}\tilde{G}_{\mathbf{k},i\omega}\tilde{W}_{\mathbf{q},i\Omega}^{\eta}}{G_{\mathrm{imp}}(i\omega+i\Omega)G_{\mathrm{imp}}(i\omega)W_{\mathrm{imp}}^{\eta}(i\Omega)}\label{eq:zeta_def}
\end{equation}
where for any $X$, 
\begin{equation}
\tilde{X}(\mathbf{k},i\omega)\equiv X(\mathbf{k},i\omega)-X_{\mathrm{loc}}(i\omega)\label{eq:X_tilde}
\end{equation}
Hence, in general, $\lambda_{\mathrm{imp}}$ is different from $\lambda$:
one has to adjust the interaction of the impurity model to satisfy
Eqs (\ref{eq:chi_3pt_sc}-\ref{eq:G_sc}-\ref{eq:W_sc}).

\subsubsection{A further simplification: reduction to density-density and spin-spin
terms\label{sub:A-further-simplification:}}

The form (\ref{eq:impurity_bare_vertex_general_case}) of the bare
impurity vertex suggests a further approximation as a preliminary
step before the full-fledged interaction term is taken into account,
namely we take:
\begin{equation}
\lambda_{\mathrm{imp}}^{\eta}(i\omega,i\Omega)\approx\lambda^{\eta}\label{eq:lambda_imp_approx}
\end{equation}
This approximation is justified when $\zeta^{\eta}(i\omega,i\Omega)$,
defined in Eq. (\ref{eq:zeta_def}), is small. Let us already notice
that $\zeta^{\eta}$ vanishes in the atomic limit (when $t\rightarrow0$,
$\tilde{G}=\tilde{W}=0$) and in the weak-coupling limit (then $W^{\eta}\rightarrow U^{\eta}$
so that $\tilde{W}^{\eta}\rightarrow0$). A corollary of this simplification
is that (using (\ref{eq:Lambda_K_imp_rel})): 
\begin{equation}
\Lambda^{\eta}(\mathbf{k},\mathbf{q},i\omega,i\Omega)=\Lambda_{\mathrm{imp}}^{\eta}(i\omega,i\Omega)\label{eq:local_approx_Lambda}
\end{equation}
We will check in subsection \ref{sub:zeta_small} that this approximation
is in practice very accurate for the Hubbard model for the parameters
we have considered. 

With (\ref{eq:lambda_imp_approx}), integrating the bosonic modes
leads to a fermionic impurity model with retarded density-density
and spin-spin interactions: 
\begin{align}
S_{\mathrm{imp}} & = & \iint_{\tau\tau'}\sum_{\sigma}\bar{c}_{\sigma}(\tau)\left[-\mathcal{G}^{-1}(\tau-\tau')\right]c_{\sigma}(\tau')\label{eq:impurity_model_TRILEX_el_only}\\
 &  & +\frac{1}{2}\iint_{\tau\tau'}\sum_{I}n_{I}(\tau)\mathcal{U}^{I}(\tau-\tau')n_{I}(\tau')\nonumber 
\end{align}
The sum $\sum_{I}$ runs on $I=0,z$ in the Ising decoupling, and
on $I=0,x,y,z$ in the Heisenberg-decoupling. We recall that $n_{x}\equiv s_{x}$,
$n_{y}\equiv s_{y}$ and $n_{z}\equiv s_{z}$ have spin commutation
rules, that is, in the Heisenberg decoupling, the spin part of the
interactions explicitly reads 
\[
S_{\mathrm{int}}^{\mathrm{sp}}=\frac{1}{2}\iint_{\tau\tau'}\mathcal{U}^{\mathrm{sp}}(\tau-\tau')\vec{s}(\tau)\cdot\vec{s}(\tau')
\]
The TRILEX method is therefore solvable with the same tools as extended
DMFT. The solution of the impurity action is elaborated on in section
\ref{sec:impurity}. We will also explain how to compute $\Lambda_{\mathrm{imp}}$
from this purely fermionic action.

\subsubsection{Choice of the decoupling channels\label{sub:Choice-of-the-decoupling-Fierz}}

Due to the freedom in rewriting the interaction term, as discussed
in subsection \ref{sub:From-an-electron-electron}, there are several
possible Hubbard-Stratonovich decoupling fields. While an exact treatment
of the mixed fermion-boson action (\ref{eq:general_action_eb}) would
lead to \emph{exact} results, any approximation to the electron-boson
action will lead to \emph{a priori} different results depending on
the choice of the decoupling. This ambiguity -- called the Fierz ambiguity
-- has been thoroughly investigated in the literature in the past\cite{Castellani1979,Cornwall1974,Gomes1977,Hamann1969,Hassing1973,Macedo1982,Macedo1991,Schulz1990,Schumann1988}
and in more recent years \cite{Baier2004,Bartosch2009,Borejsza2003,Borejsza2004,Dupuis2002}
in the context of functional renormalization group (fRG) methods. 

There is no \emph{a priori} heuristics to find an optimal decoupling
without previous knowledge of the physically relevant instabilities
of the system, except when it comes to symmetries. Optimally, the
decoupling should fulfill the symmetry of the original Hamiltonian,
for instance spin-rotational symmetry. Apart from pure symmetry reasons,
in most cases of physical interest, where several degrees of freedom
-- charge, spin, superconducting fluctuations... -- are competing
with one another, many decoupling channels must be taken into account.
This ambiguity can only be dispelled by an \emph{a posteriori} control
of the error with respect to the exact solution. 

Yet, the TRILEX method can actually take advantage of this freedom
to find the physically most relevant decoupling. It can be extended
to cluster impurity problems in the spirit of cluster DMFT methods.
By going to larger and larger cluster sizes and finding the decoupling
which minimizes cluster corrections, one can identify the dominant
physical fluctuations. In this perspective, the single-site TRILEX
method presented here should be seen only as a starting point of a
systematic cluster extension.

\subsection{The TRILEX loop\label{sub:Equations-scheme}}

In this section, we summarize the TRILEX set of equations, show how
to solve it self-consistently, and finally touch on some technical
details of the computation.

\subsubsection{Summary of the equations}

We recall the Dyson equations:\begin{subequations}

\begin{eqnarray}
G(\mathbf{k},i\omega) & = & \frac{1}{i\omega+\mu-\varepsilon(\mathbf{k})-\Sigma(\mathbf{k},i\omega)}\label{eq:G_dyson}\\
W^{\eta}(\mathbf{q},i\omega) & = & \frac{U^{\eta}}{1-U^{\eta}P^{\eta}(\mathbf{q},i\Omega)}\label{eq:W_dyson}
\end{eqnarray}
\end{subequations}They are merely are Fourier transforms of the equations
(\ref{eq:Dyson_Sigma}-\ref{eq:Dyson_P}). The relation between the
bare interaction value $U^{\eta}$ on the Hubbard $U$ depends on
the choice of decoupling. It has been discussed in subsection \ref{sub:From-an-electron-electron}.

The Weiss fields are given by:\begin{subequations}
\begin{eqnarray}
\mathcal{G}(i\omega) & = & \left[G_{\mathrm{loc}}^{-1}(i\omega)+\Sigma_{\mathrm{loc}}(i\omega)\right]^{-1}\label{eq:G_weiss}\\
\mathcal{U}^{\eta}(i\Omega) & = & \left[\left[W_{\mathrm{loc}}^{\eta}(i\Omega)\right]^{-1}+P_{\mathrm{loc}}^{\eta}(i\Omega)\right]^{-1}\label{eq:U_weiss_W_sc}
\end{eqnarray}
\end{subequations}The ``$\mathrm{loc}$'' suffix denotes summation
over the Brillouin zone.

The momentum-dependent lattice self-energies are given by (see Eqs
(\ref{eq:Sigma_Hedin_homogeneous}-\ref{eq:P_Hedin_homogeneous})):\begin{subequations}
\begin{eqnarray}
\Sigma{}_{\mathbf{k},i\omega} & = & -\sum_{\eta,\mathbf{q},i\Omega}m_{\eta}G{}_{\mathbf{k}+\mathbf{q},i\omega+i\Omega}W^{\eta}{}_{\mathbf{q},i\Omega}\Lambda_{\mathrm{imp}}^{\eta}{}_{i\omega,i\Omega}\label{eq:Sigma_GWL}\\
P^{\eta}{}_{\mathbf{q},i\Omega} & = & 2\sum_{\mathbf{k},i\omega}G{}_{\mathbf{k}+\mathbf{q},i\omega+i\Omega}G{}_{\mathbf{k},i\omega}\Lambda_{\mathrm{imp}}^{\eta}{}_{i\omega,i\Omega}\label{eq:P_GGL}
\end{eqnarray}
\end{subequations}$\Lambda_{\mathrm{imp}}$ is given by the solution
of the impurity model, Eq.(\ref{eq:impurity_model_TRILEX_el_only}).
The factor $m_{\eta}$ depends on the decoupling. In the case of the
Heisenberg decoupling, $m_{\mathrm{sp}}=3$ and $m_{\mathrm{ch}}=1$,
while in the Ising decoupling, $m_{\mathrm{sp}}=m_{\mathrm{ch}}=1$
(see subsection \ref{sub:From-an-electron-electron}).

\subsubsection{Summary of the self-consistent loop\label{sub:self-consistent-loop}}

\begin{figure}
\begin{centering}
\includegraphics[width=0.8\columnwidth]{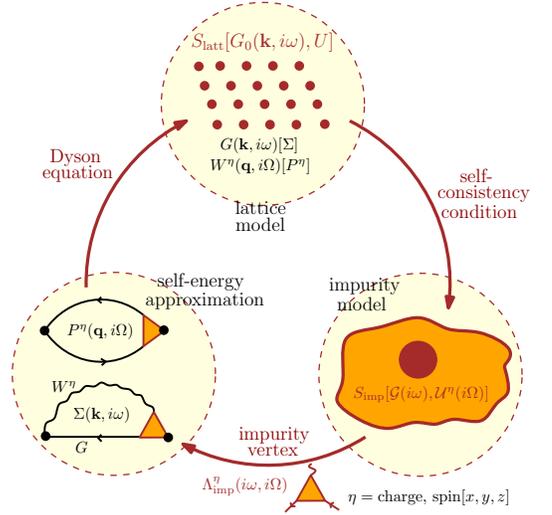}
\par\end{centering}

\caption{\label{fig:Sketch}(color online) The TRILEX self-consistency loop}
\end{figure}

The equations above can be solved self-consistently. The self-consistent
TRILEX loop consists in the following steps (as illustrated in Fig
\ref{fig:Sketch}):
\begin{enumerate}
\item \emph{Initialization}. The initialization consists in finding initial
guesses for the self-energy and polarization. Usually, converged EDMFT
self-energies provide suitable starting points for $\Sigma(\mathbf{k},i\omega)$
and $P^{\eta}(\mathbf{q},i\Omega)$. 
\item \emph{Dyson equations}. Compute lattice observables through Dyson
equations Eqs (\ref{eq:G_dyson}-\ref{eq:W_dyson}).
\item \emph{Weiss fields. }Update the Weiss fields using Eqs (\ref{eq:G_weiss}-\ref{eq:U_weiss_W_sc}).
\item \emph{Impurity model. }Solve the impurity action (\ref{eq:impurity_model_TRILEX_el_only})
for $\Lambda_{\mathrm{imp}}^{\eta}(i\omega,i\Omega)$, $\Sigma_{\mathrm{imp}}(i\omega)$
and $P_{\mathrm{imp}}(i\Omega)$
\item \emph{Self-energies. }Construct momentum-dependent lattice self-energies
using (\ref{eq:Sigma_GWL}-\ref{eq:P_GGL}).
\item Go back to step 2 until convergence
\end{enumerate}

\subsubsection{Bubble with local vertices\label{sub:Bubble-with-local-vertices}}

The calculation of the self-energies (\ref{eq:Sigma_GWL}-\ref{eq:P_GGL})
has to be carried out carefully for reasons of accuracy and speed. 

In order to avoid the infinite summation of slowly decaying summands,
we decompose (see Appendix \ref{sub:Decomposition-of-sigma_and_P})
this computation in the following way:\begin{subequations}

\begin{eqnarray}
\Sigma(\mathbf{k},i\omega) & = & \Sigma^{\mathrm{nonloc}}(\mathbf{k},i\omega)+\Sigma_{\mathrm{imp}}(i\omega)\label{eq:Sigma_decomp}\\
P^{\eta}(\mathbf{q},i\Omega) & = & P^{\eta,\mathrm{nonloc}}(\mathbf{q},i\Omega)+P_{\mathrm{imp}}^{\eta}(i\Omega)\label{eq:P_decomp}
\end{eqnarray}
\end{subequations}with\begin{subequations}
\begin{eqnarray}
\Sigma{}_{\mathbf{k},i\omega}^{\mathrm{nonloc}} & \equiv & -\sum_{\eta,\mathbf{q},i\Omega}m_{\eta}\tilde{G}{}_{\mathbf{k}+\mathbf{q},i\omega+i\Omega}\tilde{W}^{\eta}{}_{\mathbf{q},i\Omega}\Lambda_{\mathrm{imp}}^{\eta}{}_{i\omega,i\Omega}\label{eq:Sigma_nonloc}\\
P^{\eta}{}_{\mathbf{q},i\Omega}^{\mathrm{nonloc}} & \equiv & \sum_{\mathbf{k},i\omega}\tilde{G}{}_{\mathbf{k}+\mathbf{q},i\omega+i\Omega}\tilde{G}{}_{\mathbf{k},i\omega}\Lambda_{\mathrm{imp}}^{\eta}{}_{i\omega,i\Omega}\label{eq:P_nonloc}
\end{eqnarray}
\end{subequations}We also perform a further decomposition at the
level of the vertex:
\begin{equation}
\Lambda_{\mathrm{imp}}^{\eta}(i\omega,i\Omega)=\Lambda_{\mathrm{imp}}^{\eta,\mathrm{reg}}(i\omega,i\Omega)+l^{\eta}(i\Omega)\label{eq:vertex_decomp}
\end{equation}
where $l^{\eta}(i\Omega)\equiv\frac{1-\mathcal{\tilde{U}}^{\eta}(i\Omega)\chi_{\mathrm{imp}}^{\eta}(i\Omega)}{1-\mathcal{U}^{\eta}(i\Omega)\chi_{\mathrm{imp}}^{\eta}(i\Omega)}$,
and $\tilde{\mathcal{U}}^{\eta}$ is computed with $U^{\eta}$ given
by Eqs. (\ref{eq:U_ch_fierz-zonly}-\ref{eq:U_sp_fierz-zonly}) with
$\alpha=1/2$. This choice corresponds to a subtraction from $\tilde{\chi}_{\mathrm{imp}}^{\eta}(i\omega,i\Omega)$
of its asymptotic behavior. 

The final expressions are:
\begin{eqnarray}
\Sigma(\mathbf{k},i\omega) & = & -\left\{ \sum_{\eta,\mathbf{q},i\Omega}m_{\eta}\tilde{G}{}_{\substack{\mathbf{k}+\mathbf{q}\\
i\omega+i\Omega
}
}\left[\tilde{W}_{\substack{\mathbf{q}\\
i\Omega
}
}^{\eta}l_{i\Omega}^{\eta}\right]\right\} \label{eq:Sigma_decomp-1}\\
 &  & -\sum_{\eta,\mathbf{q},i\Omega}m_{\eta}\tilde{G}_{\substack{\mathbf{k}+\mathbf{q}\\
i\omega+i\Omega
}
}\tilde{W}^{\eta}{}_{\substack{\mathbf{q}\\
i\Omega
}
}\left[\Lambda_{\mathrm{imp}}^{\eta,\mathrm{reg}}\right]{}_{\substack{i\omega\\
i\Omega
}
}+\Sigma_{\mathrm{imp}}(i\omega)\nonumber \\
P^{\eta}(\mathbf{q},i\Omega) & = & 2\left\{ \sum_{\mathbf{k},i\omega}\tilde{G}_{\substack{\mathbf{k}+\mathbf{q}\\
i\omega+i\Omega
}
}\tilde{G}{}_{\substack{\mathbf{k}\\
i\omega
}
}\right\} l_{i\Omega}^{\eta}\label{eq:P_decomp-1}\\
 &  & +2\sum_{\mathbf{k},i\omega}\tilde{G}_{\substack{\mathbf{k}+\mathbf{q}\\
i\omega+i\Omega
}
}\tilde{G}_{\substack{\mathbf{k}\\
i\omega
}
}\left[\Lambda_{\mathrm{imp}}^{\eta,\mathrm{reg}}\right]_{\substack{i\omega\\
i\Omega
}
}+P_{\mathrm{imp}}^{\eta}(i\Omega)\nonumber 
\end{eqnarray}
The first term of each expression (in curly braces) is computed as
a simple product in time and space instead of a convolution in frequency
and momentum. The second term, which contains factors decaying fast
in frequencies ($\tilde{G}$, $\tilde{W}$, $\Lambda_{\mathrm{imp}}^{\mathrm{reg}}$),
is computed as a product in space and convolution in frequencies.
The spatial Fourier transforms are performed using Fast Fourier Transforms
(FFT), so that the computational expense of such calculations scales
as $N_{\omega}^{2}N_{k}\log N_{k}$, where $N_{\omega}$ is the number
of Matsubara frequencies and $N_{k}$ the number of discrete points
in the Brillouin zone.

The formulae (\ref{eq:Sigma_decomp}-\ref{eq:P_decomp}) are reminiscent
of the form of $\Sigma$ and $P$ in the GW+EDMFT approximation (see
e.g. Ref. \onlinecite{Ayral2013}). The main difference is that in
GW+EDMFT, (a) there is no local vertex correction in (\ref{eq:Sigma_nonloc}-\ref{eq:P_nonloc}),
and (b) so far GW+EDMFT has been formulated for the charge channel
only.

\subsubsection{Self-consistencies and alternative schemes\label{sub:Self-consistencies-and-alternati}}

At this point, it should be pointed out that this choice of self-consistency
conditions is not unique. In particular, inspired by the sum rules
imposed in the two-particle self-consistent approximation (TPSC\cite{Tremblay2012})
or by the ``Moriya corrections'' of the ladder version of D$\Gamma$A\cite{Katanin2009},
one may replace Eq (\ref{eq:W_sc}) by:
\begin{equation}
\chi_{\mathrm{loc}}^{\eta}(i\Omega)=\chi_{\mathrm{imp}}^{\eta}(i\Omega)\label{eq:chi_sc}
\end{equation}
where $\chi^{\eta}$ (with one frequency, not to be confused with
the three-point function) denotes the (connected) susceptibility in
channel $\eta$: 
\begin{equation}
\chi_{ij}^{\eta}\equiv\langle\left(n_{i}^{\eta}-\langle n_{i}^{\eta}\rangle\right)\left(n_{j}^{\eta}-\langle n_{j}^{\eta}\rangle\right)\rangle\label{eq:chi_def}
\end{equation}
This relation enforces sum rules on the double occupancy (among others)
and has been shown to yield good results in the TPSC and ladder-D$\Gamma$A
context, namely good agreement with exact Monte-Carlo results as well
as the fulfillment of the Mermin-Wagner theorem.\cite{Vilk1994,Vilk1997,Schafer2014}

Even when using Eq (\ref{eq:W_sc}), however, we have shown that the
sum rules are not violated for parameters where stable solutions can
be obtained.\cite{Ayral2015}

\section{Solution of the Impurity Model\label{sec:impurity}}

The impurity model (\ref{eq:impurity_model_TRILEX_el_only}) with
dynamical interactions in the charge and vector spin channel can be
solved exactly with a continuous-time quantum Monte-Carlo (CTQMC)
algorithm\cite{Rubtsov2011} either in the hybridization expansion
or in the interaction expansion.

In this paper, we use the hybridization expansion algorithm\cite{Werner2006,Werner2007}.
Retarded vector spin-spin interactions are implemented as described
in Ref. \onlinecite{Otsuki2013}. Our implementation is based on the
TRIQS toolbox.\cite{Parcollet2014} 

In this section, we give an alternative derivation of the algorithm
presented in Ref. \onlinecite{Otsuki2013}. It uses a path integral
approach, thereby allowing for a more concise presentation.

\subsection{Overview of the CTQMC algorithm}

Eq. (\ref{eq:impurity_model_TRILEX_el_only}) can be decomposed as
$S_{\mathrm{imp}}=S_{\mathrm{loc}}+S_{\mathrm{hyb}}+S_{\perp}$, with:\begin{subequations}
\begin{align*}
S_{\mathrm{loc}} & \equiv\int_{0}^{\beta}\mathrm{d}\tau\sum_{\sigma}\bar{c}_{\sigma}(\tau)\left(\partial_{\tau}-\mu\right)c_{\sigma}(\tau)\\
 & +\frac{1}{2}\iint_{0}^{\beta}\mathrm{d}\tau\mathrm{d}\tau'\sum_{\sigma\sigma'}n_{\sigma}(\tau)\mathcal{U}_{\sigma\sigma'}(\tau-\tau')n_{\sigma'}(\tau')\\
S_{\mathrm{hyb}} & \equiv\int_{0}^{\beta}\mathrm{d}\tau\int_{0}^{\beta}\mathrm{d}\tau'\sum_{\sigma}\bar{c}_{\sigma}(\tau)\Delta_{\sigma}(\tau-\tau')c_{\sigma}(\tau')\\
S_{\perp} & \equiv\frac{1}{2}\iint_{0}^{\beta}\mathrm{d}t\mathrm{d}t'\mathcal{J}_{\perp}(t-t')s_{+}(t)s_{-}(t')
\end{align*}
\end{subequations}where $\Delta$ is related to $\mathcal{G}$ through
$\mathcal{G}_{\sigma}^{-1}(i\omega)\equiv i\omega+\mu-\Delta_{\sigma}(i\omega)$,
$\mathcal{U}_{\sigma\sigma'}(\tau)\equiv\mathcal{U}^{\mathrm{ch}}(\tau)+(-)^{\sigma\sigma'}\mathcal{U}^{\mathrm{sp}}(\tau)$,
$s_{\pm}\equiv\left(n^{x}\pm in^{y}\right)/2$, and $\mathcal{J}_{\perp}(\tau)\equiv4\mathcal{U}^{\mathrm{sp}}(\tau)$.
Note that $S_{\perp}$ is absent in the $z$-decoupling case. 

We expand the partition function $Z\equiv\int\mathcal{D}[\bar{c}c]e^{-S_{\mathrm{loc}}-S_{\mathrm{hyb}}-S_{\perp}}$
in powers of $S_{\mathrm{hyb}}$ and $S_{\perp}$, which yields:

\begin{align*}
Z_{\mathrm{imp}}= & \sum_{k_{\sigma}=0}^{\infty}\sum_{m=0}^{\infty}\int_{>}\mathrm{d}\bm{\tau}^{\sigma}\int_{>}\mathrm{d}\bm{\tau'}^{\sigma}\int_{>}\mathrm{d}\bm{t}\int_{>}\mathrm{d}\bm{t'}\\
\times & \prod_{\sigma}\det\mathbf{\Delta}_{\sigma}\sum_{p\in\mathfrak{S}_{m}}\prod_{i=1}^{m}\left\{ \frac{-\mathcal{J}_{\perp}(t_{p(i)}-t_{i}^{'})}{2}\right\} \\
\times & \mathrm{Tr}\left\{ Te^{-S_{\mathrm{loc}}}\prod_{\sigma}\prod_{i=1}^{k_{\sigma}}c_{\sigma}(\tau_{i}^{'\sigma})c_{\sigma}^{\dagger}(\tau_{i}^{\sigma})\prod_{j=1}^{m}s_{+}(t_{i})s_{-}(t_{i}^{'})\right\} 
\end{align*}
where $k_{\sigma}$ (resp. $m$) denotes the expansion order in powers
of $S_{\mathrm{hyb}}$ (resp $S_{\perp}$), $\int_{>}$ denotes integration
over times sorted in decreasing order, $\bm{\tau}^{\sigma}\equiv(\tau_{1}^{\sigma}\dots\tau_{k_{\sigma}}^{\sigma})$,
$\mathbf{t}\equiv(t_{1}\dots t_{m})$. Using permutations of the $c$
and $c^{\dagger}$ operators in the time-ordered product, we have
grouped the hybridization terms into a determinant ($\mathbf{\Delta}_{\sigma}$
is the matrix $\left(\mathbf{\Delta}_{\sigma}\right)_{kl}\equiv\Delta_{\sigma}(\tau_{k}^{\sigma}-\tau_{l}^{'\sigma})$).
The term $\sum_{p\in\mathfrak{S}_{m}}\prod_{i=1}^{m}\left\{ -\frac{1}{2}\mathcal{J}_{\perp}(t_{p(i)}-t_{i}^{'})\right\} $
(where $\mathscr{\mathfrak{S}}_{m}$ is the group of permutations
of order $m$) is the permanent of the matrix $[\tilde{\bm{\mathcal{J}}}_{\perp}]_{ij}=-\frac{1}{2}\mathcal{J}_{\perp}(t_{i}-t_{j}^{'})$,
but since there is no efficient way of computing the permanent\cite{Valiant1979}
(contrary to the determinant), we will sample it. Finally, for any
$X$, we define $\mathrm{Tr}\left[X\right]=\sum_{\bm{\gamma}}\langle\bm{\gamma}|X|\bm{\gamma}\rangle$,
where $\bm{\gamma}$ is an eigenstate of the local action. 

We express this multidimensional sum as a sum over configurations,
namely 
\begin{equation}
Z_{\mathrm{imp}}=\sum_{\mathscr{C}}w_{\mathscr{C}}\label{eq:MC_sum}
\end{equation}
with 
\begin{equation}
\mathscr{C}\equiv\left\{ \mathbf{\bm{\tau}}^{\sigma},\mathbf{\mathbf{\bm{\tau'}^{\sigma}}},\bm{\gamma},\mathbf{t},\mathbf{t}',p\right\} \label{eq:MC_config}
\end{equation}
 This sum is computed using Monte-Carlo sampling in the space of configurations.
The weight $w_{\mathscr{C}}$ used to compute the acceptance probabilities
of each Monte-Carlo update is given by: 
\begin{equation}
w_{\mathscr{C}}=w_{\Delta}(\mathbf{\bm{\tau}}^{\sigma},\mathbf{\mathbf{\bm{\tau'}^{\sigma}}})w_{\perp}(\mathbf{t},\mathbf{t}',p)w_{\mathrm{loc}}(\mathbf{\bm{\tau}}^{\sigma},\mathbf{\mathbf{\bm{\tau'}^{\sigma}}},\bm{\gamma},\mathbf{t},\mathbf{t}')\label{eq:MC_weight_total}
\end{equation}
with\begin{subequations}
\begin{align}
w_{\Delta}(\mathbf{\bm{\tau}}^{\sigma},\mathbf{\mathbf{\bm{\tau'}^{\sigma}}})\equiv & \prod_{\sigma}\det\mathbf{\Delta}_{\sigma}\label{eq:w_Delta}\\
w_{\perp}(\mathbf{t},\mathbf{t}',p)\equiv & \prod_{i=1}^{m}\left\{ \frac{-\mathcal{J}_{\perp}(t_{p(i)}-t_{i}^{'})}{2}\right\} \label{eq:w_perp}\\
w_{\mathrm{loc}}(\mathbf{\bm{\tau}}^{\sigma},\mathbf{\mathbf{\bm{\tau'}^{\sigma}}},\bm{\gamma},\mathbf{t},\mathbf{t}')\equiv & \langle\bm{\gamma}|e^{-S_{\mathrm{loc}}}\prod_{\sigma}\prod_{i=1}^{k_{\sigma}}c_{\sigma}(\tau_{i}^{'\sigma})\bar{c}_{\sigma}(\tau_{i}^{\sigma})\nonumber \\
 & \prod_{j=1}^{m}s_{+}(t_{i})s_{-}(t_{i}^{'})|\bm{\gamma}\rangle\label{eq:w_loc}
\end{align}
\end{subequations}
\begin{figure}
\begin{centering}
\includegraphics[width=1\columnwidth]{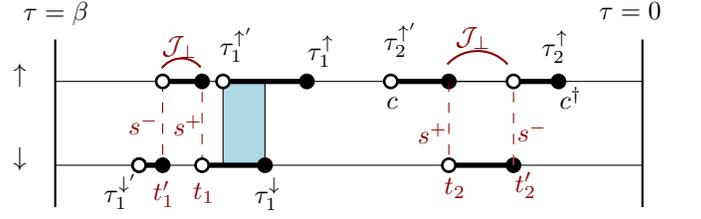}
\par\end{centering}

\caption{(color online) Pictorial representation of a Monte-Carlo configuration
$\mathscr{C}$. Full (empty) circles stand for creation (annihilation)
operators. The occupied portions of the imaginary time axis are represented
by bold segments. The red lines represent $\mathcal{J}_{\perp}$ lines.
The blue region represent an overlap between two segments.\label{fig:segment_picture_spin_spin}}
\end{figure}
Since the local action $S_{\mathrm{loc}}$ commutes with $n_{\sigma}$,
the configuration can be represented as a collection of time-ordered
``segments''\cite{Werner2006}, as illustrated in Fig. \ref{fig:segment_picture_spin_spin}.
In this segment picture, the local weight $w_{\mathrm{loc}}$ can
be simplified to:
\begin{align*}
 & w_{\mathrm{loc}}(\mathbf{\bm{\tau}}^{\sigma},\mathbf{\mathbf{\bm{\tau'}^{\sigma}}},\bm{\gamma},\mathbf{t},\mathbf{t}')\\
 & \;=e^{-\sum_{\sigma<\sigma'}\overline{U}_{\sigma\sigma'}O_{\sigma'\sigma}(\bm{\gamma})+\bar{\mu}_{\sigma}l_{\sigma}(\bm{\gamma})}w_{\mathrm{dyn}}(\mathbf{\bm{\tau}}^{\sigma},\mathbf{\mathbf{\bm{\tau'}^{\sigma}}},\mathbf{t},\mathbf{t}')
\end{align*}
with
\begin{align*}
\overline{U}_{\sigma\sigma'} & \equiv U_{\sigma\sigma'}-2\partial_{\tau}K_{\sigma\sigma'}(0^{+})\\
\bar{\mu}_{\sigma} & \equiv\mu+\partial_{\tau}K_{\sigma\sigma}(0^{+})
\end{align*}
The dynamical kernel $K(\tau)$ is defined as $\partial_{\tau}^{2}K_{\sigma\sigma'}(\tau)=\mathcal{U}_{\sigma\sigma'}(\tau)$
and $K_{\sigma\sigma'}(0^{+})=K_{\sigma\sigma'}(\beta^{-})=0$. $O_{\sigma\sigma'}(\bm{\gamma})$
denotes the total overlap between lines $\sigma$ and $\sigma'$ (blue
region in Fig. \ref{fig:segment_picture_spin_spin}), and $l_{\sigma}(\bm{\gamma})$
the added length of the segments of line $\sigma$. Both depend on
$\bm{\gamma}$ if there are lines devoid of operators (if we note
$\bm{\gamma}\equiv|n_{\uparrow},n_{\downarrow}\rangle$ in the number
representation, with $n_{\sigma}=0$ or 1, whenever a ``line'' $\sigma$
has at least one operator, only one $n_{\sigma}$ yields a nonzero
contribution, which sets its value: $n_{\sigma}$ must be specified
only for lines with no operators). Finally, the contribution to the
weight stemming from dynamical interactions is given by:\cite{Werner2007}
\begin{align*}
\ln w_{\mathrm{dyn}}(\mathbf{\bm{\tau}}^{\sigma},\mathbf{\mathbf{\bm{\tau'}^{\sigma}}},\mathbf{t},\mathbf{t}') & \equiv\sum_{\substack{\mathrm{ops}\\
a<b
}
}\zeta_{a}\zeta_{b}K_{\sigma_{1}(a)\sigma_{2}(b)}(\tilde{\tau}_{a}-\tilde{\tau}_{b})
\end{align*}
where $\zeta_{a}$ is positive (resp. negative) if $a$ corresponds
to a creation (resp. annihilation) operator, and $\tilde{\tau}$ stands
for $\tau$ ($\tau'$) for a creation (annihilation) operator. $\sum_{\substack{\mathrm{ops}\\
a<b
}
}$ denotes summation over all operator pairs in a configuration (there
are $\sum_{\sigma}2k_{\sigma}+4m$ such operators in a configuration). 

The Monte-Carlo updates required for ergodicity in the regimes of
parameters studied in this paper are (a) the insertion and removal
of segments $\{c,c^{\dagger}\}$, (b) the insertion and removal of
``spin'' segments $\{s_{+},s_{-}\}$, (c) the permutation of the
end points $\mathcal{J}_{\perp}$ lines ($p\rightarrow p'$). They
are described in more detail in Ref. \onlinecite{Otsuki2013}. In
the insulating phase at low temperatures, an additional update consisting
in moving a segment from one line to another prevents spurious spin
polarizations from appearing.

In the absence of vector spin-spin interactions, the sign of a configuration
is positive, \emph{i.e} the sign of $w_{\mathrm{loc}}w_{\Delta}$
is positive in the absence of $s_{\pm}$ operators. The introduction
of the latter does not change this statement for $w_{\mathrm{loc}}w_{\Delta}$.
The sign of $w_{\mathscr{C}}$ thus reduces to that of $w_{\perp}$:
from Eq. (\ref{eq:w_perp}), one sees that $w_{\mathscr{C}}$ is positive
if and only if $\mathcal{J}_{\perp}(\tau)<0$. In practice, $\mathcal{J}_{\perp}(\tau)$
is always negative in the self-consistency introduced in subsection
\ref{sub:self-consistent-loop}. By contrast, it is usually positive
for the alternative self-consistency introduced in subsection \ref{sub:Self-consistencies-and-alternati},
leading to a severe Monte-Carlo sign problem.

\subsection{Computation of the vertex}

The vertex is defined in Eq. (\ref{eq:Lambda_def}) as the amputated,
connected electron-boson correlation function $\chi(i\omega,i\Omega)$
(itself defined in Eq. (\ref{eq:chi3_nc_def})). Yet, since the impurity
action is written in terms of fermionic variables only, $\Lambda_{\mathrm{imp}}(i\omega,i\Omega)$
is computed from the \emph{fermionic} three-point correlation function
$\tilde{\chi}_{\mathrm{imp}}$ through the relation (see Eqs (\ref{eq:W_chi}-\ref{eq:chi_chi_tilde})
of Appendix \ref{sub:Link-between-bosonic_and_fermionic} for a general
derivation):
\begin{align}
 & \Lambda_{\mathrm{imp}}^{\eta}(i\omega,i\Omega)\label{eq:Lambda_imp_from_chi}\\
 & =\frac{\tilde{\chi}_{\mathrm{imp}}^{\eta}(i\omega,i\Omega)}{G_{\mathrm{imp}}(i\omega)G_{\mathrm{imp}}(i\omega+i\Omega)\left(1-\mathcal{U}^{\eta}(i\Omega)\chi_{\mathrm{imp}}^{\eta}(i\Omega)\right)}\nonumber 
\end{align}
where:
\begin{equation}
\tilde{\chi}_{\mathrm{imp}}^{\eta}(i\omega,i\Omega)\equiv\tilde{\chi}_{\mathrm{imp}}^{\eta,\mathrm{nc}}(i\omega,i\Omega)+\beta G_{\mathrm{imp}}(i\omega)\langle n_{\mathrm{imp}}^{\eta}\rangle\delta_{i\Omega}\label{eq:chi_tilde_conn}
\end{equation}
where:

\begin{eqnarray*}
\tilde{\chi}_{\mathrm{imp}}^{\mathrm{ch},\mathrm{nc}}(i\omega,i\Omega) & = & \tilde{\chi}_{\mathrm{imp}}^{\uparrow\uparrow,\mathrm{nc}}(i\omega,i\Omega)+\tilde{\chi}_{\mathrm{imp}}^{\uparrow\downarrow,\mathrm{nc}}(i\omega,i\Omega)\\
\tilde{\chi}_{\mathrm{imp}}^{\mathrm{sp,\mathrm{nc}}}(i\omega,i\Omega) & = & \tilde{\chi}_{\mathrm{imp}}^{\uparrow\uparrow,\mathrm{nc}}(i\omega,i\Omega)-\tilde{\chi}_{\mathrm{imp}}^{\uparrow\downarrow,\mathrm{nc}}(i\omega,i\Omega)
\end{eqnarray*}
$\tilde{\chi}_{\mathrm{imp}}^{\sigma\sigma',\mathrm{nc}}(i\omega,i\Omega)$
is the Fourier transform of 
\begin{equation}
\tilde{\chi}_{\mathrm{imp}}^{\sigma\sigma',\mathrm{nc}}(\tau,\tau')\equiv\langle Tc_{\sigma}(\tau)c_{\sigma}^{\dagger}(0)n_{\sigma'}(\tau')\rangle\label{eq:chi_tilde_tau_def}
\end{equation}
(see Eq.(\ref{eq:FT_fermion_boson}) for a definition of the Fourier
transform). 

The measurement of $\tilde{\chi}_{\mathrm{imp}}^{\sigma\sigma',\mathrm{nc}}(i\omega,i\Omega)$,
$G_{\mathrm{imp}}(i\omega)$ and $\chi_{\mathrm{imp}}^{\eta}(i\Omega)$
(defined in Eq. (\ref{eq:chi_def})) are carried out as described
in Ref. \onlinecite{Hafermann2013}.

\subsection{Computation of the self-energies}

Although only the three-leg vertex $\Lambda_{\mathrm{imp}}(i\omega,i\Omega)$
is in principle required to compute the momentum-dependent self-energies
through (\ref{eq:Sigma_GWL}-\ref{eq:P_GGL}), the impurity self-energy
and polarization may be needed for numerical stability reasons, as
explained in Section \ref{sub:Bubble-with-local-vertices}. $\Sigma_{\mathrm{imp}}(i\omega)$
is computed using improved estimators (see Ref. \onlinecite{Hafermann2013}),
namely $\Sigma_{\mathrm{imp}}(i\omega)$ is not computed from Dyson's
equation (local version of Eq. (\ref{eq:Dyson_Sigma})) but using
equations of motion (see Eq. (\ref{eq:Sigma_G})). Combined with (\ref{eq:chi_chi_tilde})
and specialized for local quantities in the paramagnetic phase, the
latter equation becomes:
\[
F_{\sigma,\mathrm{imp}}(\tau)=\sum_{I,\sigma'}\sigma_{\sigma\sigma'}^{I}\int_{0}^{\beta}d\bar{\tau}\mathcal{U}^{I}(\tau-\bar{\tau})\langle c_{\sigma'}(\tau)\bar{c}_{\sigma}(0)n^{I}(\bar{\tau})\rangle
\]
where $F_{\sigma,\mathrm{imp}}(\tau)\equiv\int d\bar{\tau}\Sigma_{\sigma,\mathrm{imp}}(\tau-\bar{\tau})G_{\sigma,\mathrm{imp}}(\bar{\tau})$.
In the Ising decoupling case ($I=0,z$), this reduces to

\begin{equation}
F_{\sigma,\mathrm{imp}}(\tau)=\sum_{\sigma'}\int_{0}^{\beta}d\bar{\tau}\mathcal{U}_{\sigma\sigma'}(\tau-\bar{\tau})\langle c_{\sigma}(\tau)\bar{c}_{\sigma}(0)n_{\sigma'}(\bar{\tau})\rangle\label{eq:F_improved_est_zonly}
\end{equation}
while in the Heisenberg decoupling case ($I=0,x,y,z$), one gets:

\begin{align}
F_{\sigma,\mathrm{imp}}(\tau) & =\int_{0}^{\beta}d\bar{\tau}\mathcal{U}^{\mathrm{ch}}(\tau-\bar{\tau})\tilde{\chi}^{\mathrm{ch}}(\tau,\bar{\tau})\nonumber \\
 & +3\int_{0}^{\beta}d\bar{\tau}\mathcal{U}^{\mathrm{sp}}(\tau-\bar{\tau})\tilde{\chi}^{\mathrm{sp}}(\tau,\bar{\tau})\label{eq:F_improved_est_xyz}
\end{align}
$G_{\sigma,\mathrm{imp}}(\tau)$ and $F_{\sigma,\mathrm{imp}}(\tau)$
are measured in the impurity solver, and the self-energy is finally
computed as
\begin{equation}
\Sigma_{\sigma,\mathrm{imp}}(i\omega)=\frac{F_{\sigma,\mathrm{imp}}(i\omega)}{G_{\sigma,\mathrm{imp}}(i\omega)}\label{eq:Sigma_improved_est}
\end{equation}
The polarization $P_{\mathrm{imp}}^{\eta}(i\Omega)$ is computed from
the correlation function $\chi_{\mathrm{imp}}^{\eta}(i\Omega)$ by
combining Eq. (\ref{eq:W_chi}) and the local version of (\ref{eq:Dyson_P}),
\emph{i.e}:
\begin{equation}
P_{\mathrm{imp}}^{\eta}(i\Omega)=\frac{-\chi_{\mathrm{imp}}^{\eta}(i\Omega)}{1-\mathcal{U}^{\eta}(i\Omega)\chi_{\mathrm{imp}}^{\eta}(i\Omega)}\label{eq:P_imp_from_chi_imp}
\end{equation}

\section{Application to the single-band Hubbard model\label{sec:Discussion}}

In this section, we elaborate on the results presented in a prior
publication (Ref. \onlinecite{Ayral2015}), where we have applied
the TRILEX method in its single-site version to the single-band Hubbard
model on a two-dimensional square lattice. 

The main conclusions of Ref. \onlinecite{Ayral2015} were the following:
\begin{itemize}
\item the TRILEX method interpolates between the spin-fluctuation regime
and the Mott regime. In the intermediate regime, both the polarization
and self-energy have a substantial momentum dependence.
\item upon doping, one finds an important variation of the spectral weight
on the Fermi surface, reminiscent of the Fermi arcs observed in angle-resolved
photoemission experiments.
\item the choice of the ratio of the charge to the spin channel does not
significantly impact the fulfillment of sum rules on the charge and
the spin susceptibility, and leads to variations only in the intermediate
regime of correlations.
\end{itemize}
In the following section, we focus on four additional aspects of the
method: (i) we show that the simplification of the impurity action
introduced in subsection \ref{sub:A-further-simplification:} is justified
\emph{a posteriori}; (ii) we show that TRILEX has, like DMFT, a first-order
Mott transition, (iii) we investigate the effect of frustration on
antiferromagnetic fluctuations in the method and, (iv) we give further
details on the influence of the decoupling choice.

\subsection{Check of the validity of $\lambda_{\mathrm{imp}}\approx\lambda$\label{sub:zeta_small}}

\begin{figure}
\begin{centering}
\includegraphics[width=1\columnwidth]{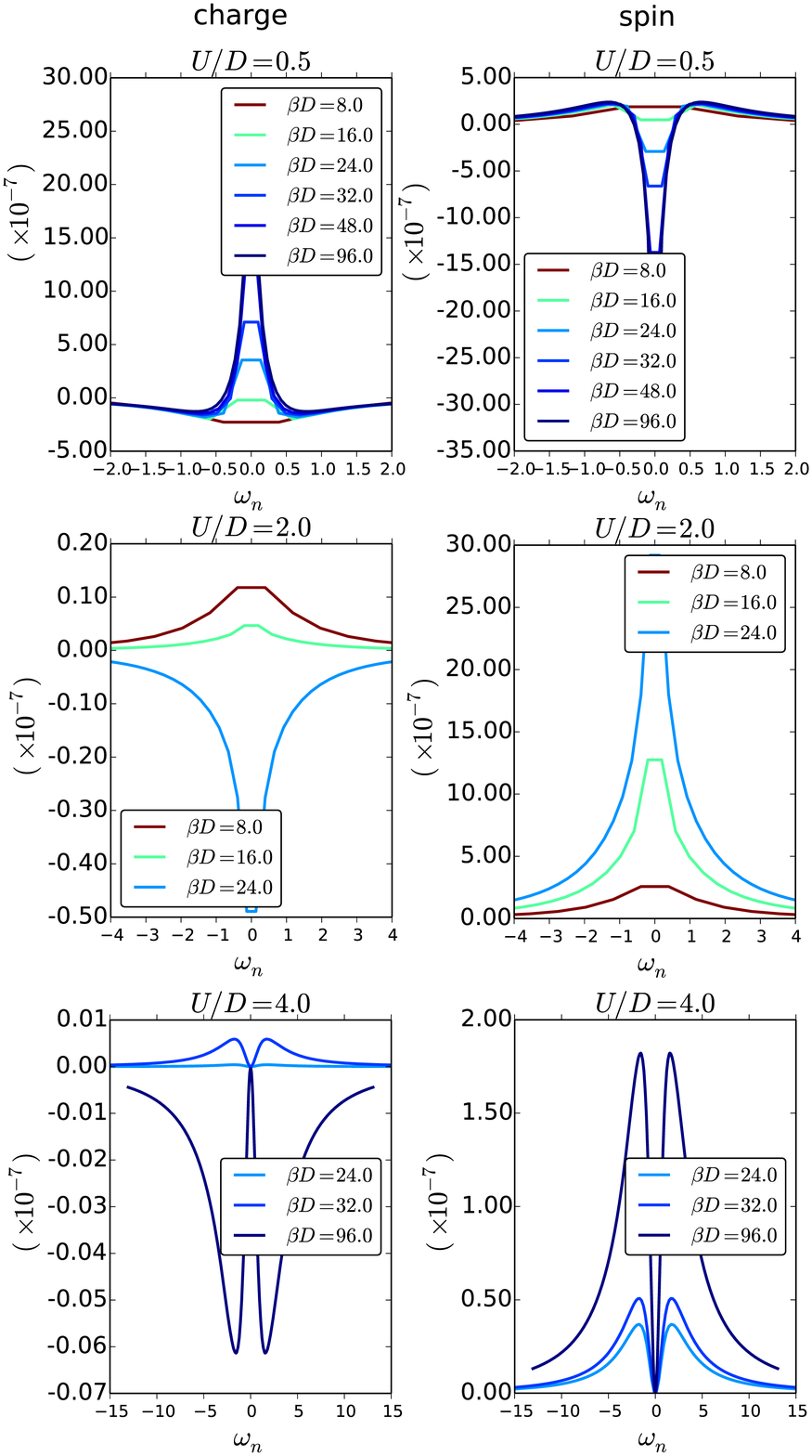}
\par\end{centering}

\caption{(color online) Evolution of $\zeta^{\eta}(i\omega_{n},i\Omega_{0})$
on the square lattice at half filling. Left column: charge channel.
Right column: spin channel. From top to bottom: $U=0.5$, $U=2.0$,
$U=4.0$.\label{fig:zeta_plots}}

\end{figure}

\begin{figure}
\begin{centering}
\includegraphics[width=1\columnwidth]{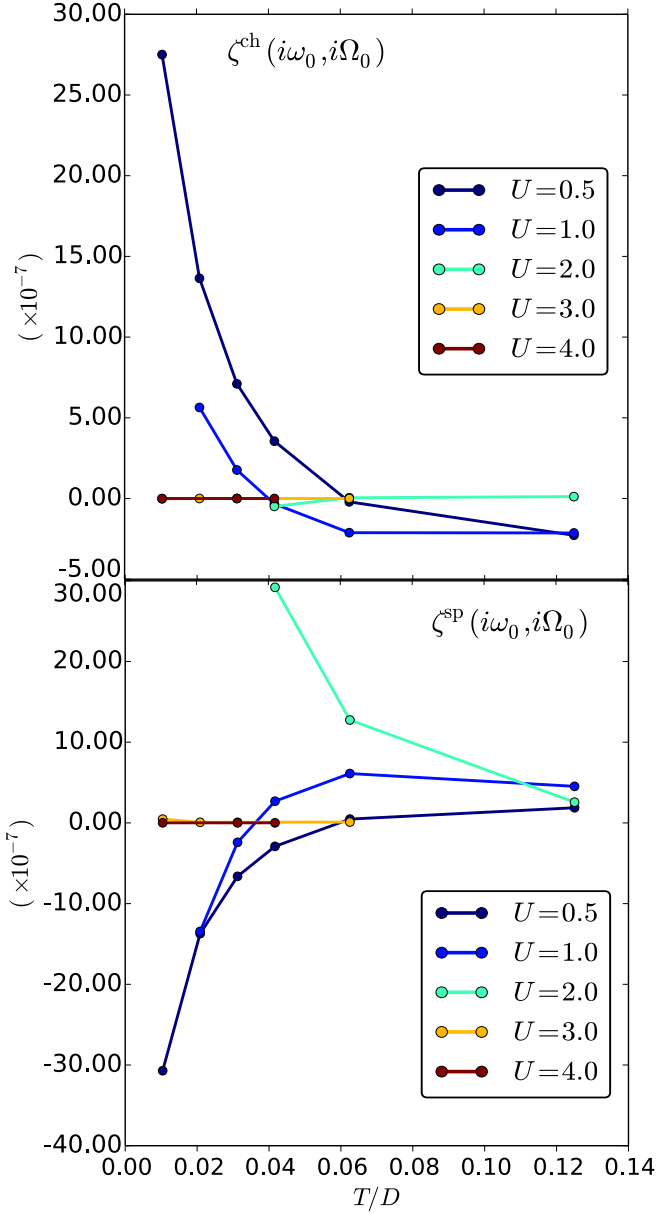}
\par\end{centering}

\caption{(color online) Dependence of $\zeta^{\eta}(i\omega_{n},i\Omega_{0})$
on temperature (square lattice, half filling). Top panel: charge channel.
Bottom panel: spin channel.\label{fig:zeta_T_dep}}
\end{figure}

The impurity action obtained after making a local expansion of the
3PI functional $\mathcal{K}$ (Eq. \ref{eq:TRILEX_approximation_functional})
contains a bare electron-boson vertex $\lambda_{\mathrm{imp}}(i\omega,i\Omega)$
which is \emph{a priori} different from $\lambda$, the bare electron-boson
vertex of the lattice action. For simplicity's sake, we have introduced
in subsection \ref{sub:A-further-simplification:} an additional approximation
where these two vertices are regarded as equal: the general case with
a frequency-dependent $\lambda_{\mathrm{imp}}(i\omega,i\Omega)$ would
require an impurity solver capable of handling retarded interaction
terms depending on three times (like the weak-coupling expansion solver). 

The deviation between both vertices is parametrized by the function
$\zeta^{\eta}(i\omega,i\Omega)$, defined in Eq. (\ref{eq:zeta_def}).
For all the converged points shown in the various phase diagrams,
we have checked that $\zeta^{\eta}(i\omega,i\Omega)$ remains very
small, giving an a posteriori justification of our choice. This is
illustrated by Figures (\ref{fig:zeta_plots}) and (\ref{fig:zeta_T_dep}).

We have also implemented an approximation where instead of neglecting
the correction to $\lambda$ altogether, we replace it with $\lambda_{\mathrm{imp}}^{\eta}(i\omega_{0},i\Omega_{0})$
(and hence the interactions become $\left(\lambda_{\mathrm{imp}}^{\eta}(i\omega_{0},i\Omega_{0})\right)^{2}\mathcal{U}^{\eta}(i\Omega)$,
which one can still handle with the impurity solver presented above).
This, however, did not lead to any visible modification of the converged
solution with respect to the simplified scheme presented throughout
this paper.

\subsection{A first-order Mott transition}

\begin{figure}
\begin{centering}
\includegraphics[width=1\columnwidth]{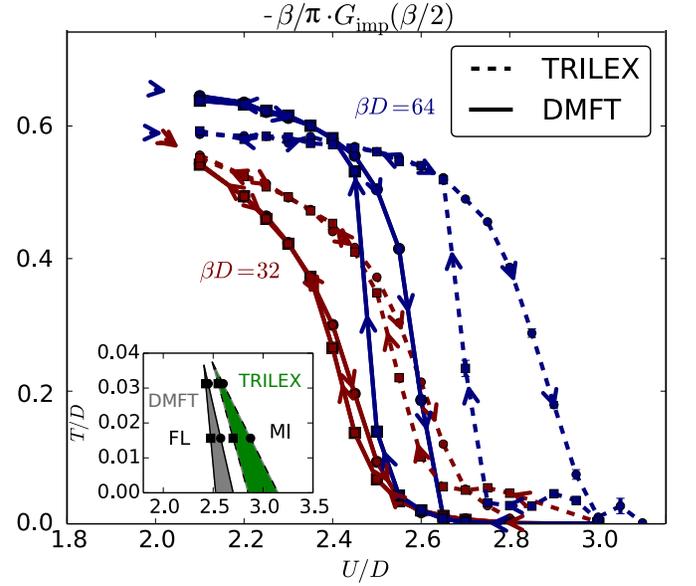}
\par\end{centering}

\caption{(color online) Evolution of $-\beta/\pi G_{\mathrm{imp}}(\tau=\beta/2)$
as a function of $U/D$ on the triangular lattice (half-filling).
Solid lines: DMFT. Dashed lines: TRILEX. Red: $\beta D=32$. Blue:
$\beta D=64$. \emph{Inset}: sketch of the coexistence regions in
DMFT (grey) and TRILEX (green) in the $(U,T)$ plane.\label{fig:G_beta_half_triangular}}
\end{figure}

In Ref. \onlinecite{Ayral2015}, several points in the phase diagram
have been studied. Due to very small denominators in $W^{\mathrm{sp}}(\mathbf{q},i\Omega=0)$,
no stable solution could be obtained at low enough temperatures to
go below the temperature of the critical end point of the Mott transition
line ($T_{\mathrm{Mott}}/D\approx0.045$ on the Bethe lattice, see
e.g. Ref \onlinecite{Vucicevic2013}). In this section, we turn to
the triangular lattice in two dimensions and at half-filling to characterize
the nature of the Mott transition. On this lattice, geometrical frustration
mitigates the low-temperature instabilities, allowing to reach lower
temperatures.

In Fig. \ref{fig:G_beta_half_triangular}, the evolution of $-\beta/\pi\cdot G_{\mathrm{imp}}(\tau=\beta/2)$
is monitored for two temperatures as a function of $U/D$. At low
enough temperatures, $-\beta/\pi\cdot G_{\mathrm{imp}}(\tau=\beta/2)$
is an accurate estimate of $A_{\mathrm{imp}}(\omega=0)$, and can
thus be used to observe the transition between a Fermi liquid ($A_{\mathrm{imp}}(\omega=0)>0$)
and a Mott insulator ($A_{\mathrm{imp}}(\omega=0)\approx0$). At low
temperatures ($\beta D=64$), both DMFT and TRILEX display a hysteretic
behavior, namely there is a coexistence region between a metallic
and insulating solution. At a higher temperature ($\beta D=32$),
the hysteretic region has shrunk. With these two estimates for $U_{c}$,
one can draw a rough sketch of the $(T,U)$ phase diagram in the triangular
lattice (see the inset).

From this study of TRILEX on the triangular lattice, two conclusions
can be drawn: (i) TRILEX, like DMFT, features a first-order Mott transition;
and (ii) in TRILEX, the critical interaction strength for the Mott
transition, $U_{c}$, is slightly enhanced with respect to the single-site
DMFT value. The latter observation is consistent with the difference
that has been observed between the local component of the TRILEX self-energy
and the single-site DMFT self-energy.\cite{Ayral2015} 

This observation contrasts with cluster methods\cite{Moukouri2001,Zhang2007,Park2008}
and diagrammatic extensions of DMFT like the D$\Gamma$A method\cite{Schafer2014}
or the dual fermion method\cite{Brener2008,Hafermann2009}. In all
these methods, $U_{c}$ is strongly reduced with respect to single-site
DMFT. This discrepancy possibly points to the partial neglect of short-range
physics in single-site TRILEX, contrary to diagrammatic and cluster
extensions of DMFT. In the former class of methods, the resummation
of ladder diagrams might explain why they seem to better capture short-range
processes. In the latter class of methods, short-range fluctuations
are treated explicitly and non-perturbatively in the extended impurity
model. This motivates the need for exploring cluster extensions of
TRILEX and comparing TRILEX with D$\Gamma$A results in more detail.

\subsection{Antiferromagnetic fluctuations: influence of frustration\label{sec:The-question-of-AF-ordering}}

\begin{figure}

\begin{centering}
\includegraphics[width=1\columnwidth]{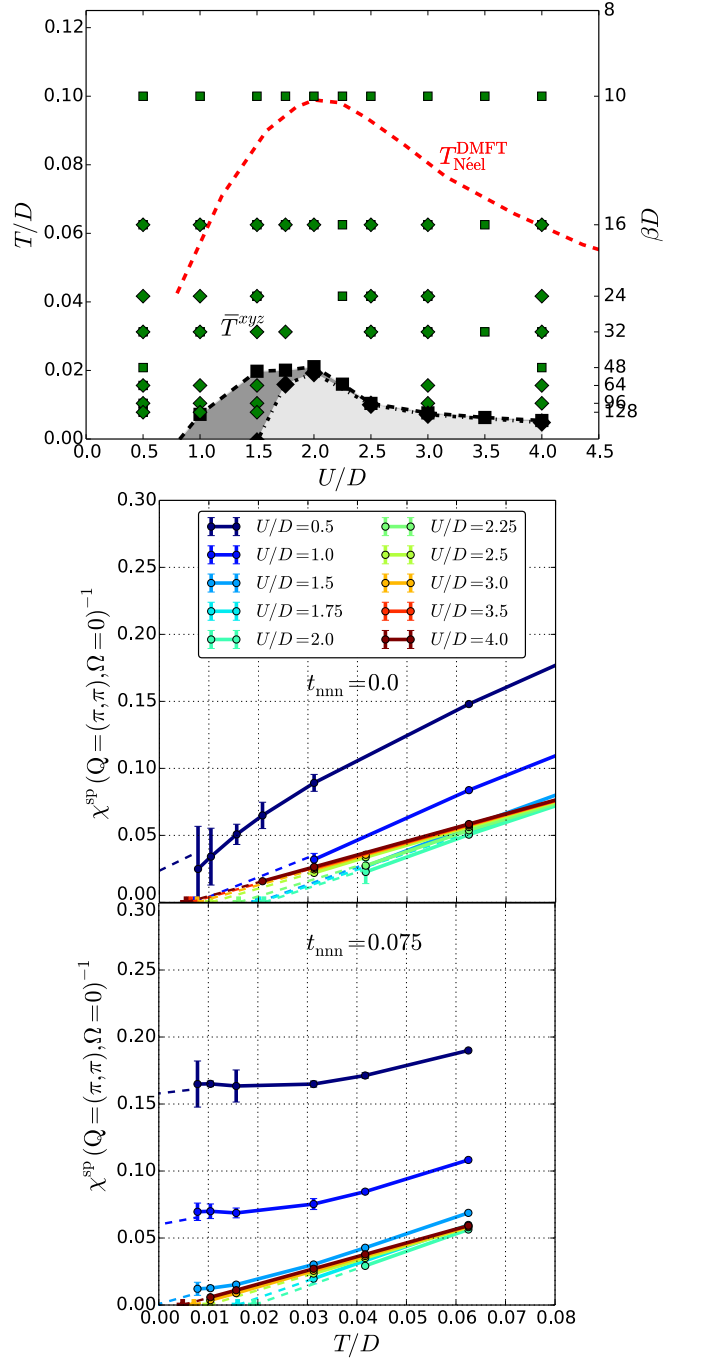}
\par\end{centering}

\caption{(color online) Influence of $t'$ on $\overline{T}^{xyz}$ (square
lattice). Phase diagram in the $(T,U$) plane at half-filling. The
green squares (resp. diamonds) denote converged TRILEX calculations
for $t'=0$ (resp. $t'=-0.3t$). The red dashed line is the Néel temperature
computed in single-site DMFT (from Ref. \onlinecite{Kunes2011}).
In the gray regions (dark gray for $t'=0$, light gray for $t'=-0.3t$)
at low temperatures, vanishing denominators in $W^{\mathrm{sp}}(\mathbf{q},i\Omega=0)$
preclude convergence.\label{fig:Influence-of-tprime-AFM} \emph{Bottom
panel}s: inverse static AF susceptibility as a function of temperature
for various $U/D$ values.\emph{ Top}: $t'=0$. \emph{Bottom}: $t'=-0.3t$.}

\end{figure}

In this section, we investigate the effect of frustration, parametrized
by a next-nearest-neighbor hopping term $t'$, on antiferromagnetic
fluctuations and on the convergence properties of the method. 

The results are gathered in Fig. \ref{fig:Influence-of-tprime-AFM}.
As shown in the lowers panels, as the temperature is decreased, the
strength of the antiferromagnetic fluctuations, parametrized by the
static inverse antiferromagnetic susceptibility $\chi^{\mathrm{sp}}(\mathbf{Q},i\Omega=0)^{-1}$,
grows, namely the product $U^{\mathrm{sp}}P^{\mathrm{sp}}(\mathbf{q},i\Omega)$
approaches the ``Stoner'' criterion $U^{\mathrm{sp}}P^{\mathrm{sp}}(\mathbf{q},i\Omega)=1$.
In the frustrated case (lower-right panel), however, the AF spin susceptibility
strongly reduced with respect to the unfrustrated case at weak values
of the local interaction $U$. It is unchanged for larger interaction
values. Consequently, the zone of unstable solutions (gray area in
the upper panel) shrinks in the weak-interaction regime and remains
unchanged in the Mott regime.

The question of the exact nature of this low-temperature phase is
still open. To decide whether at low temperatures, the inverse AF
susceptibility indeed intercepts the $x$-axis at a finite $T_{\mathrm{N\acute{e}el}}$,
as the high-temperature behavior seems to indicate, or if it displays
a bending (as observed in the correlation length in experiments --
see \emph{e.g. }Ref. \onlinecite{Keimer1992} -- or in theory -- see
\emph{e.g. }Ref. \onlinecite{Schafer2014}), requires a more refined
study which is beyond the scope of this paper. The issue could e.g.
be settled by allowing for a symmetry breaking with two sublattices.
This idea is straightforward to implement, but requires another impurity
solver, since in the AF phase the longitudinal ($z$) and perpendicular
($x,y$) spin components are no longer equivalent. In this phase,
one has to measure the perpendicular components $\Lambda_{\mathrm{imp}}^{x/y}$
of the vertex instead of $\Lambda_{\mathrm{imp}}^{z}$ only.

\subsection{Ising versus Heisenberg decoupling\label{sec:Influence-of-the-decoupling-TRILEX}}

\begin{figure}
\begin{centering}
\includegraphics[width=1\columnwidth]{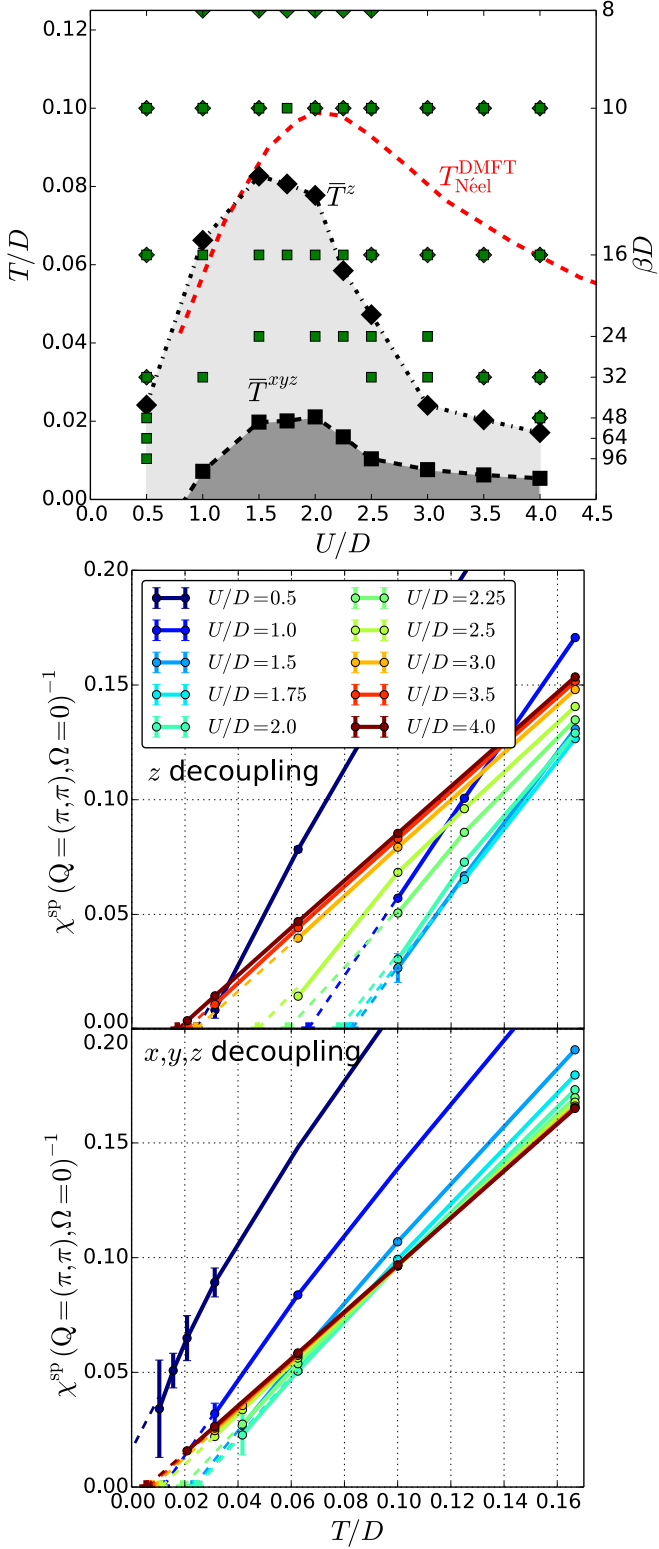}
\par\end{centering}

\caption{\label{fig:TU_Phase-diagram-Fierz}(color online) \emph{Top panel}:
Phase diagram in the $(T,U$) plane at half-filling (square lattice,
$t'=0$). The dashed red line is the Néel temperature computed in
single-site DMFT (from \cite{Kunes2011}). \emph{Left}: Heisenberg
($\overline{T}^{xyz}$) vs. Ising ($\overline{T}^{z}$) decoupling
(squares: Heisenberg; diamonds: Ising). $\overline{T}$ is determined
by extrapolating the inverse AF static susceptibility. \emph{Bottom
panels}: Inverse AF static susceptibility as a function of temperature
(square lattice). \emph{Top}: Ising or ``$z$'' decoupling. \emph{Bottom}:
Heisenberg or ``$xyz$'' decoupling}
\end{figure}

In this subsection, we discuss the practical implications of the way
the Hubbard interaction term is decoupled in terms of Hubbard-Stratonovich
terms. 

Already at the single-site level, we have investigated the influence
of the ratio of charge to spin channel and shown that it does not
impact the fulfillment of sum rules\cite{Ayral2015}. 

Here, we focus on the difference between the ``Ising'' and ``Heisenberg''
decouplings introduced in subsection \ref{sub:From-an-electron-electron}.
We show, in Fig. \ref{fig:TU_Phase-diagram-Fierz} (upper panel),
the phase diagram for both choices of decoupling. As before, the boundary
of the region of unstable solutions, shown in gray, has been obtained
by following the evolution of the inverse static AF susceptibility
as a function of temperature for both decouplings. The extrapolated
$\overline{T}$ strongly depends on the decoupling: it is much larger
for the Ising decoupling than for the Heisenberg decoupling. 

This can be understood in the following intuitive way: in the Ising
decoupling, the spin has fewer degrees of freedom to fluctuate than
in the Heisenberg decoupling. Thus, correlation lengths are much larger
in the Ising decoupling than in the Heisenberg decoupling. In either
case, $\overline{T}$ is lower than the Néel temperature computed
in single-site DMFT (except for a few points in the Ising decoupling
at weak coupling, but the difference is within error bars): TRILEX
contains spatial fluctuations beyond (dynamical) mean field theory.

\section{Conclusions and perspectives\label{sec:Conclusions-and-perspectives}}

In this paper, we have presented the TRILEX formalism, which consists
in making a local expansion of the 3PI functional $\mathcal{K}$.
This approximation entails the locality of the three-leg vertex which
is self-consistently computed by solving an impurity model with dynamical
charge and spin interactions. 

By construction, this method interpolates between two major approaches
to high-temperature superconductors, namely, fluctuation-exchange
approximations such as spin fluctuation theory, and dynamical mean-field
theory and its cluster extensions. The central quantity of TRILEX,
the impurity three-leg vertex $\Lambda_{\mathrm{imp}}(i\omega,i\Omega)$,
encodes the passage from both limits. It can be used to construct
momentum-dependent self-energies and polarizations \emph{at a reduced
cost} compared to cluster DMFT and diagrammatic extensions of DMFT.
More specifically, it requires the solution of a \emph{single-site}
local impurity model with dynamical interactions. 

Contrary to spin fluctuation theory, the method explicitly captures
Mott physics \emph{via} the frequency-dependent vertex. Contrary to
recent diagrammatic extensions of DMFT attempting to incorporate long-range
physics such as D$\Gamma$A\cite{Toschi2007,Katanin2009} and the
dual fermion method\cite{Rubtsov2008}, it deals with functions of
two (instead of three) frequencies, which makes it more easily extendable
to a cluster and/or multiorbital implementation. Indeed, four-leg
vertices are a major computational burden in those methods, owing
to their sheer size in memory and also to the appearance of divergencies
in some of these vertices already for moderate interaction values\cite{Schafer2013},
as well as divergencies when inverting the Bethe-Salpether equations
in a given channel\cite{Katanin2009}.

Here, the TRILEX method in its single-site version has been applied
to the single-band Hubbard model, on the square and on the triangular
lattice. As expected from the construction of the method, TRILEX interpolates
between (a) the fluctuation-exchange limit, where the self-energy
is given by the one-loop diagram computed with the propagator associated
to long-range fluctuations in channel $\eta$, $W^{\eta}(\mathbf{q},i\Omega)$,
and (b) the dynamical mean field limit which approximates the self-energy
by a local, but frequency-dependent impurity self-energy which reduces,
in the strong-coupling regime, to the atomic limit self-energy. At
intermediate coupling, upon doping, strong AF fluctuations cause a
sizable momentum differentiation of the Fermi surface, as observed
in photoemission in cuprate materials.

There are many open issues:
\begin{enumerate}
\item \emph{Low-temperature phase}. The issue of the instabilities in the
low-temperature regime, which is related to the fulfillment or not
of the Mermin-Wagner theorem and the associated Fierz ambiguity, deserve
further studies. This is all the more interesting as related methods
such as TPSC and ladder-D$\Gamma$A with the additional Moriya correction
fulfill the Mermin-Wagner theorem; a better understanding of the minimal
ingredients to enforce this property is needed. 
\item \emph{Extension to cluster schemes}. The accuracy of the TRILEX method
can be assessed quantitatively by extending it to clusters. Due to
the inclusion of long-range fluctuations, one may anticipate that
cluster TRILEX will converge faster than cluster DMFT with respect
to the cluster size $N_{c}$ in the physically relevant channel. Moreover,
when convergence with respect to $N_{c}$ is reached, the results
will be totally independent of the choice of channels. As a result,
the channel dependence for a given cluster (e.g. a given size) is
an indication of the degree of its convergence which does not necessitate
the computation of larger clusters. This property has no analog in
cluster DMFT methods. 
\item \emph{Extension to multiorbital systems}. Thanks to the simplicity
of solving the single-site impurity model, single-site TRILEX can
be applied to multiorbital systems to study momentum-dependent self-energy
effects. Such an endeavor is currently out of the reach of cluster
DMFT due to the sheer size of the corresponding Hilbert space (3 bands
times a $2\times2$ cluster is effectively a 12-site calculation,
already a large numerical effort). Yet, this extension may be crucial
for multiorbital systems where long-range spin physics as well as
correlations are thought to play an important role. For instance,
the pnictide superconductors, where bosonic spin-density-wave fluctuations
are sizable but correlations effects are not so strong, may prove
an ideal playing ground for TRILEX.
\item \emph{Extension to ``anomalous'' phases}. TRILEX can be straightforwardly
extended to study charge-ordered phases (as shown by its relation
to $GW$+EDMFT). Moreover, its application to superconducting phases
is also possible: in this context, it interpolates between generalized
Migdal-Eliashberg theory (or spin-fermion superconductivity) and the
superconducting version of DMFT. As such, it can capture $d$-wave
superconductivity at the cost of solving a single-site impurity problem
(which is not possible in single-site DMFT).
\item \emph{Nonlocal extensions. }One natural route beyond the local approximation
of the vertex is to construct the lattice vertex as the sum of the
impurity vertex with a nonlocal diagrammatic correction, in the same
way as the GW+EDMFT method extends EDMFT by adding nonlocal diagrams
to the impurity self-energies.
\item \emph{Extension to the three-boson vertex}. In the functional construction
presented in this paper, we have only considered a three-point source
term with one bosonic field and two fermionic fields. In principle,
for the sake of 3PI-completeness, one could also introduce an additional
three-point source term coupling three bosonic fields. We have not
explored this path here since it requires an impurity solver handling
both fermionic and bosonic fields.\end{enumerate}
\begin{acknowledgments}
We acknowledge useful discussions with S. Andergassen, J.-P. Blaizot,
S. Biermann, M. Capone, A. Eberlein, M. Ferrero, A. Georges, H. Hafermann,
A. Millis, G. Misguich, J. Otsuki, A. Toschi. This work is supported
by the FP7/ERC, under Grant Agreement No. 278472-MottMetals. Part
of this work was performed using HPC resources from GENCI-TGCC (Grant
No. 2015-t2015056112).

\bibliographystyle{apsrev4-1}
\bibliography{/home/tayral/Documents/library}

\end{acknowledgments}

\appendix

\section{Symmetry Properties of the Vertex and Fourier Conventions}

\subsection{Fourier conventions}

We follow the following Fourier conventions, depending on whether
we want to work with a fermionic and a bosonic Matsubara frequency,
or two fermionic frequencies:\begin{subequations}

\begin{eqnarray}
A_{\mathbf{R}_{1}\mathbf{R}_{2}\mathbf{R}_{3}}(i\omega,i\Omega) & \equiv & \iint_{0}^{\beta}\mathrm{d}\tau\mathrm{d}\tau'e^{i\omega\tau+i\Omega\tau'}A_{\mathbf{R}_{1}\mathbf{R}_{2}\mathbf{R}_{3}}(\tau,0,\tau')\label{eq:FT_fermion_boson}\\
\hat{A}_{\mathbf{R}_{1}\mathbf{R}_{2}\mathbf{R}_{3}}(i\omega_{1},i\omega_{2}) & \equiv & \iint_{0}^{\beta}\mathrm{d}\tau\mathrm{d}\tau'e^{i\omega_{1}\tau+i\omega_{2}\tau'}A_{\mathbf{R}_{1}\mathbf{R}_{2}\mathbf{R}_{3}}(\tau,\tau',0)\label{eq:FT_fermion_fermion}
\end{eqnarray}
\end{subequations}for any three-point function $A(\mathbf{R}_{1},\tau_{1};\mathbf{R}_{2},\tau_{2};\mathbf{R}_{3},\tau_{3})$,
$e.g$ $A_{\mathbf{R}_{1}\mathbf{R}_{2}\mathbf{R}_{3}}(\tau_{1},\tau_{2},\tau_{3})=\langle Tc_{\mathbf{R}_{1}}(\tau_{1})c_{\mathbf{R}_{2}}^{\dagger}(\tau_{2})\phi_{\mathbf{R}_{3}}(\tau_{3})\rangle$.
Both functions are related:

\begin{equation}
A_{\mathbf{R}_{1}\mathbf{R}_{2}\mathbf{R}_{3}}(i\omega,i\Omega)=\hat{A}_{\mathbf{R}_{1}\mathbf{R}_{2}\mathbf{R}_{3}}(i\omega,-i\omega-i\Omega)\label{eq:symmetry_conventions}
\end{equation}
In the main text, we only use the first form $A_{\mathbf{R}_{1}\mathbf{R}_{2}\mathbf{R}_{3}}(i\omega,i\Omega)$.

\subsection{Lehmann representation of the three-leg vertex\label{sub:Lehmann-representation}}

Using the identity $\int_{0}^{\beta}\int_{0}^{\beta}dt_{1}dt_{2}Tf_{1}(t_{1})f_{2}(t_{2})=\int_{0}^{\beta}\int_{0}^{t_{1}}dt_{1}dt_{2}\sum_{p\in\mathfrak{S}_{2}}\sigma(p)f_{p1}(t_{1})f_{p2}(t_{2})$,
we can write, using the definition of $\tilde{\chi}$ (Eq. \ref{eq:def_chi3_tilde})
and of its Fourier transform (Eq. \ref{eq:FT_fermion_fermion})

\begin{eqnarray}
 &  & \hat{\tilde{\chi}}_{123}(i\omega_{1},i\omega_{2})\nonumber \\
 &  & \equiv\sum_{p\in\mathfrak{S}_{2}}\int_{0}^{\beta}\mathrm{d}\tau\int_{0}^{\tau}\mathrm{d}\tau'\sigma(p)\langle O_{p1}(\tau)O_{p2}(\tau')n_{3}(0)\rangle e^{i\omega_{p1}\tau}e^{i\omega_{p2}\tau'}\nonumber \\
 &  & =\frac{1}{Z}\sum_{ijk}\sum_{p\in\mathfrak{S}_{2}}\sigma(p)\langle i|O_{p1}|j\rangle\langle j|O_{p2}|k\rangle\langle k|n_{3}|i\rangle f_{ijk}(\omega_{p1},\omega_{p2})\label{eq:Lehmann_3leg}
\end{eqnarray}
with $O_{1}=c_{1}^{\dagger}$ and $O_{2}=c_{2}$, and: 
\begin{eqnarray*}
 &  & f_{ijk}(\omega_{1},\omega_{2})\\
 &  & =e^{-\beta\epsilon_{i}}\int_{0}^{\beta}\mathrm{d}\tau e^{\tau(i\omega_{1}+\epsilon_{i}-\epsilon_{j})}\int_{0}^{\tau}\mathrm{d}\tau'e^{\tau'(i\omega_{2}+\epsilon_{j}-\epsilon_{k})}\\
 &  & =e^{-\beta\epsilon_{i}}\int_{0}^{\beta}\mathrm{d}\tau e^{\tau(i\omega_{1}+\epsilon_{i}-\epsilon_{j})}\frac{e^{\tau(i\omega_{2}+\epsilon_{j}-\epsilon_{k})}-1}{i\omega_{2}+\epsilon_{j}-\epsilon_{k}}\\
 &  & =\frac{e^{-\beta\epsilon_{i}}}{i\omega_{2}+\epsilon_{j}-\epsilon_{k}}\int_{0}^{\beta}\mathrm{d}\tau\left(e^{\tau(i\omega_{1}+i\omega_{2}+\epsilon_{i}-\epsilon_{k})}-e^{\tau(i\omega_{1}+\epsilon_{i}-\epsilon_{j})}\right)\\
 &  & =\frac{e^{-\beta\epsilon_{i}}}{i\omega_{2}+\epsilon_{j}-\epsilon_{k}}\Bigg(\frac{e^{\beta(i\omega_{1}+i\omega_{2}+\epsilon_{i}-\epsilon_{k})}-1}{i\omega_{1}+i\omega_{2}+\epsilon_{i}-\epsilon_{k}}\left(1-\delta_{ik}\right)\\
 &  & \;\;-\frac{e^{\beta(i\omega_{1}+\epsilon_{i}-\epsilon_{j})}-1}{i\omega_{1}+\epsilon_{i}-\epsilon_{j}}\Bigg)+\frac{e^{-\beta\epsilon_{i}}}{i\omega_{2}+\epsilon_{j}-\epsilon_{i}}\beta\delta_{i\omega_{1}+i\omega_{2}}\delta_{ik}\\
\\
 &  & =\frac{1}{i\omega_{2}+\epsilon_{j}-\epsilon_{k}}\Bigg(\frac{e^{-\beta\epsilon_{k}}-e^{-\beta\epsilon_{i}}}{i\omega_{1}+i\omega_{2}+\epsilon_{i}-\epsilon_{k}}\left(1-\delta_{ik}\right)\\
 &  & \;\;+\frac{e^{-\beta\epsilon_{j}}+e^{-\beta\epsilon_{i}}}{i\omega_{1}+\epsilon_{i}-\epsilon_{j}}\Bigg)+\frac{e^{-\beta\epsilon_{i}}}{i\omega_{2}+\epsilon_{j}-\epsilon_{i}}\beta\delta_{i\omega_{1}+i\omega_{2}}\delta_{ik}
\end{eqnarray*}
We have used the fact that both $i\omega_{1}$ and $i\omega_{2}$
are fermionic Matsubara frequencies ($e^{\beta i\omega_{1}}=-1$).

\subsection{Symmetries of the three-point vertex}

\begin{figure}

\begin{centering}
\includegraphics[width=0.8\columnwidth]{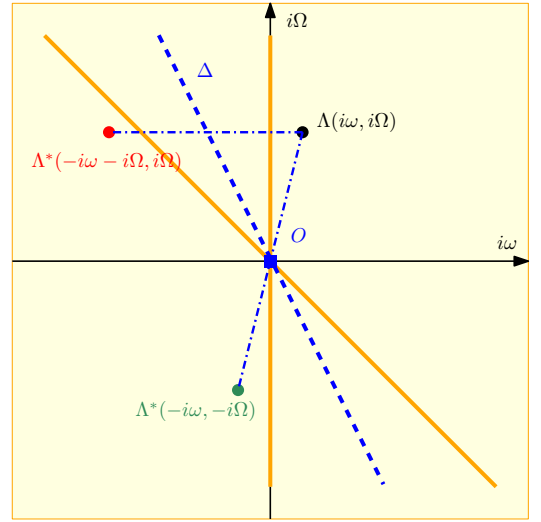}
\par\end{centering}

\caption{(color online) Vertex symmetries\label{fig:Vertex-symmetries}}

\end{figure}

In this section, we derive the main symmetries of the three-point
vertex in a simple limit. We consider the most simple fermionic model,
namely a single fermionic level, $O_{1}=c^{\dagger}$, $O_{2}=c$.
$O_{1}^{\dagger}=O_{2}$ (in the notations of Section \ref{sub:Lehmann-representation}).
The Hilbert space consists in two states: $|0\rangle$ and $|1\rangle$
with respective energies $0$ and $\epsilon$. Starting from \ref{eq:Lehmann_3leg},
we have:
\begin{eqnarray*}
\hat{\tilde{\chi}}(i\omega_{1},i\omega_{2}) & = & \frac{1}{Z}\sum_{ijk}\langle i|O_{1}|j\rangle\langle j|O_{2}|k\rangle\langle k|n|i\rangle f(\omega_{1},\omega_{2})\\
 &  & +\underbrace{\sum_{ijk}\langle i|O_{2}|j\rangle\langle j|O_{1}|k\rangle\langle k|n|i\rangle f(\omega_{2},\omega_{1})}_{=0}\\
 & = & \frac{1}{Z}\langle1|c^{\dagger}|0\rangle\langle0|c|1\rangle\langle1|n|1\rangle f(\omega_{1},\omega_{2})\\
 & = & \frac{1}{Z}f_{101}(\omega_{1},\omega_{2})\\
 & = & \frac{1}{Z}\frac{1}{i\omega_{2}-\epsilon}\left(\frac{1+e^{-\beta\epsilon}}{i\omega_{1}+\epsilon}\right)+\frac{1}{Z}\frac{e^{-\beta\epsilon}}{i\omega_{2}-\epsilon}\delta_{i\omega_{1}+i\omega_{2}}
\end{eqnarray*}
Hence,

\begin{align*}
\tilde{\chi}(i\omega,i\Omega) & \propto\frac{1}{-i\omega-i\Omega-\epsilon}\frac{1}{i\omega+\epsilon}+\frac{1}{-i\omega-\epsilon}\delta_{i\Omega}\\
 & \propto\frac{1}{i\omega+i\Omega+\epsilon}\frac{1}{i\omega+\epsilon}+\frac{1}{i\omega+\epsilon}\delta_{i\Omega}
\end{align*}
One can notice:
\begin{align*}
\tilde{\chi}(i\omega-i\Omega,i\Omega) & \propto\frac{1}{i\omega+\epsilon}\frac{1}{i\omega-i\Omega+\epsilon}+\frac{1}{i\omega-i\Omega+\epsilon}\delta_{i\Omega}\\
 & =\chi(i\omega,-i\Omega)
\end{align*}
and:
\begin{align*}
\tilde{\chi}^{*}(i\omega,-i\Omega) & \propto\left(\frac{1}{i\omega-i\Omega+\epsilon}\frac{1}{i\omega+\epsilon}+\frac{1}{i\omega+\epsilon}\delta_{i\Omega}\right)^{*}\\
 & =\frac{1}{-i\omega+\epsilon}\frac{1}{i\Omega-i\omega+\epsilon}+\frac{1}{-i\omega+\epsilon}\delta_{i\Omega}\\
 & =\tilde{\chi}(-i\omega,i\Omega)
\end{align*}
Thus, we obtain the following symmetry relations:\begin{subequations}
\begin{align}
\tilde{\chi}(i\omega-i\Omega,i\Omega) & =\tilde{\chi}(i\omega,-i\Omega)\label{eq:sym1}\\
\tilde{\chi}^{*}(i\omega,-i\Omega) & =\tilde{\chi}(-i\omega,i\Omega)\label{eq:sym2}
\end{align}
\end{subequations}One can check that these symmetry relations hold
in the general case and carry over to the vertex $\Lambda(i\omega,i\Omega)$.
A pictorial representation of these symmetries is given in Fig. \ref{fig:Vertex-symmetries}.

\section{Link between bosonic correlation functions and fermionic correlation
functions\label{sub:Link-between-bosonic_and_fermionic}}

In this appendix, we prove the following relations between observables
of the mixed fermion-boson action (\ref{eq:general_action_eb}) and
observables of the fermionic action: \begin{subequations}

\begin{eqnarray}
\varphi_{\alpha}^{\eta} & = & U_{\alpha\beta}^{\eta}\langle n_{\beta}^{\eta}\rangle\label{eq:varphi_n}\\
W_{\alpha\beta}^{\eta,\mathrm{nc}} & = & U_{\alpha\beta}^{\eta}-U_{\alpha\gamma}^{\eta}\chi_{\gamma\delta}^{\eta,\mathrm{nc}}U_{\delta\beta}^{\eta}\label{eq:W_chi}\\
W_{\alpha\beta}^{\eta} & = & U_{\alpha\beta}^{\eta}-U_{\alpha\gamma}^{\eta}\chi_{\gamma\delta}^{\eta}U_{\delta\beta}^{\eta}\label{eq:W_conn_chi}\\
\chi_{u\bar{v}\alpha}^{\eta} & = & U_{\alpha\beta}^{\eta}\tilde{\chi}_{u\bar{v}\beta}^{\eta}\label{eq:chi_chi_tilde}
\end{eqnarray}
\end{subequations}$W_{\alpha\beta}^{\eta}$, $\chi_{\alpha\beta}^{\eta}$
and $\chi_{uv\alpha}^{\eta}$ have been defined in Eqs (\ref{eq:def_W}),
(\ref{eq:chi_def}) and (\ref{eq:chi3_nc_def}) respectively, and

\begin{equation}
\tilde{\chi}_{u\bar{v}\alpha}^{\mathrm{nc}}\equiv\langle c_{u}\bar{c}_{\bar{v}}n_{\alpha}\rangle\label{eq:def_chi3_tilde}
\end{equation}
 Let us recall the definition of the partition function in the presence
of sources 
\begin{equation}
Z[h,F,B]\equiv\int\mathcal{D}[\bar{c},c,\phi]e^{-S_{\mathrm{eb}}+h_{\alpha}\phi_{\alpha}-F_{\bar{u}v}\bar{c}_{\bar{u}}c_{v}-\frac{1}{2}\phi_{\alpha}B_{\alpha\beta}\phi_{\beta}}\label{eq:generating_functional_eb}
\end{equation}
Integrating out the bosonic fields yields:

\begin{align}
Z[h,F,B] & =\mathrm{Det}\left[\bar{U}^{-1}\right]^{-1/2}\label{eq:generating_functional_ee}\\
 & \;\times\int\mathcal{D}[\bar{c},c]e^{-\bar{c}_{\bar{u}}\left\{ -G_{0,\bar{u}v}^{-1}+F_{\bar{u}v}\right\} c_{v}+\frac{1}{2}\bar{U}_{\alpha\beta}(h_{\alpha}-\bar{c}_{\bar{u}}\lambda_{\bar{u}v\alpha}c_{v})^{2}}\nonumber \\
 & =e^{\frac{1}{2}\mathrm{Tr}\log\left[\bar{U}\right]}\nonumber \\
 & \;\times\int\mathcal{D}[\bar{c},c]e^{-\bar{c}_{\bar{u}}\left\{ -G_{0,\bar{u}v}^{-1}+F_{\bar{u}v}\right\} c_{v}+\frac{1}{2}\bar{U}_{\alpha\beta}(h_{\alpha}-\bar{c}_{\bar{u}}\lambda_{\bar{u}v\alpha}c_{v})^{2}}\nonumber 
\end{align}
with $\bar{U}_{\alpha\beta}=\left[\left(-U^{-1}+B\right)^{-1}\right]_{\alpha\beta}$.
Hence
\begin{align*}
\Omega & =\frac{1}{2}\mathrm{Tr}\log\left[-U^{-1}+B\right]\\
 & \;\;-\log\int\mathcal{D}[\bar{c},c]\Bigg[e^{-\bar{c}_{\bar{u}}\left\{ -G_{0,\bar{u}v}^{-1}+F_{\bar{u}v}\right\} c_{v}}\\
 & \;\;\times e^{\frac{1}{2}\bar{U}_{\alpha\beta}\left(h_{\alpha}-\bar{c}_{\bar{u}}\lambda_{\bar{u}v\alpha}c_{v}\right)\left(h_{\beta}-\bar{c}_{\bar{u}}\lambda_{\bar{u}v\beta}c_{v}\right)}\Bigg]
\end{align*}

Relation (\ref{eq:varphi_n}) follows from computing $\varphi_{\alpha}$
by successively using (\ref{eq:generating_functional_eb}) and (\ref{eq:generating_functional_ee}):

\[
\varphi_{\alpha}=\frac{1}{Z}\frac{\partial Z}{\partial h_{\alpha}}=U_{\alpha\beta}\langle\bar{c}_{\bar{u}}\lambda_{\bar{u}v\beta}c_{v}\rangle
\]
Similarly, one has:

\begin{eqnarray*}
W_{\alpha\beta}^{\mathrm{nc}} & = & -2\frac{\partial\Omega}{\partial B_{\alpha\beta}}\\
 & = & -2\left[\frac{1}{2}\left(-U_{\alpha\beta}\right)\right]\\
 &  & -2\left[-\frac{1}{2}\left(\frac{\partial\bar{U}_{\gamma\delta}}{\partial B_{\alpha\beta}}\right)\left(h_{\gamma}-\bar{c}_{\bar{u}}\lambda_{\bar{u}v\gamma}c_{v}\right)\left(h_{\delta}-\bar{c}_{\bar{u}}\lambda_{\bar{u}v\delta}c_{v}\right)\right]\\
 & = & U_{\alpha\beta}-U_{\alpha\delta}\langle\left(\bar{c}_{u}\lambda_{\bar{u}v\delta}c_{v}\right)\left(\bar{c}_{\bar{u}}\lambda_{\bar{u}v\gamma}c_{v}\right)\rangle U_{\gamma\beta}\\
 & = & U_{\alpha\beta}-U_{\alpha\delta}\langle n_{\delta}n_{\gamma}\rangle U_{\gamma\beta}
\end{eqnarray*}
which proves (\ref{eq:W_chi}-\ref{eq:W_conn_chi}) and:
\begin{eqnarray*}
\chi_{uv\alpha}^{\mathrm{nc}} & = & \frac{1}{Z}\left.\frac{\partial^{2}Z}{\partial F_{\bar{v}u}\partial h_{\alpha}}\right|_{h=0}\\
 & = & \frac{1}{Z}\frac{\partial}{\partial F_{\bar{v}u}}\int\mathcal{D}[\bar{c},c]\left(U_{\alpha\beta}\left(n_{\beta}-h_{\beta}\right)\right)e^{-S}\\
 & = & U_{\alpha\beta}\langle c_{u}\bar{c}_{\bar{v}}n_{\beta}\rangle
\end{eqnarray*}
which proves (\ref{eq:chi_chi_tilde}).

\section{Non-interacting free energy\label{sec:Details-of-Legendre}}

The non-interacting free energy in the presence of sources reads:

\begin{eqnarray*}
 &  & \Omega[h,B,F,\lambda=0]\\
 &  & =-\log\int\mathcal{D}[\bar{c},c,\phi]e^{-S_{\mathrm{eb}}-\bar{c}_{\bar{u}}F_{\bar{u}v}c_{v}+h_{\alpha}\phi_{\alpha}-\frac{1}{2}\phi_{\alpha}B_{\alpha\beta}\phi_{\beta}}\\
 &  & =-\log\left\{ \mathrm{Det}\left(-G_{0}^{-1}+F\right)\mathrm{Det}\left(-W_{0}^{-1}+B\right)^{-1/2}\right\} \\
 &  & \;\;-\frac{1}{2}h_{\alpha}\left[-W_{0}^{-1}+B\right]_{\alpha\beta}^{-1}h_{\beta}\\
 &  & =-\mathrm{Tr}\log\left(G_{0}^{-1}-F\right)+\frac{1}{2}\mathrm{Tr}\log\left(W_{0}^{-1}-B\right)\\
 &  & \;\;+\frac{1}{2}h_{\alpha}\left[W_{0}^{-1}-B\right]_{\alpha\beta}^{-1}h_{\beta}
\end{eqnarray*}
Hence, applying Eqs (\ref{eq:def_varphi}-\ref{eq:def_G}-\ref{eq:def_W})
in the case $\lambda=0$ lead to

\begin{eqnarray*}
\varphi_{\alpha} & = & -h_{\beta}\left[W_{0}^{-1}-B\right]_{\beta\alpha}^{-1}\\
W_{\alpha\beta}^{\mathrm{nc}} & = & \left(W_{0}^{-1}-B\right)_{\alpha\beta}^{-1}-h_{\delta}\left[W_{0}^{-1}-B\right]_{\delta\alpha}^{-1}\left[W_{0}^{-1}-B\right]_{\beta\gamma}^{-1}h_{\gamma}\\
G_{u\bar{v}} & = & \left(G_{0}^{-1}-F\right)_{u\bar{v}}^{-1}
\end{eqnarray*}
yielding the following inversion relations:

\begin{eqnarray*}
h_{\alpha} & = & -\varphi_{\beta}\left(W_{0}^{-1}-B\right)_{\beta\alpha}\\
F_{\bar{u}v} & = & G_{0,\bar{u}v}^{-1}-G_{\bar{u}v}^{-1}\\
B_{\alpha\beta} & = & W_{0,\alpha\beta}^{-1}-W_{\alpha\beta}^{-1}
\end{eqnarray*}
and the final expression:
\begin{eqnarray}
\Omega[h,B,F,\lambda=0] & = & -\mathrm{Tr}\log\left[G^{-1}\right]\label{eq:noninteracting_free_energy}\\
 &  & +\frac{1}{2}\mathrm{Tr}\log\left[W^{-1}\right]+\frac{1}{2}\varphi_{\alpha}W_{\alpha\beta}^{-1}\varphi_{\beta}\nonumber 
\end{eqnarray}

\section{Alternative derivation using the Equations of Motions\label{sec:Alternative-derivation-EOMS}}

In this section, we derive Eqs. (\ref{eq:Sigma_Hedin}-\ref{eq:P_Hedin})
using equations of motions.

\subsection{Prerequisite: Schwinger-Dyson equations\label{sub:Schwinger-Dyson-equations}}

\subsubsection{Fermionic fields}

For any conjugate Grassmann fields $c_{i}$ and $\bar{c}_{\bar{i}}$,
matrix $\left[G_{0}\right]_{\bar{i}j}$ and function $f$, we can
define:
\[
A\equiv\int\mathcal{D}\left[\bar{c}c\right]e^{\bar{c}_{\bar{k}}\left[G_{0}^{-1}\right]_{\bar{k}l}c_{l}}\frac{\partial f[\bar{c}c]}{\partial\bar{c}_{\bar{i}}}
\]
Then, by integration by parts:

\begin{eqnarray*}
A & = & -\int\mathcal{D}\left[\bar{c}c\right]\frac{\partial}{\partial\bar{c}_{\bar{i}}}e^{\bar{c}_{\bar{k}}\left[G_{0}^{-1}\right]_{\bar{k}l}c_{l}}f[\bar{c}c]\\
 & = & -\left[G_{0}^{-1}\right]_{\bar{i}l}\int\mathcal{D}\left[\bar{c}c\right]c_{l}e^{\bar{c}_{\bar{k}}\left[G_{0}^{-1}\right]_{\bar{k}l}c_{l}}f[\bar{c}c]
\end{eqnarray*}
For $f[\bar{c}c]\equiv h[\bar{c}c]e^{-V}$:

\begin{align*}
A & =\int\mathcal{D}\left[\bar{c}c\right]e^{\bar{c}_{\bar{k}}\left[G_{0}^{-1}\right]_{\bar{k}l}c_{l}}\left(\frac{\partial h[\bar{c}c]}{\partial\bar{c}_{\bar{i}}}+\frac{\partial V}{\partial\bar{c}_{\bar{i}}}\right)e^{-V}\\
 & =-\left[G_{0}^{-1}\right]_{\bar{i}l}\int\mathcal{D}\left[\bar{c}c\right]c_{l}e^{\bar{c}_{\bar{k}}\left[G_{0}^{-1}\right]_{\bar{k}l}c_{l}}h[\bar{c}c]e^{-V}
\end{align*}
i.e. for any functions $h$ and $V$:
\begin{equation}
\left\langle \frac{\partial h[\bar{c}c]}{\partial\bar{c}_{\bar{i}}}+h[\bar{c}c]\frac{\partial V}{\partial\bar{c}_{\bar{i}}}\right\rangle =-\left[G_{0}^{-1}\right]_{\bar{i}l}\left\langle c_{l}h[\bar{c}c]\right\rangle \label{eq:SD_fermionic-full}
\end{equation}

\subsubsection{Bosonic fields}

Similarly to the previous section, for any bosonic field $\phi_{\alpha}$,
matrix $U_{\alpha\beta}$ and function $f$, let us define:
\[
A\equiv\int\mathcal{D}\left[\phi\right]e^{\frac{1}{2}\phi_{\alpha}\left[U^{-1}\right]_{\alpha\beta}\phi_{\beta}}\frac{\partial f[\phi]}{\partial\phi_{\gamma}}
\]
By integration by parts, we have
\begin{eqnarray*}
A & = & -\int\mathcal{D}\left[\phi\right]\frac{\partial}{\partial\phi_{\gamma}}e^{\frac{1}{2}\phi_{\alpha}\left[U^{-1}\right]_{\alpha\beta}\phi_{\beta}}f[\phi]\\
 & = & -\left[U^{-1}\right]_{\gamma\beta}\int\mathcal{D}\left[\phi\right]\phi_{\beta}e^{\frac{1}{2}\phi_{\alpha}\left[U^{-1}\right]_{\alpha\beta}\phi_{\beta}}f[\phi]
\end{eqnarray*}
and taking $f[\phi]\equiv h[\phi]e^{-\lambda_{\bar{u}v\delta}\phi_{\delta}\bar{c}_{\bar{u}}c_{v}}$,
one has:
\begin{align*}
A & =\int\mathcal{D}\left[\phi\right]\Bigg[e^{\frac{1}{2}\phi_{\alpha}\left[U^{-1}\right]_{\alpha\beta}\phi_{\beta}}\\
 & \;\;\;\left\{ \frac{\partial h[\phi]}{\partial\phi_{\gamma}}-\lambda_{\bar{u}v\gamma}\bar{c}_{\bar{u}}c_{v}h[\phi]\right\} e^{-\lambda_{\bar{u}v\delta}\phi_{\delta}\bar{c}_{\bar{u}}c_{v}}\Bigg]\\
 & =-\left[U^{-1}\right]_{\gamma\beta}\int\mathcal{D}\left[\phi\right]\phi_{\beta}e^{\frac{1}{2}\phi_{\alpha}\left[U^{-1}\right]_{\alpha\beta}\phi_{\beta}}h[\phi]e^{-\lambda_{\bar{u}v\delta}\phi_{\delta}\bar{c}_{\bar{u}}c_{v}}
\end{align*}
i.e, for any function $h$:
\begin{equation}
\left\langle \frac{\partial h[\phi]}{\partial\phi_{\gamma}}-\lambda_{\bar{u}v\gamma}\bar{c}_{\bar{u}}c_{v}h[\phi]\right\rangle =-\left[U^{-1}\right]_{\gamma\beta}\left\langle \phi_{\beta}h[\phi]\right\rangle \label{eq:SD_bosonic}
\end{equation}

\subsection{Equations of motion for $G$ and $W$}

\subsubsection{Fermionic propagator $G$}

Specializing Eq (\ref{eq:SD_fermionic-full}) for $h[\bar{c}c]\equiv\bar{c}_{\bar{p}}$
and $V=\frac{1}{2}n_{\alpha}U_{\alpha\beta}n_{\beta}=\frac{1}{2}U_{\alpha\beta}\bar{c}_{\bar{u}}\lambda_{\bar{u}v\alpha}c_{v}\bar{c}_{\bar{w}}\lambda_{\bar{w}l\beta}c_{l}$,
and noting that:
\begin{eqnarray*}
\frac{\partial V}{\partial\bar{c}_{\bar{i}}} & = & \frac{1}{2}U_{\alpha\beta}\lambda_{\bar{i}v\alpha}c_{v}\bar{c}_{\bar{w}}\lambda_{\bar{w}l\beta}c_{l}+\frac{1}{2}U_{\alpha\beta}\bar{c}_{\bar{u}}\lambda_{\bar{u}v\alpha}c_{v}\lambda_{\bar{i}l\beta}c_{l}\\
 & = & U_{\alpha\beta}\lambda_{\bar{i}v\alpha}\lambda_{\bar{w}l\beta}c_{v}\bar{c}_{\bar{w}}c_{l}
\end{eqnarray*}
(we have used $U_{\alpha\beta}=U_{\beta\alpha}$), one has:

\begin{eqnarray*}
-\left[G_{0}^{-1}\right]_{\bar{i}l}\left\langle c_{l}\bar{c}_{\bar{p}}\right\rangle  & = & \delta_{\bar{i}\bar{p}}+U_{\alpha\beta}\lambda_{\bar{i}v\alpha}\lambda_{\bar{w}l\beta}\langle\bar{c}_{\bar{p}}c_{v}\bar{c}_{\bar{w}}c_{l}\rangle
\end{eqnarray*}
\emph{i.e.}, multiplying by $\left[G_{0}\right]_{m\bar{i}}$ and using
definitions (\ref{eq:def_G}) and (\ref{eq:def_chi3_tilde}): 
\begin{equation}
G_{m\bar{p}}=\left[G_{0}\right]_{m\bar{p}}-\left[G_{0}\right]_{m\bar{i}}U_{\alpha\beta}\lambda_{\bar{i}v\alpha}\tilde{\chi}_{v\bar{p}\beta}\label{eq:EOM_G}
\end{equation}
Using (\ref{eq:chi_chi_tilde}), we can rewrite this as:
\begin{equation}
G_{m\bar{p}}=\left[G_{0}\right]_{m\bar{p}}-\left[G_{0}\right]_{m\bar{i}}\lambda_{\bar{i}v\alpha}\chi_{v\bar{p}\alpha}\label{eq:EOM_G_chi}
\end{equation}

\subsubsection{Bosonic propagator $W$}

Specializing Eq. (\ref{eq:SD_bosonic}) for $h[\phi]\equiv\phi_{\alpha}-\varphi_{\alpha}$,
we find:

\begin{align*}
 & \left\langle \delta_{\gamma\alpha}-\lambda_{\bar{u}v\gamma}\bar{c}_{\bar{u}}c_{v}\left(\phi_{\alpha}-\varphi_{\alpha}\right)\right\rangle \\
 & =-\left[U^{-1}\right]_{\gamma\beta}\left\langle \left(\phi_{\beta}-\varphi_{\beta}\right)\left(\phi_{\alpha}-\varphi_{\alpha}\right)\right\rangle 
\end{align*}
whence:
\[
-\langle\left(\phi_{\delta}-\varphi_{\delta}\right)\left(\phi_{\alpha}-\varphi_{\alpha}\right)\rangle=U_{\delta\alpha}-U_{\delta\gamma}\lambda_{\bar{u}v\gamma}\langle\bar{c}_{\bar{u}}c_{v}\left(\phi_{\alpha}-\varphi_{\alpha}\right)\rangle
\]
\emph{i.e.}, using definitions (\ref{eq:def_W}-\ref{eq:chi3_nc_def}):

\begin{equation}
W_{\delta\alpha}=U_{\delta\alpha}+U_{\delta\gamma}\lambda_{\bar{u}v\gamma}\chi_{v\bar{u}\alpha}\label{eq:EOM_W}
\end{equation}

\subsubsection{General formulae for the self-energy and polarization}

Identifying $\Sigma$ and $P$ from the Dyson equations (\ref{eq:Dyson_Sigma}-\ref{eq:Dyson_P})
and (\ref{eq:EOM_G_chi}-\ref{eq:EOM_W}) yields:\begin{subequations}
\begin{align}
\Sigma_{\bar{i}j}G_{j\bar{p}} & =-\lambda_{\bar{i}v\alpha}\chi_{v\bar{p}\alpha}^{\mathrm{nc}}\label{eq:Sigma_G}\\
P_{\gamma\beta}W_{\beta\alpha} & =\lambda_{\bar{u}v\gamma}\chi_{v\bar{u}\alpha}\label{eq:PW}
\end{align}
\end{subequations}whence:\begin{subequations}
\begin{align}
\Sigma_{\bar{i}k} & =-\lambda_{\bar{i}v\alpha}\chi_{v\bar{p}\alpha}^{\mathrm{nc}}\left[G^{-1}\right]_{\bar{p}k}\label{eq:Sigma_EOM}\\
P_{\gamma\delta} & =\lambda_{\bar{u}v\gamma}\chi_{v\bar{u}\alpha}\left[W^{-1}\right]_{\alpha\delta}\label{eq:P_EOM}
\end{align}
\end{subequations}Using the definition of the three-leg vertex, Eq.
(\ref{eq:Lambda_def}), we find:

\begin{align*}
\Sigma_{\bar{i}j} & =-\lambda_{\bar{i}k\alpha}G_{k\bar{l}}W_{\alpha\beta}\Lambda_{\bar{l}j\beta}+\lambda_{\bar{i}j\alpha}\varphi_{\alpha}\\
P_{\alpha\beta} & =\lambda_{\bar{i}k\alpha}G_{j\bar{i}}G_{k\bar{l}}\Lambda_{\bar{l}j\beta}
\end{align*}
which are the formulae (\ref{eq:Sigma_Hedin}-\ref{eq:P_Hedin}) we
have derived using functionals in section \ref{sub:Functional-derivation}.

\section{Details of some calculations}

\subsection{Simplification of $\Sigma$ and $P$ in the homogeneous phase\label{sub:Simplification-of-Sigma_and_P_homog}}

In the normal, paramagnetic phase, Eqs (\ref{eq:Sigma_Hedin}-\ref{eq:P_Hedin})
can be simplified, namely

\begin{eqnarray*}
\Sigma_{\bar{u}v} & = & -\left(\sigma_{\sigma_{u}\sigma_{w}}^{I_{\alpha}}\delta_{i_{u}i_{\alpha}}\delta_{i_{u}i_{w}}\right)\left(G_{i_{w}i_{x}}\delta_{\sigma_{w}\sigma_{x}}\right)W_{i_{\alpha}i_{\beta}}^{\eta(I_{\alpha})}\Lambda_{i_{x}i_{v}i_{\beta}}^{\eta(I_{\beta})}\sigma_{\sigma_{x}\sigma_{v}}^{I_{\beta}}\\
 &  & +\left(\sigma_{\sigma_{u}\sigma_{v}}^{I_{\alpha}}\delta_{i_{u}i_{\alpha}}\delta_{i_{u}i_{v}}\right)\varphi_{i_{\alpha}}^{\eta(I_{\alpha})}\\
 & = & -\left(\sigma_{\sigma_{u}\sigma_{w}}^{I_{\alpha}}\sigma_{\sigma_{w}\sigma_{v}}^{I_{\alpha}}\right)G_{i_{w}i_{x}}W_{i_{\alpha}i_{\beta}}^{\eta(I_{\alpha})}\Lambda_{i_{x}i_{v}i_{\beta}}^{\eta(I_{\beta})}+\sigma_{\sigma_{u}\sigma_{v}}^{I_{\alpha}}\varphi_{i_{u}}^{\eta(I_{\alpha})}\delta_{i_{u}i_{v}}\\
 & = & -\left(\delta_{\sigma_{u}\sigma_{w}}\delta_{\sigma_{w}\sigma_{v}}\right)G_{i_{w}i_{x}}W_{i_{\alpha}i_{\beta}}^{\eta(I_{\alpha})}\Lambda_{i_{x}i_{v}i_{\beta}}^{\eta(I_{\beta})}\\
 &  & -\left(2\delta_{\sigma_{u}\sigma_{v}}\delta_{\sigma_{w}\sigma_{w}}-\delta_{\sigma_{u}\sigma_{w}}\delta_{\sigma_{w}\sigma_{v}}\right)G_{i_{w}i_{x}}W_{i_{\alpha}i_{\beta}}^{\eta(I_{\alpha})}\Lambda_{i_{x}i_{v}i_{\beta}}^{\eta(I_{\beta})}\\
 &  & +\sigma_{\sigma_{u}\sigma_{v}}^{I_{\alpha}}\varphi_{i_{u}}^{\eta(I_{\alpha})}\delta_{i_{u}i_{v}}\\
 & = & \Sigma_{i_{u}i_{v}}\delta_{\sigma_{u}\sigma_{v}}
\end{eqnarray*}
which yields Eq. (\ref{eq:Sigma_Hedin_homogeneous}). Similarly:

\begin{eqnarray*}
P_{\alpha\beta} & = & \sigma_{\sigma_{u}\sigma_{w}}^{I_{\alpha}}\delta_{i_{u}i_{w}}\delta_{i_{u}i_{\alpha}}G_{i_{v}i_{u}}\delta_{\sigma_{v}\sigma_{u}}G_{i_{w}i_{x}}\delta_{\sigma_{w}\sigma_{x}}\sigma_{\sigma_{x}\sigma_{v}}^{I_{\beta}}\Lambda_{i_{x}i_{v}i_{\beta}}^{\eta(I_{\beta})}\\
 & = & \mathrm{tr}\left[\left(\sigma^{I_{\alpha}}\right)^{2}\right]G_{i_{v}i_{u}}G_{i_{w}i_{x}}\Lambda_{i_{x}i_{v}i_{\beta}}^{\eta(I_{\beta})}\delta_{I_{\alpha}I_{\beta}}\\
 & = & P_{i_{\alpha}i_{\beta}}^{\eta(I_{\alpha})}\delta_{I_{\alpha}I_{\beta}}
\end{eqnarray*}
which yields Eq. (\ref{eq:P_Hedin_homogeneous}).

\subsection{Decomposition of $\Sigma$ and $P$\label{sub:Decomposition-of-sigma_and_P}}

Starting from (\ref{eq:Sigma_GWL}), one can rewrite:\begin{widetext}

\begin{eqnarray*}
\Sigma(\mathbf{k},i\omega) & = & -\sum_{\eta}m_{\eta}\sum_{\mathbf{q},i\Omega}\left(\tilde{G}(\mathbf{k}+\mathbf{q},i\omega+i\Omega)+G_{\mathrm{loc}}(i\omega)\right)\left(\tilde{W}^{\eta}(\mathbf{q},i\Omega)+W_{\mathrm{loc}}^{\eta}(i\Omega)\right)\Lambda_{\mathrm{imp}}^{\eta}(i\omega,i\Omega)\\
 & = & -\sum_{\eta}m_{\eta}\sum_{\mathbf{q},i\Omega}\tilde{G}(\mathbf{k}+\mathbf{q},i\omega+i\Omega)\tilde{W}^{\eta}(\mathbf{q},i\Omega)\Lambda_{\mathrm{imp}}^{\eta}(i\omega,i\Omega)-\sum_{\eta}m_{\eta}\sum_{i\Omega}G_{\mathrm{loc}}(i\omega+i\Omega)W_{\mathrm{loc}}^{\eta}(i\Omega)\Lambda_{\mathrm{imp}}^{\eta}(i\omega,i\Omega)\\
 & = & -\sum_{\eta}m_{\eta}\sum_{\mathbf{q},i\Omega}\tilde{G}(\mathbf{k}+\mathbf{q},i\omega+i\Omega)\tilde{W}^{\eta}(\mathbf{q},i\Omega)\Lambda_{\mathrm{imp}}^{\eta}(i\omega,i\Omega)+\Sigma_{\mathrm{imp}}(i\omega)
\end{eqnarray*}
\end{widetext}This yields (\ref{eq:Sigma_decomp}). An analogous
calculation yields (\ref{eq:P_decomp}).

\section{Atomic Limit\label{sec:Atomic-Limit}}

In this section, we derive the expression for the three-leg vertex
in the atomic limit. We proceed in two steps. First, we use the Lehmann
representation of the three-point correlation function in the case
of a single atomic site to compute the expression for the three-point
correlation function in the atomic limit. We then amputate the legs
to find the expression of the vertex function.

\begin{widetext}

\subsection{Three-point correlation function in the atomic limit}

\subsubsection{Full correlator $\chi$}

We use Lehmann's representation (Eq (\ref{eq:Lehmann_3leg})) to compute
the exact three-point correlation function $\hat{\tilde{\chi}}_{\sigma_{1}\sigma_{2}\sigma_{3}}(i\omega_{1},i\omega_{2})$
in the atomic limit, \emph{i.e} when the eigenvectors and corresponding
eigenenergies are (at half-filling)
\begin{eqnarray*}
|0\rangle\; & \rightarrow & \;\epsilon_{0}=0\\
|\uparrow\rangle\; & \rightarrow & \;\epsilon_{\uparrow}=-U/2\\
|\downarrow\rangle\; & \rightarrow & \;\epsilon_{\downarrow}=-U/2\\
|\uparrow\downarrow\rangle\; & \rightarrow & \;\epsilon_{\uparrow\downarrow}=0
\end{eqnarray*}
If $n_{3}$ is a ``particle-hole'' term (\emph{i.e} of the form
$c_{\sigma}^{\dagger}c_{\sigma}$), then the matrix element $\langle k|n_{3}|i\rangle$
selects states with the same occupation and same spin, so that:
\begin{eqnarray*}
\hat{\tilde{\chi}}_{\sigma_{1}\sigma_{2}\sigma_{3}}(i\omega_{1},i\omega_{2}) & = & \frac{1}{Z}\sum_{ij}\sum_{p}\sigma(p)\langle i|O_{p\sigma_{1}}|j\rangle\langle j|O_{p\sigma_{2}}|i\rangle\langle i|n_{\sigma_{3}}|i\rangle f_{iji}(\omega_{p1},\omega_{p2})\\
 & = & \frac{1}{Z}\sum_{ij}\langle i|c_{\sigma_{1}}|j\rangle\langle j|c_{\sigma_{2}}^{\dagger}|i\rangle\langle i|n_{\sigma_{3}}|i\rangle f_{iji}(\omega_{1},\omega_{2})-\sum_{ij}\langle i|c_{\sigma_{2}}^{\dagger}|j\rangle\langle j|c_{\sigma_{1}}|i\rangle\langle i|n_{\sigma_{3}}|i\rangle f_{iji}(\omega_{2},\omega_{1})
\end{eqnarray*}
Furthermore,
\begin{eqnarray*}
f_{iji}(\omega_{2},\omega_{1}) & = & \frac{1}{i\omega_{1}+\epsilon_{j}-\epsilon_{i}}\frac{e^{-\beta\epsilon_{j}}+e^{-\beta\epsilon_{i}}}{i\omega_{2}+\epsilon_{i}-\epsilon_{j}}+\beta\frac{e^{-\beta\epsilon_{i}}}{i\omega_{1}+\epsilon_{j}-\epsilon_{i}}\delta_{i\omega_{1}+i\omega_{2}}\\
f_{jij}(\omega_{1},\omega_{2}) & = & \frac{1}{i\omega_{2}+\epsilon_{i}-\epsilon_{j}}\frac{e^{-\beta\epsilon_{j}}+e^{-\beta\epsilon_{i}}}{i\omega_{1}+\epsilon_{j}-\epsilon_{i}}+\beta\frac{e^{-\beta\epsilon_{j}}}{-i\omega_{1}+\epsilon_{i}-\epsilon_{j}}\delta_{i\omega_{1}+i\omega_{2}}
\end{eqnarray*}
whence

\[
f_{iji}(\omega_{2},\omega_{1})=f_{jij}(\omega_{1},\omega_{2})+\beta\frac{e^{-\beta\epsilon_{i}}+e^{-\beta\epsilon_{j}}}{i\omega_{1}+\epsilon_{j}-\epsilon_{i}}\delta_{i\omega_{1}+i\omega_{2}}
\]
Using this identity and swapping the dummy indices in the second term,
one gets:

\begin{eqnarray*}
\hat{\tilde{\chi}}_{\sigma_{1}\sigma_{2}\sigma_{3}}(i\omega_{1},i\omega_{2}) & = & \frac{1}{Z}\sum_{ij}\langle i|c_{\sigma_{1}}|j\rangle\langle j|c_{\sigma_{2}}^{\dagger}|i\rangle\left\{ \langle i|n_{\sigma_{3}}|i\rangle-\langle j|n_{\sigma_{3}}|j\rangle\right\} f_{iji}(\omega_{1},\omega_{2})\\
 &  & -\beta\sum_{ij}\langle j|c_{\sigma_{2}}^{\dagger}|i\rangle\langle i|c_{\sigma_{1}}|j\rangle\langle j|n_{\sigma_{3}}|j\rangle\frac{e^{-\beta\epsilon_{i}}+e^{-\beta\epsilon_{j}}}{i\omega_{1}+\epsilon_{i}-\epsilon_{j}}\delta_{i\omega_{1}+i\omega_{2}}
\end{eqnarray*}
Obviously, $\sigma_{1}=\sigma_{2}$, and $i=|\uparrow\downarrow\rangle$
and $j=|0\rangle$ do not contribute, i.e, after defining $\hat{\tilde{\chi}}_{\sigma\sigma'}\equiv\hat{\tilde{\chi}}_{\text{\ensuremath{\sigma\sigma\sigma}'}}$
and $f_{ij}\equiv f_{iji}=f_{ij}^{reg}+\beta\frac{e^{-\beta\epsilon_{i}}}{i\omega_{2}+\epsilon_{j}-\epsilon_{i}}\delta_{i\omega_{1}+i\omega_{2}}$:
\begin{eqnarray*}
\hat{\tilde{\chi}}_{\sigma\sigma'}(i\omega_{1},i\omega_{2}) & = & \hat{\tilde{\chi}}_{\sigma\sigma'}^{1}(i\omega_{1},i\omega_{2})+\hat{\tilde{\chi}}_{\sigma\sigma'}^{2}(i\omega_{1},i\omega_{2})
\end{eqnarray*}
with

\begin{eqnarray*}
\hat{\tilde{\chi}}_{\sigma\sigma'}^{1}(i\omega_{1},i\omega_{2}) & \equiv & \frac{1}{Z}\sum_{i=|0\rangle,|\uparrow\rangle,|\downarrow\rangle}\sum_{j=|\uparrow\rangle,|\downarrow\rangle,|\uparrow\downarrow\rangle}|\langle i|c_{\sigma}|j\rangle|^{2}\left\{ \langle i|n_{\sigma'}|i\rangle-\langle j|n_{\sigma'}|j\rangle\right\} f_{ij}(\omega_{1},\omega_{2})\\
\hat{\tilde{\chi}}_{\sigma\sigma'}^{1}(i\omega_{1},i\omega_{2}) & \equiv & -\beta\frac{1}{Z}\sum_{ij}|\langle i|c_{\sigma}|j\rangle|^{2}\langle j|n_{\sigma'}|j\rangle\frac{e^{-\beta\epsilon_{i}}+e^{-\beta\epsilon_{j}}}{i\omega_{1}+\epsilon_{i}-\epsilon_{j}}\delta_{i\omega_{1}+i\omega_{2}}
\end{eqnarray*}
One also sees that: $\hat{\tilde{\chi}}_{\uparrow\downarrow}=\hat{\tilde{\chi}}_{\downarrow\uparrow}$
and $\hat{\tilde{\chi}}_{\uparrow\uparrow}=\hat{\tilde{\chi}}_{\downarrow\downarrow}$.
Out of the nine remaining terms, we can see that only the terms where
$i$ and $j$ are states with a difference of occupation of one electron
are nonzero:
\begin{eqnarray*}
Z\hat{\tilde{\chi}}_{\sigma\sigma'}^{1}(i\omega_{1},i\omega_{2}) & = & |\langle0|c_{\sigma}|\uparrow\rangle|^{2}\left\{ \langle0|n_{\sigma'}|0\rangle-\langle\uparrow|n_{\sigma'}|\uparrow\rangle\right\} f_{0\uparrow}(\omega_{1},\omega_{2})\\
 & + & |\langle0|c_{\sigma}|\downarrow\rangle|^{2}\left\{ \langle0|n_{\sigma'}|0\rangle-\langle\downarrow|n_{\sigma'}|\downarrow\rangle\right\} f_{0\downarrow}(\omega_{1},\omega_{2})\\
 & + & |\langle\uparrow|c_{\sigma}|\uparrow\downarrow\rangle|^{2}\left\{ \langle\uparrow|n_{\sigma'}|\uparrow\rangle-\langle\uparrow\downarrow|n_{\sigma'}|\uparrow\downarrow\rangle\right\} f_{\uparrow,\uparrow\downarrow}(\omega_{1},\omega_{2})\\
 & + & |\langle\downarrow|c_{\sigma}|\uparrow\downarrow\rangle|^{2}\left\{ \langle\downarrow|n_{\sigma'}|\downarrow\rangle-\langle\uparrow\downarrow|n_{\sigma'}|\uparrow\downarrow\rangle\right\} f_{\downarrow,\uparrow\downarrow}(\omega_{1},\omega_{2})
\end{eqnarray*}

and 
\begin{eqnarray*}
Z\hat{\tilde{\chi}}_{\sigma\sigma'}^{2}(i\omega_{1},i\omega_{2}) & = & -\beta|\langle0|c_{\sigma}|\uparrow\rangle|^{2}\langle\uparrow|n_{\sigma'}|\uparrow\rangle\frac{1+e^{-\beta\epsilon}}{i\omega_{1}-\epsilon}\delta_{i\omega_{1}+i\omega_{2}}\\
 & - & \beta|\langle0|c_{\sigma}|\downarrow\rangle|^{2}\langle\downarrow|n_{\sigma'}|\downarrow\rangle\frac{1+e^{-\beta\epsilon}}{i\omega_{1}-\epsilon}\delta_{i\omega_{1}+i\omega_{2}}\\
 & - & \beta|\langle\uparrow|c_{\sigma}|\uparrow\downarrow\rangle|^{2}\langle\uparrow\downarrow|n_{\sigma'}|\uparrow\downarrow\rangle\frac{1+e^{-\beta\epsilon}}{i\omega_{1}+\epsilon}\delta_{i\omega_{1}+i\omega_{2}}\\
 & - & \beta|\langle\downarrow|c_{\sigma}|\uparrow\downarrow\rangle|^{2}\langle\uparrow\downarrow|n_{\sigma'}|\uparrow\downarrow\rangle\frac{1+e^{-\beta\epsilon}}{i\omega_{1}+\epsilon}\delta_{i\omega_{1}+i\omega_{2}}
\end{eqnarray*}
Thus, on the one hand:

\begin{eqnarray}
Z\hat{\tilde{\chi}}_{\uparrow\downarrow}^{1}(i\omega_{1},i\omega_{2}) & = & 0\label{eq:chi_up_down_1}\\
Z\hat{\tilde{\chi}}_{\uparrow\downarrow}^{2}(i\omega_{1},i\omega_{2}) & = & -\beta\left(1+e^{-\beta\epsilon}\right)\delta_{i\omega_{1}+i\omega_{2}}\frac{1}{i\omega_{1}+\epsilon}\label{eq:chi_up_down_2}
\end{eqnarray}

\emph{i.e}, switching back from $\hat{\tilde{\chi}}(i\omega_{1},i\omega_{2})$
to $\tilde{\chi}(i\omega,i\Omega)$:
\begin{equation}
\chi_{\uparrow\downarrow}(i\omega,i\Omega)=-\beta\langle n_{\sigma}\rangle\frac{1}{i\omega-U/2}\delta_{i\Omega}\label{eq:chi_up_down}
\end{equation}
On the other hand:

\begin{eqnarray*}
Z\hat{\tilde{\chi}}_{\uparrow\uparrow}^{1}(i\omega_{1},i\omega_{2}) & = & -f_{0\uparrow}(\omega_{1},\omega_{2})-f_{\downarrow,\uparrow\downarrow}(\omega_{1},\omega_{2})\\
 & = & -\frac{e^{-\beta\epsilon_{0}}+e^{-\beta\epsilon_{\uparrow}}}{\left(i\omega_{1}+\epsilon_{\uparrow}-\epsilon_{0}\right)\left(i\omega_{2}+\epsilon_{0}-\epsilon_{\uparrow}\right)}-\beta\frac{1}{i\omega_{2}+\epsilon}\delta_{i\omega_{1}+i\omega_{2}}\\
 &  & -\frac{e^{-\beta\epsilon_{\downarrow}}+e^{-\beta\epsilon_{\uparrow\downarrow}}}{\left(i\omega_{1}+\epsilon_{\uparrow\downarrow}-\epsilon_{\downarrow}\right)\left(i\omega_{2}+\epsilon_{\downarrow}-\epsilon_{\uparrow\downarrow}\right)}-\beta\frac{e^{-\beta\epsilon}}{i\omega_{2}-\epsilon}\delta_{i\omega_{1}+i\omega_{2}}\\
 & = & \underbrace{-\frac{1+e^{\beta U/2}}{\left(i\omega_{1}-U/2\right)\left(i\omega_{2}+U/2\right)}-\frac{1+e^{\beta U/2}}{\left(i\omega_{1}+U/2\right)\left(i\omega_{2}-U/2\right)}}_{\equiv Z\tilde{\chi}_{\uparrow\uparrow}^{1,reg}}-\left[\frac{1}{i\omega_{2}+\epsilon}+\frac{e^{-\beta\epsilon}}{i\omega_{2}-\epsilon}\right]\beta\delta_{i\omega_{1}+i\omega_{2}}
\end{eqnarray*}

\emph{i.e}
\begin{eqnarray*}
\hat{\tilde{\chi}}_{\uparrow\uparrow}^{1reg}(i\omega_{1},i\omega_{2}) & = & -\frac{1}{2}\frac{\left(i\omega_{1}+U/2\right)\left(i\omega_{2}-U/2\right)+\left(i\omega_{1}-U/2\right)\left(i\omega_{2}+U/2\right)}{\left(\left(i\omega_{1}\right)^{2}-U^{2}/4\right)\left(\left(i\omega_{2}\right)^{2}-U^{2}/4\right)}\\
 & = & -\frac{1}{2}\frac{i\omega_{1}i\omega_{2}+U/2\left(i\omega_{2}-i\omega_{1}\right)-U^{2}/4+\left[i\omega_{1}i\omega_{2}+U/2\left(-i\omega_{2}+i\omega_{1}\right)-U^{2}/4\right]}{\left(\left(i\omega_{1}\right)^{2}-U^{2}/4\right)\left(\left(i\omega_{2}\right)^{2}-U^{2}/4\right)}\\
 & = & -\frac{i\omega_{1}i\omega_{2}-U^{2}/4}{\left(\left(i\omega_{1}\right)^{2}-U^{2}/4\right)\left(\left(i\omega_{2}\right)^{2}-U^{2}/4\right)}
\end{eqnarray*}
and

\[
\hat{\tilde{\chi}}_{\uparrow\uparrow}^{2}(i\omega_{1},i\omega_{2})=-\frac{\beta}{2}\left(\frac{1}{i\omega_{1}+U/2}+\frac{1}{i\omega_{1}-U/2}\right)\delta_{\Omega}
\]
Thus,
\begin{eqnarray}
\hat{\tilde{\chi}}_{\uparrow\uparrow}(i\omega_{1},i\omega_{2}) & = & -\frac{i\omega_{1}i\omega_{2}-U^{2}/4}{\left(\left(i\omega_{1}\right)^{2}-U^{2}/4\right)\left(\left(i\omega_{2}\right)^{2}-U^{2}/4\right)}-\frac{\beta}{2}\left(\frac{1}{i\omega_{1}+U/2}+\frac{1}{i\omega_{1}-U/2}\right)\delta_{\Omega}\nonumber \\
 &  & +\frac{1}{2\left(1+e^{-\beta\epsilon}\right)}\left[\frac{1}{i\omega_{1}+U/2}+\frac{e^{-\beta\epsilon}}{i\omega_{1}-U/2}\right]\beta\delta_{i\Omega}\nonumber \\
 & = & -\frac{i\omega_{1}i\omega_{2}-U^{2}/4}{\left(\left(i\omega_{1}\right)^{2}-U^{2}/4\right)\left(\left(i\omega_{2}\right)^{2}-U^{2}/4\right)}-\frac{\beta}{2\left(1+e^{-\beta\epsilon}\right)}\left[\frac{e^{-\beta\epsilon}}{i\omega_{1}+U/2}+\frac{1}{i\omega_{1}-U/2}\right]\delta_{i\Omega}\nonumber \\
 & = & -G(i\omega_{1})G(i\omega_{2})+\frac{U^{2}/4}{\left(\left(i\omega_{1}\right)^{2}-U^{2}/4\right)\left(\left(i\omega_{2}\right)^{2}-U^{2}/4\right)}-\frac{\beta}{2\left(1+e^{-\beta\epsilon}\right)}\left[\frac{e^{-\beta\epsilon}}{i\omega_{1}+U/2}+\frac{1}{i\omega_{1}-U/2}\right]\delta_{i\Omega}\label{eq:chi_upup_expr}
\end{eqnarray}

\subsubsection{Connected part $\chi^{\mathrm{c}}$}

The connected part is defined as:
\begin{eqnarray*}
\tilde{\chi}_{\sigma\sigma'}^{\mathrm{c}}(\tau,\tau') & \equiv & \tilde{\chi}_{\sigma\sigma'}(\tau,\tau')-\langle c_{\sigma}(\tau)c_{\sigma}^{\dagger}(\tau')n_{\sigma'}\rangle_{\mathrm{disc}}\\
 & = & \tilde{\chi}_{\sigma\sigma'}(\tau,\tau')-\langle c_{\sigma}(\tau)c_{\sigma}^{\dagger}(\tau')\rangle\langle n_{\sigma'}\rangle\\
 & = & \tilde{\chi}_{\sigma\sigma'}(\tau,\tau')+G_{\sigma}(\tau-\tau')\langle n_{\sigma'}\rangle
\end{eqnarray*}
whence

\begin{equation}
\tilde{\chi}_{\sigma\sigma'}^{\mathrm{c}}(\omega,\Omega)=\tilde{\chi}_{\sigma\sigma'}(\omega,\Omega)+\beta G_{\sigma}(i\omega)\langle n_{\sigma'}\rangle\delta_{\Omega}\label{eq:chi_conn_expr}
\end{equation}
This yields:\begin{subequations}

\begin{align}
\tilde{\chi}_{\uparrow\downarrow}^{\mathrm{c}}(i\omega,\Omega) & =\frac{\beta\langle n_{\downarrow}\rangle}{2}\left[\frac{1}{i\omega+U/2}-\frac{1}{i\omega-U/2}\right]\delta_{\Omega}\label{eq:chi_conn_up_down}\\
\hat{\tilde{\chi}}_{\uparrow\uparrow}^{\mathrm{c}}(i\omega_{1},i\omega_{2}) & =-G(i\omega_{1})G(i\omega_{2})+\frac{U^{2}/4}{\left\{ \left(i\omega_{1}\right)^{2}-U^{2}/4\right\} \left\{ \left(i\omega_{2}\right)^{2}-U^{2}/4\right\} }+A(\omega_{1})\delta_{i\Omega}\label{eq:chi_conn_up_up}
\end{align}
\end{subequations}with
\begin{eqnarray*}
A(\omega_{1}) & \equiv & \beta\langle n_{\sigma}\rangle\left\{ G_{\uparrow}(i\omega_{1})-\frac{1}{1+e^{\beta U/2}}\left[\frac{e^{\beta U/2}}{i\omega_{1}+U/2}+\frac{1}{i\omega_{1}-U/2}\right]\right\} \\
 & = & \beta\langle n_{\sigma}\rangle\left\{ \frac{1}{i\omega_{1}-U/2}\left\{ \frac{1}{2}-\frac{1}{1+e^{\beta U/2}}\right\} +\frac{1}{i\omega_{1}+U/2}\left\{ \frac{1}{2}-\frac{e^{\beta U/2}}{1+e^{\beta U/2}}\right\} \right\} \\
 & = & \frac{\beta\langle n_{\sigma}\rangle}{2}\tanh\left(\beta U/4\right)\left[\frac{1}{i\omega_{1}-U/2}-\frac{1}{i\omega_{1}+U/2}\right]\\
 & = & \frac{\beta\langle n_{\sigma}\rangle}{2}\tanh\left(\beta U/4\right)\frac{U}{\left(i\omega_{1}\right)^{2}-U^{2}/4}
\end{eqnarray*}
We can check expression (\ref{eq:chi_conn_up_down}) and get some
physical intuition by computing the self-energy from the equation
of motion for $G$ (\emph{i.e} Eq. (\ref{eq:Sigma_G}) specialized
for the atomic limit). The self-energy $\Sigma$ can be decomposed
into a Hartree contribution and a contribution beyond Hartree:

\[
\Sigma(i\omega)\equiv\frac{U}{2}+\frac{U^{2}}{4i\omega}=\Sigma_{H}(i\omega)+\Sigma_{\mathrm{bH}}(i\omega)
\]
On the one hand, one can notice that:

\begin{equation}
\Sigma(i\omega)G(i\omega)=-U\frac{1}{\beta}\sum_{\Omega}\tilde{\chi}_{\uparrow\downarrow}(\omega,\Omega)\label{eq:Sigma_G_chi_AL}
\end{equation}
On the other hand:

\begin{eqnarray*}
\Sigma_{\mathrm{bH}}(i\omega) & \equiv & \Sigma(i\omega)-\Sigma_{H}=-U\frac{1}{\beta}\sum_{\Omega}\frac{\tilde{\chi}_{\uparrow\downarrow}(\omega,\Omega)}{G(i\omega)}-U\langle n_{\sigma}\rangle\\
 & = & -U\frac{1}{\beta}\sum_{\Omega}\left\{ \frac{\tilde{\chi}_{\uparrow\downarrow}(\omega,\Omega)}{G(i\omega)}+\langle n_{\sigma}\rangle\beta\delta_{\Omega}\right\} \\
 & = & -U\frac{1}{\beta}\sum_{\Omega}\left\{ \frac{\tilde{\chi}_{\uparrow\downarrow}(\omega,\Omega)+G(i\omega)\langle n_{\sigma}\rangle\beta\delta_{\Omega}}{G(i\omega)}\right\} \\
 & = & -U\frac{1}{\beta}\sum_{\Omega}\frac{\tilde{\chi}_{\uparrow\downarrow}^{\mathrm{c}}(\omega,\Omega)}{G(i\omega)}
\end{eqnarray*}
\emph{i.e}:
\begin{equation}
G(i\omega)\Sigma_{\mathrm{bH}}(i\omega)=-U\frac{1}{\beta}\sum_{\Omega}\tilde{\chi}_{\uparrow\downarrow}^{\mathrm{c}}(\omega,\Omega)\label{eq:Sigma_bH}
\end{equation}
which is to be contrasted with (\ref{eq:Sigma_G_chi_AL}).

\subsubsection{Expressions in charge and spin channels}

Let us now transform from the $(\uparrow,\downarrow)$ space to the
$(\mathrm{ch},$$\mathrm{sp})$ space:\begin{subequations}

\begin{eqnarray*}
\tilde{\chi}^{\mathrm{ch},\mathrm{c}}(i\omega_{1},i\omega_{2}) & \equiv & \tilde{\chi}_{\uparrow\uparrow}^{\mathrm{c}}+\tilde{\chi}_{\uparrow\downarrow}^{\mathrm{c}}\\
 & = & -G(i\omega_{1})G(i\omega_{2})+\frac{U^{2}/4}{\left\{ \left(i\omega_{1}\right)^{2}-U^{2}/4\right\} \left\{ \left(i\omega_{2}\right)^{2}-U^{2}/4\right\} }\\
 &  & +\frac{\beta\langle n_{\sigma}\rangle}{2}\left\{ \tanh\left(\beta U/4\right)\left[\frac{1}{i\omega_{1}-U/2}-\frac{1}{i\omega_{1}+U/2}\right]+\left[\frac{1}{i\omega_{1}+U/2}-\frac{1}{i\omega_{1}-U/2}\right]\right\} \delta_{i\Omega}\\
\tilde{\chi}^{\mathrm{sp},\mathrm{c}}(i\omega_{1},i\omega_{2}) & \equiv & \tilde{\chi}_{\uparrow\uparrow}^{\mathrm{c}}-\tilde{\chi}_{\uparrow\downarrow}^{\mathrm{c}}\\
 & = & -G(i\omega_{1})G(i\omega_{2})+\frac{U^{2}/4}{\left\{ \left(i\omega_{1}\right)^{2}-U^{2}/4\right\} \left\{ \left(i\omega_{2}\right)^{2}-U^{2}/4\right\} }\\
 &  & +\frac{\beta\langle n_{\sigma}\rangle}{2}\left\{ \tanh\left(\beta U/4\right)\left[\frac{1}{i\omega_{1}-U/2}-\frac{1}{i\omega_{1}+U/2}\right]-\left[\frac{1}{i\omega_{1}+U/2}-\frac{1}{i\omega_{1}-U/2}\right]\right\} \delta_{i\Omega}
\end{eqnarray*}
\end{subequations}Simplifying and transposing to $i\omega,i\Omega$
variables, one gets (using $G^{\mathrm{at}}(-i\omega)=-G^{\mathrm{at}}(i\omega)$):\begin{subequations}

\begin{eqnarray}
\tilde{\chi}^{\mathrm{ch},\mathrm{c}}(i\omega,i\Omega) & = & G(i\omega)G(i\Omega)+\frac{U^{2}/4}{\left\{ \left(i\omega\right)^{2}-U^{2}/4\right\} \left\{ \left(i\omega+i\Omega\right)^{2}-U^{2}/4\right\} }\nonumber \\
 &  & +\frac{\beta\langle n_{\sigma}\rangle}{2}\frac{U}{\left(i\omega\right)^{2}-U^{2}/4}\left\{ \tanh\left(\beta U/4\right)-1\right\} \delta_{i\Omega}\label{eq:chi_ch_conn_AL}\\
\tilde{\chi}^{\mathrm{sp},\mathrm{c}}(i\omega,i\Omega) & = & G(i\omega)G(i\Omega)+\frac{U^{2}/4}{\left\{ \left(i\omega\right)^{2}-U^{2}/4\right\} \left\{ \left(i\omega+i\Omega\right)^{2}-U^{2}/4\right\} }\nonumber \\
 &  & +\frac{\beta\langle n_{\sigma}\rangle}{2}\frac{U}{\left(i\omega\right)^{2}-U^{2}/4}\left\{ \tanh\left(\beta U/4\right)+1\right\} \delta_{i\Omega}\label{eq:chi_sp_conn_AL}
\end{eqnarray}

\end{subequations}

\subsection{Vertex $\Lambda$}

The vertex is defined as the amputated connected correlation function
(see Eq. \ref{eq:Lambda_imp_from_chi}). We can easily compute the
``legs'' in the atomic limit:\begin{subequations}

\begin{eqnarray*}
G(i\omega)G(i\omega+i\Omega)\left(1-U^{\mathrm{ch}}\chi^{\mathrm{ch,c}}(i\Omega)\right) & = & \frac{i\omega\left(i\omega+i\Omega\right)}{\left(\left(i\omega\right)^{2}-U^{2}/4\right)\left(\left(i\omega+i\Omega\right)^{2}-U^{2}/4\right)}\left(1-\beta U\frac{1}{4}\frac{e^{-\beta U/4}}{\cosh(\beta U/4)}\delta_{\Omega}\right)\\
G(i\omega)G(i\omega+i\Omega)\left(1-U^{\mathrm{sp}}\chi^{\mathrm{sp,c}}(i\Omega)\right) & = & \frac{i\omega\left(i\omega+i\Omega\right)}{\left(\left(i\omega\right)^{2}-U^{2}/4\right)\left(\left(i\omega+i\Omega\right)^{2}-U^{2}/4\right)}\left(1+\beta U\frac{1}{4}\frac{e^{\beta U/4}}{\cosh(\beta U/4)}\delta_{\Omega}\right)
\end{eqnarray*}
\end{subequations}Hence:

\begin{eqnarray*}
\Lambda^{\mathrm{ch}}(i\omega,i\Omega) & = & \frac{\frac{U^{2}/4}{\left\{ \left(i\omega\right)^{2}-U^{2}/4\right\} \left\{ \left(i\omega+i\Omega\right)^{2}-U^{2}/4\right\} }+\frac{\beta\langle n_{\sigma}\rangle}{2}\frac{U}{\left(i\omega\right)^{2}-U^{2}/4}\left\{ \tanh\left(\beta U/4\right)-1\right\} \delta_{i\Omega}}{\frac{i\omega\left(i\omega+i\Omega\right)}{\left(\left(i\omega\right)^{2}-U^{2}/4\right)\left(\left(i\omega+i\Omega\right)^{2}-U^{2}/4\right)}\left(1-\beta U\frac{1}{4}\frac{e^{-\beta U/4}}{\cosh(\beta U/4)}\delta_{\Omega}\right)}\\
 &  & +\frac{1}{1-\beta U\frac{1}{4}\frac{e^{-\beta U/4}}{\cosh(\beta U/4)}\delta_{\Omega}}\\
 & = & \frac{U^{2}/4}{i\omega\left(i\omega+i\Omega\right)\left(1-\beta U\frac{1}{4}\frac{e^{-\beta U/4}}{\cosh(\beta U/4)}\delta_{\Omega}\right)}+\frac{\beta\langle n_{\sigma}\rangle}{2}\left\{ \frac{\left(i\omega\right)^{2}-U^{2}/4}{i\omega}\right\} ^{2}\frac{U}{\left(i\omega\right)^{2}-U^{2}/4}\frac{\tanh\left(\beta U/4\right)-1}{\left(1-\beta U\frac{1}{4}\frac{e^{-\beta U/4}}{\cosh(\beta U/4)}\right)}\delta_{i\Omega}\\
 &  & +\frac{1}{1-\beta U\frac{1}{4}\frac{e^{-\beta U/4}}{\cosh(\beta U/4)}\delta_{\Omega}}
\end{eqnarray*}
Simplifying, one finds the results: \begin{subequations}

\begin{eqnarray}
\Lambda^{\mathrm{ch}}(i\omega,i\Omega) & = & \frac{U^{2}/4}{i\omega\left(i\omega+i\Omega\right)\left(1-\beta U\frac{1}{4}\frac{e^{-\beta U/4}}{\cosh(\beta U/4)}\delta_{\Omega}\right)}\nonumber \\
 &  & +\frac{U\beta\langle n_{\sigma}\rangle}{2}\left\{ 1-\frac{U^{2}}{4\left(i\omega\right)^{2}}\right\} \frac{\tanh\left(\beta U/4\right)-1}{\left(1-\beta U\frac{1}{4}\frac{e^{-\beta U/4}}{\cosh(\beta U/4)}\right)}\delta_{i\Omega}\label{eq:Lambda_ch}\\
 &  & +\frac{1}{1-\beta U\frac{1}{4}\frac{e^{-\beta U/4}}{\cosh(\beta U/4)}\delta_{\Omega}}
\end{eqnarray}

\begin{eqnarray}
\Lambda^{\mathrm{sp}}(i\omega,i\Omega) & = & \frac{U^{2}/4}{i\omega\left(i\omega+i\Omega\right)\left(1+\beta U\frac{1}{4}\frac{e^{\beta U/4}}{\cosh(\beta U/4)}\delta_{\Omega}\right)}\nonumber \\
 &  & +\frac{U\beta\langle n_{\sigma}\rangle}{2}\left\{ 1-\frac{U^{2}}{4\left(i\omega\right)^{2}}\right\} \frac{\tanh\left(\beta U/4\right)+1}{\left(1+\beta U\frac{1}{4}\frac{e^{\beta U/4}}{\cosh(\beta U/4)}\right)}\delta_{i\Omega}\label{eq:Lambda_sp}\\
 &  & +\frac{1}{1+\beta U\frac{1}{4}\frac{e^{\beta U/4}}{\cosh(\beta U/4)}\delta_{\Omega}}
\end{eqnarray}

\end{subequations}\end{widetext}
\end{document}